\documentclass[manuscript]{aastex63}

\defcitealias{Clemens12a}{Paper~I}
\defcitealias{Clemens12b}{Paper~II}
\defcitealias{Clemens12c}{Paper~III}

\newcommand{\microns}{~$\mu$m}
\newcommand{\um}{$\mu$m}
\newcommand{\planck}{{\it Planck}\ }
\newcommand{\Gaia}{{\it Gaia}\ }
\newcommand{\gaia}{{\it Gaia}\ }
\newcommand{\wise}{{\it WISE}\ }
\newcommand{\WISE}{{\it WISE}\ }
\newcommand{\msx}{{\it MSX}\ }
\newcommand{\glimpse}{GLIMPSE\ }

\shorttitle{GPIPS DR4}
\shortauthors{Clemens et al.}

\begin{document}

\title{The Galactic Plane Infrared Polarization Survey (GPIPS): Data Release 4}

\author[0000-0002-9947-4956]{Dan P. Clemens}
\affiliation{Institute for Astrophysical Research, Boston University,
    725 Commonwealth Ave, Boston, MA 02215}

\author[0000-0003-3185-6066]{L.R. Cashman}
\affiliation{Institute for Astrophysical Research, Boston University,
    725 Commonwealth Ave, Boston, MA 02215}

\author[0000-0002-8261-9098]{C. Cerny}
\affiliation{Institute for Astrophysical Research, Boston University,
    725 Commonwealth Ave, Boston, MA 02215}

\author[0000-0002-4958-1382]{A. M. El-Batal}
\affiliation{Institute for Astrophysical Research, Boston University,
    725 Commonwealth Ave, Boston, MA 02215}
    
\author[0000-0001-7105-0994]{K. E. Jameson}
\affiliation{Institute for Astrophysical Research, Boston University,
    725 Commonwealth Ave, Boston, MA 02215}
\affiliation{CSIRO Astronomy and Space Science, 26 Dick Perry Avenue, Kensington, WA 6151, Australia}

\author[0000-0001-9662-3875]{R. Marchwinski}
\affiliation{Institute for Astrophysical Research, Boston University,
    725 Commonwealth Ave, Boston, MA 02215}

\author[0000-0003-4609-0630]{J. Montgomery}
\affiliation{Institute for Astrophysical Research, Boston University,
    725 Commonwealth Ave, Boston, MA 02215}

\author[0000-0002-4064-3436]{M. Pavel}
\affiliation{Institute for Astrophysical Research, Boston University,
    725 Commonwealth Ave, Boston, MA 02215}

\author{A. Pinnick}
\affiliation{Institute for Astrophysical Research, Boston University,
    725 Commonwealth Ave, Boston, MA 02215}

\author[0000-0003-2886-9341]{B. W. Taylor}
\affiliation{Institute for Astrophysical Research, Boston University,
    725 Commonwealth Ave, Boston, MA 02215}
\affiliation{Advanced Instrumentation Technology Centre,
The Australian National University, Cotter Road, Weston,
ACT 2611, Australia}

\correspondingauthor{Dan P. Clemens}
\email{clemens@bu.edu}

\begin{abstract}
The Galactic Plane Infrared Polarization Survey (GPIPS) seeks to characterize the magnetic field in the dusty Galactic disk using near-infrared stellar polarimetry. All GPIPS observations were completed using the 1.83~m Perkins telescope and Mimir instrument.  GPIPS observations surveyed 76~deg$^2$ of the northern Galactic plane, from Galactic longitudes 18 to 56\degr\ and latitudes $-1$ to $+$1\degr, in the $H$-band (1.6~$\mu$m).  
Surveyed stars span 7th to 16th mag, resulting in nearly 10~million stars with measured linear polarizations. Of these stars, ones with $m_H < 12.5$ mag and polarization percentage uncertainties under 2\% were judged to be high quality and number over one million. GPIPS data reveal plane-of-sky magnetic field orientations for numerous interstellar clouds for A$_V$ values to $\sim$30~mag. 
The average sky separation of stars with $m_H < 12.5$~mag is about 30~arcsec, or 
about 60 per {$\it Planck$} polarization resolution element. 
Matching to \Gaia DR2 showed the brightest GPIPS stars are red giants with distances in the 0.6--7.5~kpc range. Polarization orientations are mostly 
parallel to the Galactic disk, with some zones showing significant orientation departures. 
Changes in orientations are stronger as a function of Galactic longitude than of latitude.
Considered at 10~arcmin angular scales, directions that show the greatest polarization fractions and narrowest polarization position angle distributions are confined to about ten large, coherent structures that are not correlated with star forming clouds. The GPIPS polarimetric and photometric data products (Data Release 4 catalogs and images) are
publicly available for over 13~million stars. 
\end{abstract}


\section{Introduction}

The magnetic field in the cool, dusty, star-forming interstellar medium (ISM) of the Milky Way has not been well-described and the roles played by the magnetic field in cloud, clump/core, star, and planet formation and evolution are rather poorly understood. This is mostly due to a lack of high-quality, high-resolution data that can reveal the
magnetic field in such settings and can be used to characterize the importance of the magnetic field. Yet knowledge of the field properties, especially strength and orientation, is vital to assessing the relative roles of gravity, turbulence, and magnetic fields in star formation \citep[see review by][]{Crutcher12}.

Magnetic fields associated with the Milky Way were first revealed in the optical
starlight linear-polarization studies of \citet{Hiltner49a,Hiltner49b} and \citet{Hall49}, and the
later surveys of \citet{Mathewson70} and \citet{Klare77}, who found strong associations of starlight polarization orientations with the Galactic plane orientation. Linear polarization of the 
light from background stars occurs because the intervening ISM contains spinning dust 
grains that are aligned with local magnetic field directions via the Radiative Aligned 
Torques (RATs) mechanism \citep{Lazarian07, Andersson15}.
This results in anisotropic absorption (dichroism) of the isotropically emitted starlight. Sky-projected magnetic field orientations are revealed as the linear polarization orientations of the electric field vector amplitudes of the stellar radiation. Published optical polarization studies were collected in the compendium of \citet{Heiles00} and studied by \citet{Fosalba02}. However, as shown by the latter, the stars in the \citet{Heiles00} collection predominantly had distances out to only $\sim$1~kpc and exhibited only up to a few magnitudes of visual extinction A$_V$.  Additionally, the average sky sampling of the stars measured for polarization was about one star per two square degrees, much too sparse to permit study of all but the closest, largest, and least-extincted molecular clouds.

Studies of radio synchrotron emission intensity and polarization have revealed much about the magnetic field in the hot, plasma phase of the Milky Way and in other galaxies \citep[e.g., reviews by][]{Beck96, Noutsos12, Beck15, Haverkorn15, Han17, Akahori18}.  Faraday rotation measurements of linear polarization from Galactic pulsars and distant AGN have been used to explore the symmetry of the Milky Way's halo magnetic field and to provide a glimpse of the nature of the disk magnetic field \citep[e.g.,][]{Rand89, Han99, Han06, Brown07, VanEck11, Han18}. However, cold molecular clouds contain little of the hot plasma traced by radio wavelength polarization Rotation Measures (RM), leaving the magnetic field in the star-forming ISM component inadequately explored.

Complex models of the Galactic magnetic field have been developed based on these radio continuum techniques  \citep[e.g.,][]{Ferriere00, Brown07, Jaffe10, Moss10, VanEck11, 
JF12, Han18}, with large numbers of model parameters used to characterize the halo and disk fields and including spiral arms. Yet agreement as to the number, pitch angles, and magnetic field directions along and between spiral arms has remained elusive for the Milky Way. This is partially due to our location within the Galactic disk and partially due to the lack of adequate magnetic field probes of the cool, star-forming molecular gas and dust that dominate the mass of the diffuse matter in the ISM.

This situation improved with the release of the \planck \citep{Planck_I_2015} polarization maps \citep{Planck_XIX_2015}, which sampled the full sky at angular resolutions as  fine as 5~arcmin for 353~GHz (850~$\mu$m) observations of the Galactic plane. Polarization at these wavelengths is due to modified thermal emission from anisotropic, spinning dust grains, which tend to align their long axes perpendicular to the local magnetic field orientation through the same
RATs mechanism that produces optical and infrared starlight polarization. For emission polarimetry,
the result is a linearly-polarized submm intensity whose maximum electric field amplitude orientation appears perpendicular to the sky-projected magnetic field orientation. These data still suffer from low angular resolution and line-of-sight confusion, however, as the optically thin submm wavelengths lead to the collection and averaging of signals from all emitting structures along the line-of-sight. This is especially problematic in the Galactic midplane, where the \planck lines of sight can extend to many tens of kpc, but the spiral arms and individual clouds whose magnetic 
properties are sought only span tiny fractions of these distances \citep{Planck_XXI_2015}. 
\planck data have been particularly effective for the study of larger, relatively nearby clouds and filaments that are located away from the confusion presented by the Galactic plane \citep{Soler16, Planck_XXXII_2016, Planck_XXXV_2016}. These studies find that the dust grain alignment, which is associated with 
magnetic field orientation, changes depending on the observed values of the gas and dust column densities, traced using the Histograms of Relative Orientation method \citep{Soler13, Planck_XXXII_2016, Planck_XXXV_2016} and other techniques \citep{Planck_XII_2018}.

On smaller angular scales, linearly-polarized thermal dust emission from the high column-density regions associated with massive star-forming cloud cores has been probed at mm, submm, and/or far-infrared (FIR) wavebands from the ground \citep{Barvainis88}, from balloons \citep{Cudlip82, Hu19}, and from airborne platforms (KAO: \citet{Hildebrand84}; SOFIA: \citet{Chuss19}). At these wavelengths, the magnetic field 
{has been} 
characterized over angular scales as fine as about ten arcsecs, but only for the small regions that are bright enough (many MJ~sr$^{-1}$) to permit polarization detection.
{Recently, the POL-2 instrument on JCMT has been used by the BISTRO team 
\citep[see][]{WT17} to map and study 
thermal dust emission polarization across a number of star-forming cloud cores \citep[e.g.,][]{Wang19,Coude19} and to fainter surface brightness levels \citep{Soam18,Liu19}.} 
Additionally, Zeeman Effect circular-polarization spectroscopy of cool molecular gas in dark clouds \citep{Crutcher09} and in masers associated with high-mass star formation regions
\citep{Fish03} has revealed magnetic properties within some molecular cloud environments, but at low angular resolution and sensitivity for the former and only in the special locations and conditions associated with high-mass star formation for the latter.

Millimeter and submm interferometers have proven ideal for examining thermal dust emission polarimetry on the finest of size scales \citep[e.g.,][]{Akeson97, Rao98, Girart06, Hull17}, down to a few tens of AU, though scattering of submm emission by protoplanetary disks also yields strong 
polarization signals \citep{Kataoka15} and may greatly alter magnetic field interpretations. The physical extents of the regions studied are limited to less than a fraction of a parsec, typically, and so are inadequate to address the nature of magnetic fields affecting cloud and core/clump formation
that takes place over larger size scales.
 
The Galactic Plane Infrared Polarization Survey \citep[GPIPS:][hereafter Paper~I]{Clemens12a} was begun in 2006 to address the lack of data available for revealing magnetic field properties for the denser, star-forming material in the disk of the Milky Way. GPIPS was a ground-based survey  covering 76~deg$^2$ of the northern Galactic plane, including the majority of the star-forming portion of the midplane, and matching the zone covered by the 
molecular gas probing $^{13}$CO ($J=1 \to 0$) Galactic Ring Survey \cite[GRS;][]{Jackson06}. GPIPS observations were collected using the 1.83~m Perkins telescope located on Anderson Mesa outside Flagstaff, AZ which was operated by Lowell Observatory until 2019 June and Boston University thereafter. The Mimir near-infrared (NIR) imaging polarimeter and spectrometer \citep{Clemens07} was used to acquire the GPIPS data, which were reduced, calibrated, and analyzed using custom IDL-based software.
The GPIPS project recently completed all observations and data processing, leading to this fourth data release (DR4\footnote{Available in full form from http://sites.bu.edu/gpips and  planned for release in more limited form from the Infrared Space Archive (IRSA).}). 

The observations comprising GPIPS are summarized in Section~\ref{observations}. The characteristics of the data and released data products in DR4 are presented in Section~\ref{data} and in the Appendices.
The Galactic latitude and longitude dependencies of polarization properties are explored in
Section~\ref{analysis}. Findings regarding the nature of the magnetic field in the Galactic disk are described in Section~\ref{discussion}. Section~\ref{summary} summarizes key GPIPS characteristics and the general findings.


\section{Observations, Data Processing, and Survey Quality Control}\label{observations}

\subsection{Observations and Basic Data Products}

Descriptions of the motivation, design, and implementation of the data collection strategies for GPIPS are detailed in the survey introduction paper \citepalias{Clemens12a}, the survey calibration paper \citep[][Paper~II]{Clemens12b}, and the first data release (DR1) paper \citep[][Paper~III]{Clemens12c}. GPIPS was conducted over more than 400 nights on the 1.83~m Perkins telescope between 2006 and 2019. 
NIR polarimetric observations were obtained for  overlapping sky fields spanning Galactic longitudes (GL) of 18 to 56\degr\ and Galactic latitudes (GB) of $-1$ to $+$1\degr\ in the northern, inner Milky Way disk. Each equatorial-oriented sky field covered one $10 \times 10$ arcmin$^2$ field-of-view (FOV) of the Mimir instrument, pointed toward one of 3,237 centers on the $9 \times 9$~arcmin spaced grid making up GPIPS. The Mimir pixel size of $0.58 \times 0.58$~arcsec$^2$ was designed to adequately sample the $\sim$1.5~arcsec average seeing at the Perkins telescope site. Each FOV was observed as 96 (or up to 119) individual images, each of 2.5~sec exposure time. 
Each image was taken through one of 16 independent rotational orientations of an internal, cold, half-wave plate (HWP) in Mimir, to modulate the incoming linear polarization for analysis by a fixed, internal, cold, wire-grid prior to detection by the $1024 \times 1024$~pixel InSb ALADDIN~III array detector. Performing a 6-position (or 7-position\footnote{GPIPS observing modes evolved from seven sky pointings, with 17 HWP images at each pointing direction for 119 total images in the earliest years, to six sky pointings with 17 HWP images (102 total images), and finally to six pointings of 16 images (96 total images per FOV).}) sky dither, with typical offsets of 15-18~arcsec, for each sky field allowed removal of {the effects due to} bad {or missing detector} pixels, boosted the signal-to-noise ratio (SNR), and provided robustness against a variety of observing, data, or data processing problems.

The 6 (or 7) dithered images of a sky field obtained through each unique HWP orientation angle
were spatially registered, filtered of bad pixels, and summed to yield one HWP image. The resulting 16 HWP images were also
summed to yield one deeper photometric image of the field. The brightnesses and positions of stars found in that deep image were measured and are listed in the photometry catalog \citepalias[PHOTCAT:][]{Clemens12c} data product. The positions of these stars served as a basis for the stellar photometry performed on each of the 16 HWP images for each field. The stellar brightness variations as a function of HWP angle (e.g., Figure 16 of \citet{Clemens07}) were calibrated and analyzed to quantify the linear polarization attributes, which are listed in the polarization catalog \citepalias[POLCAT:][]{Clemens12c} data product. Photometric in-band calibration and astrometry were performed using the $\sim$200 2MASS \citep{Skrutskie06} stars detected in each Mimir FOV \citepalias{Clemens12a}. 

Stellar photometry was obtained using a method based on point spread function (PSF) modeling and stellar neighbor removal followed by multi-aperture photometry, as described in \citetalias{Clemens12a}. This method used stars in an image as the basis to create a detector-location dependent variable-PSF model for the image. 
The model was used to fit and subtract the brightness contributions from all stars within about 15~arcsec of each target star prior to the application of aperture photometry on the target star. This process was found to be necessary to achieve polarization SNR values similar to those expected from the photon noise of each target star \citepalias{Clemens12a}. The fitting and subtraction of nearest neighbors and subsequent aperture photometry of the target star was repeated for every PHOTCAT star appearing in each of the summed images.

Calibration of polarization efficiency, instrumental polarization contributions, and polarization position angle fiducials were based on observations of previously known polarized stars
\citep{Whittet92} as well as (mostly unpolarized) globular cluster stars, particularly of globular clusters located far from the disk of the Milky Way. Both calibration procedures are described in the GPIPS calibration paper \citepalias{Clemens12b}. The stars in the globular cluster fields were used to map the instrumental polarization across the detector FOV of Mimir. This took the form of central (boresight) globular cluster placements as well as tens of placements of the globular clusters across the full Mimir FOV. The resulting polarization percentage calibration uncertainty was at, or under, 0.10\% \citepalias{Clemens12b}. {Subsequent checks of \citet{Whittet92} standards and monitoring of 
instrumental polarization for each observation indicated a lack of temporal evolution of calibration values.}

\subsection{Survey Quality Control}

The GPIPS introduction paper \citepalias{Clemens12a} listed three principal quality goals that the 
data needed to meet in order to be included in public data releases. The first
was a seeing goal, which required the combined deep photometric image for an 
observed field to have 2~arcsec or smaller mean full-width-at-half-maximum
(FWHM) stellar PSF profiles. In DR4, 99.4\% of the FOVs 
meet this goal, with no FOV exhibiting a FWHM exceeding 2.25~arcsec. The second goal required the axis ratio of the PSF shape in the deep photometric image for each FOV to be smaller than 1.5, in order to prevent wind-driven telescope motion from corrupting PSF construction and photometry. All DR4 data products meet this goal.
The final goal permitted rejection of no more than five images, of the 96 to 119 images comprising the observation of a single field, due to poorly formed PSF shapes or bad detector readouts. 
This goal was met by 99.5\% of the final DR4 FOVs, with all FOVs having at least 89 good images that sampled all 16~HWP orientations. Note that some faint stars might not be 
detected in all 16~HWP images. The number of independent HWP image detections for each 
star is contained in the DR4 data products.

Additional data quality goals were found to be necessary and led to many of the GPIPS fields being re-observed. New goals included upper limits on the true pointing accuracy of each FOV relative to its target grid center direction, the amplitude of slow time-variations in sky transmission during the course of an observation, and the amplitude of 
fast time-variations in sky transmission (sky noise). Goals also included an absence of zero-phase reset errors \citepalias{Clemens12c} in the HWP rotation unit and a positive assessment of data quality with respect to the resulting polarization pattern as it appeared in each FOV. This latter set of tests were triggered by the discovery of  
artificial polarization patterns that appeared sometimes. The goal-identifying keywords and values for each FOV are listed in the Field Properties Summary Table in the DR4 distribution. These are described in Appendix~\ref{quality_checks}, with a shortened version of the Field Properties Summary Table presented as Table~\ref{tab_sum}.

The overall high quality of the DR4 data, with respect to the many quality control checks and characterizations, was judged to have met the GPIPS project goals, leading to cessation of new observations on 27 June 2019.


\section{GPIPS DR4 Data Products and Characterizations}\label{data}

The data processing steps involved in transforming the 96 to 119 raw instrument images that comprise 
an observation of a single FOV to the final photometry, polarimetry, and combined deep photometric image for that FOV are described in \citetalias{Clemens12a}. 
No significant deviations from these steps occurred for DR4, with the exception of the 
inclusion of improved 2MASS and GLIMPSE \citep{Benjamin03} stellar matching, as noted in Appendix~\ref{gaia_match}.

The resulting data products released as DR4 take the form of a set of FOV-based products, a single data file containing unique star entries, supporting metadata files, and the Summary Table. Because of the overlapping FOV design of GPIPS, many stars appear in more than one FOV \citepalias{Clemens12c}. Duplicate stars were identified, their polarization properties averaged, and a single data file \citep{Clemens20} with unique star entries was developed from these self-matches and from the non-overlapping stars. Both forms of these data products, FOV-based and unique star based, are described below.

\subsection{FOV-Based DR4 Data Products}

The FOV-based data products follow the DR1 forms described in \citetalias{Clemens12c}. They are, for each of the 3,237 FOVs, a file of stellar polarimetry (a POLCAT), a file of stellar photometry (a PHOTCAT), the deep photometric image (a FITS file), and printable image files (in PDF and Postscript) showing the deep image of the stars in a FOV overlaid with the 
{UF1 (see Section~\ref{sec_UF} and Appendix~\ref{Appendix_PSNR})}
high-reliability linear polarization orientation lines. The GPIPS stars were also matched to 2MASS stars 
and to GLIMPSE stars, as described in \citetalias{Clemens12c} (but modified
as described in Appendix~\ref{gaia_match}), and updated photometry from those data sets were included in the POLCAT and PHOTCAT files. 
The characterizing properties of stars in each of the \Gaia DR2 \citep{Gaia, Gaia_DR2}, 2MASS, GLIMPSE, and \WISE \citep{Wright10} data sets that positionally matched to PHOTCAT and POLCAT stars, as described in  Appendix~\ref{gaia_match}, are contained in additional DR4 files for each GPIPS FOV.
In addition, copies of the hand-written observing logs for each night, the Field Properties Summary Table (shown in short form as Table~\ref{tab_sum}), and a text file of explanatory Data Release Notes are included in DR4.  

Information in the FOV-based PHOTCAT files include metadata related to the observations
and data calibration as well as entries reporting the in-band photometric properties
measured for each star. Similar information, augmented by the $H$-band polarization data for each star in the field, is contained in the POLCAT files. The set of data values for each star include an RA-ordered serial number, an equatorial coordinate based designation, the X (column) and Y (row) pixel locations where the star appears in the registered and combined deep photometric image, the mean X and Y  pixel location of the star as it appears in detector coordinates (used for instrumental polarization correction), the J2000 R.A. and decl., the $H$-band magnitude (calibrated to the 2MASS $H$-band magnitudes for the stars in the FOV, but {\it not} corrected for color effects), internal and external uncertainties in the $H$-band magnitude, sky count level, debiased linear polarization percentage\footnote{GPIPS debiased linear polarization percentages follow \citetalias{Clemens12a} such that $P^\prime = (P_{RAW}^2 - \sigma_P^2)^{0.5}$ for the case where the measured $P_{RAW}$ is greater than its uncertainty $\sigma_P$ and zero otherwise.} $P^\prime$ and its uncertainty $\sigma_P$, the equatorial polarization position angle (EPA, or PA), the Galactic polarization position angle (GPA), the uncertainty in position angle ($\sigma_{PA}$), and the Stokes $Q$ and $U$ fractional values and uncertainties (as normalized by Stokes $I$). 

\subsection{Unique-Star DR4 Data Products}\label{sec_unique}

The FOV-based PHOTCATs were examined to find duplicate stars appearing in neighboring FOVs. A master catalog of unique stars was constructed from the stars that had no duplicates plus the duplicate stars. For each set of duplicates, their polarization properties were averaged using inverse variance weighting of their Stokes $U$ and $Q$ values, followed by computation of the raw polarization percentage, its uncertainty, the debiased polarization percentage, the polarization position angle, and its uncertainty. The usage flag (UF) designation (see Section~\ref{sec_UF}, below) was redetermined and could differ from the FOV-based designations due to lowered uncertainties in $\sigma_P$. 

The total number of PHOTCAT stars, after resolving duplicates, is nearly 14 million. The differences in the numbers of FOV-based star counts and unique star 
counts is 10.6\% of the former. This indicates that the effective Mimir FOV size, accounting for dithering and 
field-to-field overlap, is about $9.5 \times 9.5$~arcmin$^2$. Analysis performed using the
unique star file will be more accurate in terms of accounting for the actual numbers of stars. 

The unique star data file also contains photometric and parallax information based on 
cross-matching GPIPS to \gaia DR2, 2MASS, GLIMPSE, and \WISE. The matching process
and statistics of the match results are described in Appendix~\ref{gaia_match}, as are the 
contents and formats of the entries in the unique star data file \citep{Clemens20}.


\subsection{Stellar Usage Flag Designations}\label{sec_UF}

As described in \citetalias{Clemens12c} and shown in Figure~16 there, a measure of reliability may be obtained from the values of polarization percentage uncertainty and $H$-band magnitude for each star. GPIPS stars that have $m_H < 12.5$~mag and $\sigma_P < 2$\% were identified as being of high quality and
were classified as UF1 (Usage Flag = 1) stars. Additionally, those UF1 stars with debiased polarization SNR (P$^\prime$SNR $\equiv$ $P^\prime / \sigma_P$) exceeding three, corresponding to $\sigma_{PA} <  9.6\degr$, became the UF0 subset. Outside of the values defining UF1, stars were classified as UF2 if they had $12.5 \le m_H < 14$~mag and $2 \le \sigma_P < 10$\%. The remainder of the stars in the POLCATs were classified as UF3. The UF designation for each star is included in the FOV-based POLCAT files and in the unique star file. Note that the particular $m_H$ and 
$\sigma_P$ values that define the three UF types are tied to GPIPS observing details. Other 
data sets require different boundary values to invoke their UF classifications.
 
\begin{deluxetable}{ccccc}
\tablecaption{GPIPS DR4 Stars and Polarization Quality \label{tab_uf}}
\tablewidth{0pt}
\tablehead{
\colhead{Usage Flag} & \colhead{$m_H$ Range} & \colhead{$\sigma_P$ Range} & \multicolumn{2}{c}{Number of Stars} \\
\colhead{Designation}&\colhead{(mag)} &\colhead{(\%)} & \colhead{FOV-based} & \colhead{Unique}\\
\colhead{(1)}&\colhead{(2)}&\colhead{(3)}&\colhead{(4)}&\colhead{(5)}
}
\startdata
UF1	& $m_H < 12.5$ & $\sigma_P < 2.0$ 	& \ 1,021,274 & \ 932,131\\
\ \ \ {\it UF0} & \multicolumn{2}{c}{\it adds $\sigma_{PA} < 9.6\degr$} & {\it \ \ 238,955} & {\it \ \ 234,530} \\
UF2	& $12.5 \le m_H < 14.0$ 	& 2.0 $\le$ $\sigma_P$ $<$ 10.0	& \ 2,876,808 & \ 2,538,898 \\
UF3	& $14.0 \le m_H$			&	10.0 $\le$ $\sigma_P$		& \ 5,809,074 & \ 5,014,999 \\
\hline
\multicolumn{2}{l}{POLCAT Total}&& \ 9,707,156 & \ 8,486,028 \\ 
\hline
\multicolumn{2}{l}{PHOTCAT Total}&& 15,512,486 & 13,861,329 \\
\enddata
\end{deluxetable}

As explored in \citetalias{Clemens12c}, the UF1 stars generally are of sufficient quality to yield polarization position angle uncertainties low enough for independent use to reveal magnetic field 
orientations. The UF2 and UF3 stars are generally only useful for probing magnetic field properties when grouped and averaged by sky zone and/or magnitude. The SNR distributions of the stars, binned by UF value, are explored in more detail in Appendix~\ref{Appendix_PSNR}.

Table~\ref{tab_uf} lists the numbers of stars in these UF designations in the FOV-based POLCATs and in the unique star data file. UF1 stars account for 10-11\% of all POLCAT stars. 
When averaged over the 76~deg$^2$ survey region, these imply average sampling of about one per 0.27-0.29~arcmin$^2$, or a mean separation of 31-33~arcsec between UF1 stars. This exceeds the GPIPS requirement \citepalias{Clemens12a} of one star per 45~arcsec and meets the GPIPS goal of one~pc mean sampling of dusty molecular clouds out to 6~kpc distance. 
UF0 values are listed in italics to highlight that it is a subset of UF1.
The total number of stars in the photometrically deeper PHOTCAT files is over 15~million
for the FOV-based sample and nearly 14~million for the unique star sample. 

\subsection{Multi-Scale Near-Infrared Polarization Overview}\label{zoom}

\begin{figure}
\includegraphics[width=8.0in, angle=90]{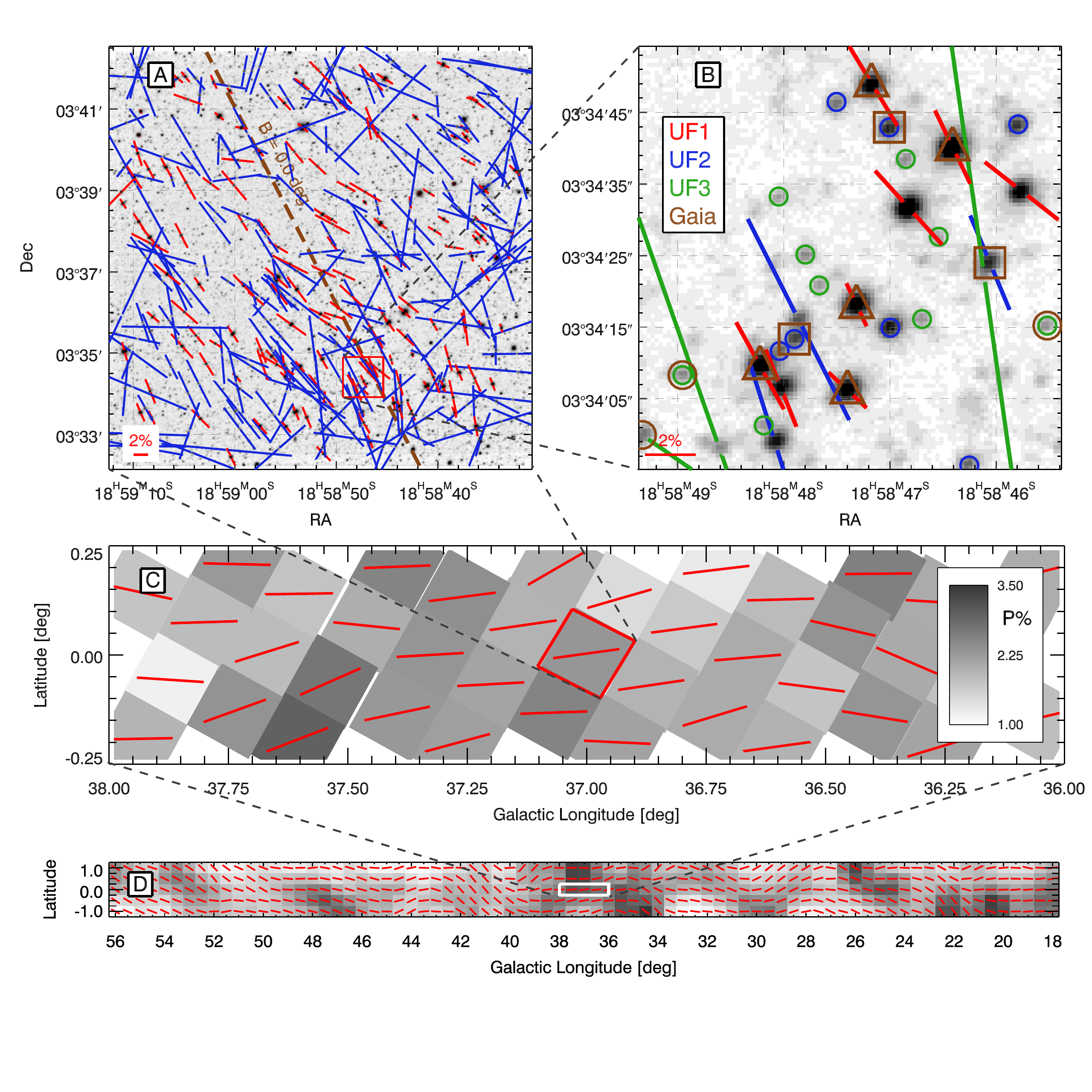}
\caption{Multi-scale overview of GPIPS data products and derived quantities.}\label{fig_mosaic}
\end{figure}

Figure \ref{fig_mosaic} presents a visual summary of GPIPS data, spanning angular scales from arcseconds to
the tens of degrees characterizing the full survey region. The upper-left, A-panel displays one Mimir FOV of data, after processing. This field (number 1619) is located at the center of the GPIPS survey, at (GL, GB)~=~ (37.0$\degr$, 0.0$\degr$). 
Shown in the reversed, gray-scale background is the deep photometric image constructed 
from the sum of the 96 individual images.
Overlaid on the gray-scale image are lines indicating the polarization properties of the starlight. 
Line lengths are proportional to $H$-band debiased linear polarization 
percentages. 
Line orientations indicate equatorial polarization position angles (EPA {- measured East from North}). Lines colored red 
represent polarization properties for stars classified as UF1. Lines colored 
blue are for UF2 stars. 
Most of the line orientations, both red and 
blue, point along a direction from southwest to northeast, revealing that 
the dominant magnetic field orientation in this FOV is somewhat parallel to the Galactic plane. 
There is a wide range in polarization percentage values, from below 1\% to well beyond 5\%.
Some lines, mostly for UF2 stars (in blue), exhibit position angles nearly perpendicular
to the dominant orientation. These may represent real magnetic field direction changes along some
lines of sight or their deviations may be due to their lower SNR values.
In the FOV covered by the A-panel, the total number of polarization-measured
stars (UF1+UF2+UF3) is 1,926 while the stars without measured polarization (PHOTCAT 
entries minus POLCAT entries for this field) number 1,481.

\clearpage
\begin{figure}
{Figure~\ref{fig_mosaic} caption, continued:} \\
(Upper left, A-panel):
Equatorial, reversed gray-scale representation of the deep photometric
image for field number 1619. Colored lines through stars encode UF1 and UF2
$P^\prime$ and $EPA$. UF3 information 
is suppressed in this panel. A red 2\% $P^\prime$ 
reference scale appears in the lower left corner. The Galactic equator is identified by the brown, dashed,
diagonal line and ``B = 0.0 deg" label. 
(Upper right, B-panel): Enlargement of selected 
portion of the A-panel. 
Colored lines identify polarization detections and
circles identify $P^\prime$ upper limits. 
\Gaia DR2 star matches are shown as the brown symbols.
(Middle, C-panel): A $2 \times 0.5\degr$ portion of the northern Galactic mid-plane, 
shown in Galactic coordinates. Individual GPIPS FOVs are shown as rotated gray squares, 
aligned to Galactic orientations. 
Gray-scale shade encodes median percentage polarization, ranging from 1\% (white) to 3.5\% (black). Red lines (of uniform 
length) represent the  median Galactic polarization position angle of the UF1 stars in 
each FOV. They reveal a mostly disk-parallel magnetic field, but also some departures from uniformity.
(Bottom, D-panel): Low-resolution representation of the median polarization percentage and 
median polarization position angles across the full GPIPS survey region. Gray-scale encodes median
polarization percentage, from 1.1\% (white) to 2.4\% (black).
\end{figure}

In Figure~\ref{fig_mosaic}, the upper-right, B-panel shows a zoomed-in view of a 
$60 \times 60$~arcsec$^2$
region drawn from the upper-left, A-panel. 
In the B-panel, the inverse gray-scale shows the presence of many tens of stars.
The mean PSF FWHM is 1.8 arcsec, which is only somewhat greater than the average 
for the entire GPIPS data set. 
Lines and circles indicate stellar polarization detections ($P^\prime > 0$\%) and 
upper limits ($P^\prime = 0$\%), respectively: 
red for UF1 stars (8 detections, 
no upper limits), blue for UF2 stars (3 detections, 7 upper limits), and green for UF3 stars 
(3 detections, 8 upper limits). Orientation line lengths encode debiased linear polarization percentage in
$H$-band. Line orientations encode polarization equatorial position angles. 
Nearly all of the stars in the B-panel have been measured for
polarization, as noted by their hosting either an associated orientation line or a circle.

In the B-panel, \Gaia DR2 stars are indicated as brown symbols.
All stars that appear in \Gaia DR2 are matched to POLCAT entries
for this $1 \times 1$~arcmin$^2$ field. Yet, the natures of the \Gaia-matched stars are complex. Only five of the eight UF1 stars
have \Gaia matches, with a couple of the most NIR-bright UF1 stars not matched to \Gaia
stars. These non-matches
are strongly reddened (mean $H-K$ $\sim$ 1.2~mag, or $A_V \sim 17$~mag), which affects 
\Gaia $g$-band, optical magnitudes
more than GPIPS $H$-band magnitudes. The effects of reddening on the ability of \Gaia
to provide distances to GPIPS POLCAT stars are explored in greater detail in 
Section~\ref{gaia_analysis}.

The middle, C-panel of Figure \ref{fig_mosaic} spans 2$\degr$ of Galactic longitude and 
presents each plotted GPIPS FOV as a single, gray-scale,
rotated square with an inset red orientation line. 
The rotation of the field from the A-panel to the C-panel is due to the relative
orientations of the equatorial orientation of Mimir and the Galactic plane. Each of the 
41 FOVs shown in the C-panel is shaded to represent a single representative polarization
percentage. These were computed as the medians of the UF1 star values for each FOV and are encoded
as gray-scale values, where black represents the greatest polarization percentage. A single 
red orientation line is oriented in each FOV
to display the similarly computed mean polarization position angle, in Galactic coordinates. 

These representative values were obtained from the
median values of polarization percentage and Galactic position angle (GPA) for the UF1 stars in each FOV, as described in Section~\ref{analysis}.
They represent a single, coarse characterization of the polarization properties for each FOV.
Across the C-panel, the mean $H$-band polarization fraction is seen to vary from
a low of 1\% to a high of 3\%, with the A-panel field exhibiting a value of 2.3\%. The red lines
show that the magnetic field being traced (at 10 arcmin resolution) is mostly
parallel to the Galactic plane, but some FOVs also exhibit significant, coherent deviations of polarization position angle.

The bottom, D-panel of Figure \ref{fig_mosaic} encompasses the entire GPIPS region with 
representations of polarization percentage and position angle that relate to those shown in the previous 
panels. The single-FOV median properties of $P^\prime$ and GPA of the UF1 stars were smoothed, 
using gaussian distance weighting (FWHM=0.75$\degr$) onto a $0.5 \times 0.5$ sq deg grid. 
The red 
position angle orientation lines are mostly parallel to the Galactic plane, but again show significant 
deviations, notably near Galactic longitudes 22, 29, 34, 41, 44, and 52-56$\degr$. 
The averages shown are coarse
representations, as no Stokes $U$, $Q$ averaging was performed - instead, straight averages
of the FOV median values were used.

GPIPS data reveal a generally disk-parallel magnetic field, with some departures likely due to interactions with the gas dynamics present in the thin, molecular disk of the Milky Way.

\clearpage

\section{Analyses}\label{analysis}

This section begins with an analysis of the natures of the GPIPS stars that match, and do
not match, to \Gaia DR2 stars. The details of the matching algorithm and the resulting match statistics may be found in Appendix~\ref{gaia_match}.
In Section~\ref{one_field}, characterizations of the data products from 
the same central FOV of the GPIPS survey (number 1619; Figure~\ref{fig_mosaic}.A)
are described and evaluated to
introduce the FOV-based stellar polarization characterizations employed throughout the 
subsequent analyses. 
These single-FOV evaluations 
provide physical insight into the nature of the region being probed and the magnetic field properties 
sampled along the sight-lines contained in each FOV.
These characterizations were then applied to the entire set of GPIPS
FOVs, as described in Section~\ref{all fields}, to establish general distribution functions 
{(i.e., marginalized over GL and GB, so zero-dimensional)}, 
one-dimensional (1-D) Galactic longitude and latitude 
distributions of the polarization properties, tests for correlations among the properties, 
and two-dimensional Galactic directional distributions of the polarization properties. 

\subsection{The Natures of the GPIPS-to-\Gaia Matching, and Non-Matching, Stars}\label{gaia_analysis}

The utility of GPIPS data products for revealing magnetic field 
properties depends on the characteristics of the stars observed, as well as the natures of the 
dust and gas distributed along the lines of sight to the stars. The GPIPS observations were designed
\citepalias{Clemens12a} to ensure that most of the stars that would have polarization detections 
would be moderately extincted 
{(}${1} \le {A_V} \le {30}${~mag)}, 
and thereby polarized, distant giants. The analysis of
the first 17\%\ of GPIPS data \citepalias[DR1;][]{Clemens12c} confirmed the expected
excess of giants over nearby dwarfs. Stellar reddening excesses $E(H-K)$ as great
as 2~mag were found in DR1. Such excesses implied that extinctions were being probed to about 30~mag of $A_V$. 

However, accurate distances to the GPIPS stars remained elusive until the release of
\Gaia DR2. One important distance to establish is the minimum needed through the diffuse ISM
to develop detectable NIR polarization signatures to GPIPS levels. This ``near horizon" for
GPIPS is likely beyond the extent of the Local Bubble \citep{Local_Bubble}, but is it close enough
to reveal magnetic fields for dark molecular clouds at 300-400~pc? The ``far horizon" would be
the distance limit beyond which GPIPS stars are too faint or
too extincted to be detected. Both distances are needed to estimate
the range of line-of-sight distances over which GPIPS data offer the most useful 
magnetic field information.

The matching of \Gaia and GPIPS stars, described in Appendix~\ref{gaia_match}, led to the creation of FOV-based files
of \Gaia star information and their corresponding GPIPS star match identifiers. Examination of the characteristics of the stars with GPIPS-\gaia matches, and those without such matches, was performed to establish stellar characterizations and to reveal
the GPIPS near- and far-horizons.

\begin{figure}
\includegraphics[width=7.25in]{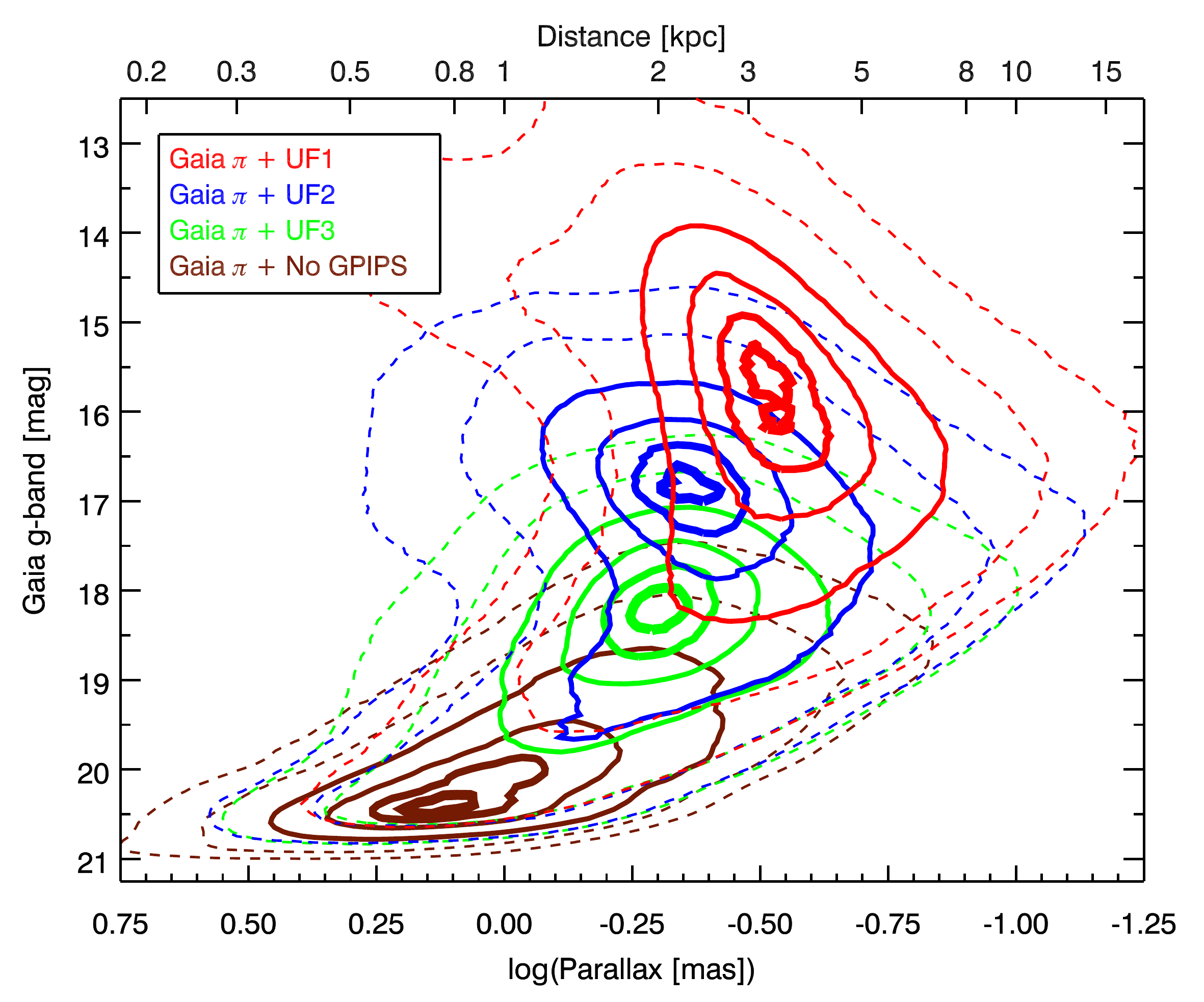}
\caption{Comparison of \Gaia $g$-band apparent magnitudes versus \Gaia parallax for stars
matched and not-matched to GPIPS POLCAT entries. Bottom horizontal axis presents base-10
logarithm of the parallax in mas. Upper horizontal axis shows corresponding distances. Densities of star counts in this plane
are indicated by the colored contours, with values of 10 and 25\% (dashed lines) and 50, 75, 90, 
and 97.5\%\ (solid lines) of the peak value for each of the four subsamples of stars. Red contours show the distribution of \Gaia stars that match to UF1 stars in POLCATs. Blue contours represent
\Gaia stars matched to UF2 stars and green contours represent UF3 stars. The brown 
contours show the distribution of \Gaia stars with parallaxes that have no matching stellar entries
in the GPIPS POLCATs. The POLCAT-matching stars tend to be brighter and more distant
than the non-matching stars.
\label{fig_G_para}}
\end{figure}

Figure~\ref{fig_G_para} displays contour representations of the stellar count density distributions 
of selected \Gaia stars matched to subsets of GPIPS stars, as functions of the
optical, $g$-band \citep{Gaia_DR2} apparent magnitude and the base-10 log of the 
\Gaia parallax $\pi$ (in mas) to
each star. \gaia stars included in this plot had $g$-band magnitude uncertainties less than
0.66~mag, parallax SNRs ($\equiv \pi / \sigma_\pi$) $\ge 0.5$, and parallaxes $\pi > -2$~mas. These liberal
limits were chosen to avoid introducing parallax bias \citep{Luri18,BailerJones18} to the sample population characteristics.
These criteria selected for just over 40\%\ of all \Gaia stars appearing in the GPIPS FOVs.

In Figure~\ref{fig_G_para}, contours shown in red represent the stellar count density of 
the selected \Gaia stars in the
mag-parallax plane that matched to POLCAT UF1 stars (``\gaia $\pi$ + UF1" in the legend). 
These matched stars account for only 3.4\%\ of all \Gaia stars but 46\%\ of all POLCAT UF1 stars. 
The portion of the Figure exhibiting 50\%\ of the peak density of counts of \Gaia parallax stars 
(``\gaia$\pi$" hereafter) matched to UF1 stars (marked by the outermost solid red contour) spans $g$-band apparent magnitudes 
of about 14--18th and log parallaxes of $-$0.26 to $-$0.85, or distances of 1.8 to 7.1~kpc.

The blue contours in Figure~\ref{fig_G_para} represent the density of counts of {\Gaia}$\pi$ stars matched to POLCAT UF2 stars, 
and the green contours represent the same for UF3 stars. They both span correspondingly fainter apparent magnitudes at somewhat closer distances than do the UF1 stars (red contours). Ignoring extinction (but see the discussion regarding Figure~\ref{fig_G3_hk_para} below), 
the $g$-band absolute magnitudes of the peaks of the UF1 (red), UF2 (blue), and 
UF3 (green) distributions, after correcting for the distance moduli implied in the parallaxes, 
are about 3.1, 5.0, and 6.7~mag at
inferred mean distances of about 3.2, 2.3, and 2.0~kpc, respectively. 
(These, and other values
are collected in Table~\ref{tab_summary})

The brown contours in Figure~\ref{fig_G_para} indicate the density of \Gaia stars that meet the magnitude and parallax SNR selection criteria but for which no POLCAT star was matched. Again, ignoring extinction, the $g$-band 
absolute magnitude for the peak of this distribution is about 10.9 mag, at a distance
of about 0.78~kpc. Hence, the \Gaia$\pi$ stars that match to POLCAT entries have brighter $g$-band apparent magnitudes and are at greater
distances, and are thereby more luminous, than the \Gaia$\pi$ stars that do not match to POLCAT entries.

\begin{figure}
\includegraphics[width=7.25in]{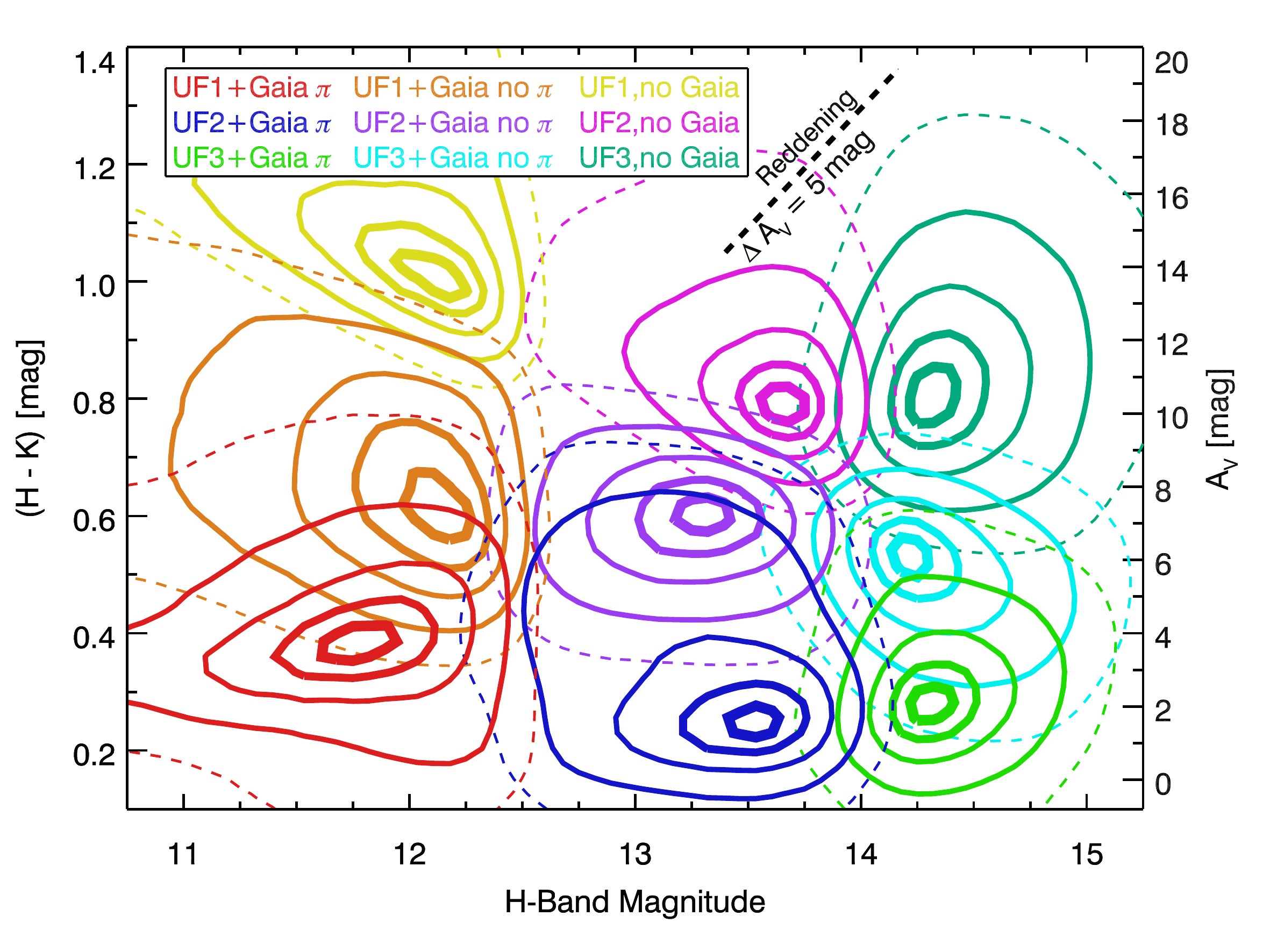}
\caption{Comparison of 2MASS $H$-band magnitudes and $(H - K)$ colors for matching 
GPIPS UF1 ($m_H \sim 12$), UF2 ($m_H \sim 13-14$), and UF3 ($m_H \sim 14-15$) stars.
A scale for visual extinction, assuming average dust properties, is shown as the right axis. 
A reddening vector, corresponding to an A$_V$ change of 5~mag, is shown in black at 
upper right.
At each UF designation, three vertically offset colored sets of contours show subsets of stars for that
designation that have different \Gaia DR2 matching properties. A legend relating contour set 
color to \Gaia DR2 matching properties is shown at upper left. The lowest contour sets (red, blue,
and light green), centered
at the least red $(H - K)$ (least A$_V$) values, represent star count densities of GPIPS stars
matching \Gaia stars that have good parallax values (e.g., ``UF1+Gaia $\pi$" in red in the legend). The middle contour sets (orange, purple, and
cyan) represent GPIPS stars that match to \Gaia stars, but for which there are no good parallax values. 
The highest contour sets (yellow, magenta, dark green) represent GPIPS stars that do not
match to \Gaia stars. All contour sets are drawn representing  
25 \% (dashed) and 50, 75, 90 and 97.5\%\ (solid) of the peak stellar counts in each distribution. 
\label{fig_G2_hk_mh}}
\end{figure}

Establishing characteristic distances to the GPIPS stars (the near and far horizons) requires correcting the \Gaia distances for the effects of dust extinction. This affects the optical, $g$-band 
magnitudes of \Gaia more than it affects the NIR magnitudes of GPIPS. In Figure~\ref{fig_G2_hk_mh},
contours of star counts in the $(H-K)$ versus $H$-band color-magnitude plane 
have been colored-coded to identify nine selected subsets of stars. Stars in POLCATs that had 
\Gaia$\pi$ matches are represented by the red, blue, and green contour distributions, based on their
UF1, UF2, or UF3 designations, as was done in Figure~\ref{fig_G_para}. 
Stars in POLCATs with \Gaia star matches that did not meet the parallax criteria (``\gaia${\rm no}$~$\pi$"), are represented by the orange, purple, and cyan colored contour distributions, based on their 
UF1, 2, and 3 designations, respectively.
GPIPS stars that did not match to any \gaia stars are shown by the yellow, magenta, and dark-green
contours.
All POLCAT NIR stellar magnitudes and colors included in
the distributions were drawn from 2MASS \citep{Skrutskie06}, subject to a $(H-K)$ color uncertainty
criterion of $\le 0.5$~mag. These 2MASS and \gaia selection criteria, taken together, tend to 
bias the distributions against faint and red stars. 
This is seen as the successively smaller fractions of the total POLCAT stars with increasing UF
number in Table~\ref{tab_summary}.

The groupings in Figure~\ref{fig_G2_hk_mh} reveal the expected magnitude boundaries 
separating UF1, 2, and 3 stars. 
These boundaries are seen as the node-like vertical contours near $m_H = 12.5$ and 14~mag, 
which are manifestations of the imposed UF definitions. 
The right axis indicates approximate values of $A_V$, from zero to 20~mag, using 
the NIR Color Excess method \citep[NICE;][]{Lada94}.

GPIPS stars with \Gaia$\pi$ matches 
tend to be less extincted than GPIPS stars of similar $H$-band brightnesses, but for which the \Gaia matches yielded no parallaxes. The GPIPS stars not matching to \Gaia stars are the most extincted 
of all, as already noted in the Figure~\ref{fig_mosaic}.B discussion.
For the POLCAT stars matched to \gaia$\pi$ stars (the red, blue, and green contours), the 
UF1 distribution peak exhibits about 2~mag of A$_V$ greater than seen at the UF2 and UF3 peaks.
The UF2+\gaia$\pi$ distribution shows a bifurcation in A$_V$, with components near 2 and 
6~A$_V$~mag.
The POLCAT stars matched to \gaia${\rm no}$~$\pi$ stars show distribution peaks offset from the 
\gaia$\pi$ stars distribution peaks by 3.5, 5.5, and 4.0~A$_V$~mag for UF1, 2, and 3,
respectively.
The POLCAT stars not matched to \gaia stars show distribution peaks offset from the \gaia$\pi$ 
stars distribution peaks by 10, 8.5, and 8.5~A$_V$~mag, respectively.
The A$_V$ steps within each UF vertical group, of about 4~mag for \gaia$\pi$ to \gaia${\rm no}$~$\pi$ and
another 4~mag to the no-\gaia stars, show that \gaia selects the lowest extinctions while GPIPS
without \gaia stars selects the most extincted stars.

The UF1 stars with \Gaia$\pi$ matches represent 45.9\%\ of all UF1 stars (FOV-based). 
The unmatched and no parallax subsets contain 14.0 and 31.9\%\ 
of all UF1 stars, respectively. 
The remaining 8.2\%\ of UF1 stars fails to meet the 2MASS color uncertainty
criterion applied. 
Moving to the fainter, UF2 stars, those with \Gaia matches and parallaxes
account for 26.0\%\ of all UF2 stars (FOV-based), the \Gaia-unmatched and no parallax 
subsets account for 24.2 and 22.5\%, respectively,
and the remaining 27.3\%\ fails the color uncertainty selection criterion. 
The faintest, UF3, stars with 
\Gaia matches and parallaxes account for only 7.6\%\ of UF3 stars, while only another
17.5 and 5.4\%\ are in the \Gaia-unmatched and no parallax subsets. 
The high fraction of UF3 stars
failing the 2MASS color uncertainty selection criterion, 69.4\%, is because GPIPS probed stars
fainter and/or redder than stars in the 2MASS catalog.

In Figure~\ref{fig_G3_hk_para}, extinctions and parallaxes for \Gaia$\pi$ stars matched to POLCAT entries are represented as the red-, blue-, and green-colored contours of 
stellar density for UF1-, UF2-, and UF3-matched stars, respectively, in the central, A-panel.
The stars contributing to Figure~\ref{fig_G3_hk_para} were required to have the same
$(H - K)$ color uncertainty $\le 0.5$~mag, parallax SNR $\ge 0.5$, and 
parallax greater than $-2$~mas as for the previous Figure.
{POLCAT stars matched to \Gaia$\pi$ stars tend to have extinctions ranging from 
0 to $>$6~mag of $A_V$, and span distances of $<$0.5 to about 15~kpc. }
The distances corresponding
to the locations of peaks in the contour distributions are similar to those described for 
Figure~\ref{fig_G_para}, for each of the three subsets of stars. 

The lower, B-panel of Figure~\ref{fig_G3_hk_para} shows the marginalization over NIR color as cumulative likelihoods (CL) of star count density with log parallax for each of the UF1, UF2, and UF3 samples that match to \gaia$\pi$ stars and also satisfy the selection criteria applied to the 2MASS values. The UF1 CL (red) curve crosses the 10\%\ and 90\%\ horizontal dotted gray lines at about 0.91~kpc and 6.3~kpc, respectively. 
That CL curve also shows quartile boundaries at 1.6, 2.63, and 4.1~kpc. The UF2 CL (blue) curve shows a median of about 1.95~kpc, with 10 and 90\%\  limits of 0.72 and 4.0~kpc. UF3 stars (green CL curve) have a median distance of 1.78~kpc, with 10 and 90\%\ limits of 0.71 and 4.22~kpc. The 2MASS selection criterion on color uncertainty added some bias against fainter stars, so the distance values quoted here should be considered upper and lower limits, respectively for the 10\% and 90\% values, though they are reasonably characteristic of the samples.

In the right, C-panel of Figure~\ref{fig_G3_hk_para}, the central panel distributions have been marginalized over parallax to create
histograms of $(H-K)$ colors of stars in the \Gaia$\pi$, 
\Gaia${\rm no}$~$\pi$, and no-\gaia groups, shown as the blue, purple, and magenta curves, respectively. 
The histograms have been normalized by the sum of the stars contained in 
the 0.02~mag wide bins in both
histograms. These histograms reveal similar extinction offsets between the $\sim$1.6~million 
\gaia$\pi$ matched, the $\sim$1.1~million \gaia${\rm no}$~$\pi$ matched, and the 0.6~million \gaia-unmatched samples of about 4.5~mag of $A_V$ between each successive group, similar to what was seen in Figure~\ref{fig_G2_hk_mh}. 

\begin{figure}
\includegraphics[width=7.25in]{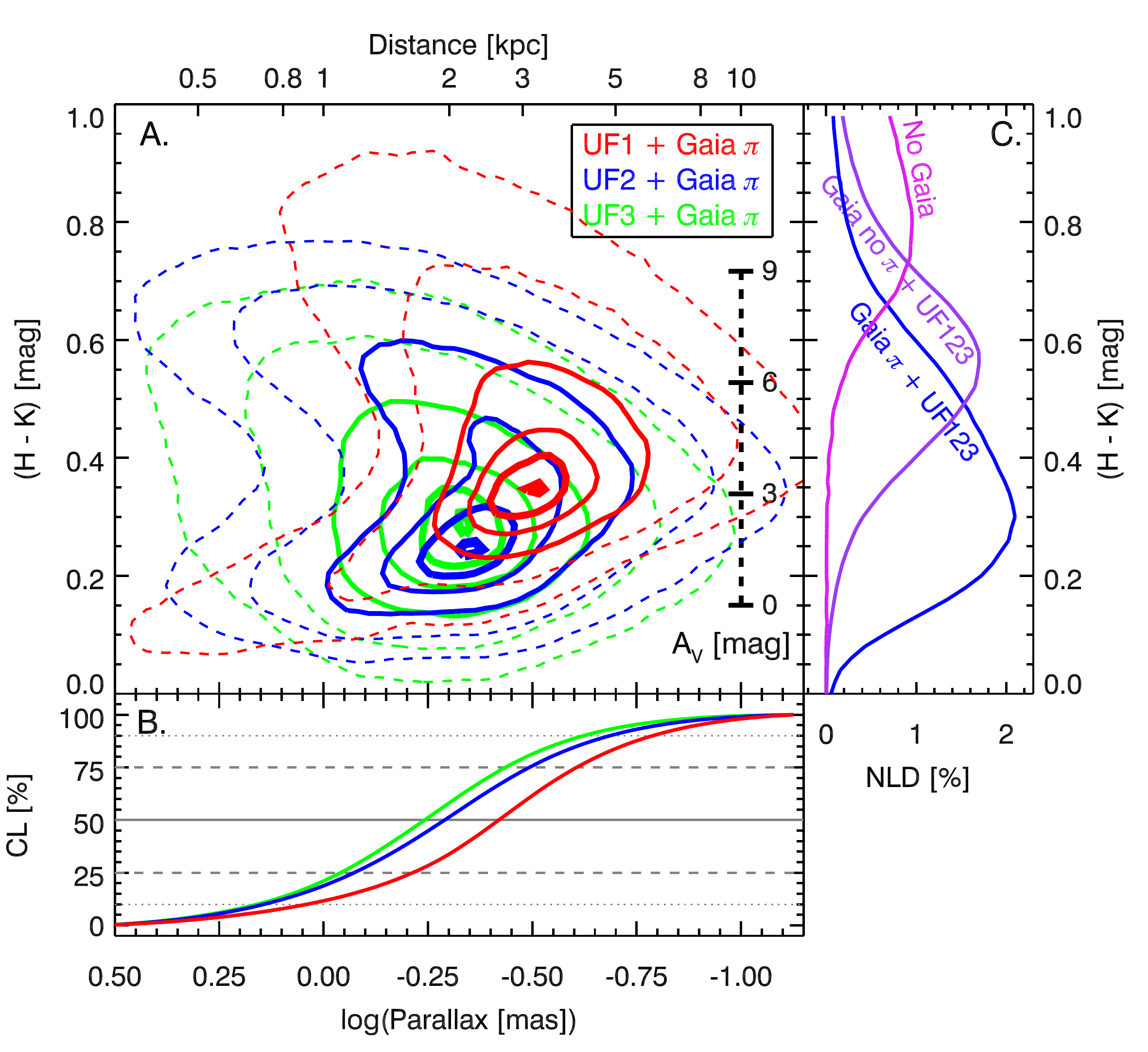}
\caption{Comparison of 2MASS $(H - K)$ colors with \Gaia parallaxes for DR4 POLCAT stars.
(Central, A-panel): Contours of normalized
star count density for \Gaia stars with parallaxes matched to POLCAT stars. Contours represent 10 and 25 (dashed) and 50, 75, 90,  and 99\%\ (solid) of the peak star count
density in each sample. Extinction scale is shown inset at right; distance scale is shown along the
top axis. (Bottom, B-panel): 
Cumulative likelihood (CL) for the \Gaia$\pi$-matching POLCAT stars. Gray dotted, dashed, and solid lines are drawn at 10, 25, 50, 75, and 90\%\ cumulative probabilities. (Right, C-panel):
Normalized likelihood distributions (NLD) histograms of $(H - K)$ colors for POLCAT 
stars matched to \Gaia$\pi$ stars (blue curve), POLCAT stars matched to \gaia${\rm no}$~$\pi$ stars (purple), and POLCAT stars not matched to \Gaia
stars (magenta). The
POLCAT stars without \gaia matches tend to exhibit greater values of extinction than the stars with \Gaia matches. 
\label{fig_G3_hk_para}}
\end{figure}

The more distant UF1 stars with \Gaia$\pi$ matches were also found to be somewhat more extincted
than the fainter UF2 and UF3 subsets, at least for the contours near the peak regions of the star count
distributions. As shown in the C-panel of Figure~\ref{fig_G3_hk_para}, the stars without \Gaia matches are even more extincted,
on average, than the \Gaia$\pi$- and \gaia${\rm no}$~$\pi$-matched UF1 stars. 
Correcting for the extinctions corresponding
to the peaks of the colored contour distributions in Figure~\ref{fig_G3_hk_para}.A. (see Table~\ref{tab_summary}), the $g$-band apparent  magnitudes found in Figure~\ref{fig_G_para} become $g$-band absolute magnitudes $M_g$ of about $-$0.1, 3.4, and 4.5~mag, for the 
UF1, UF2, and UF3 samples, respectively. 

Table~\ref{tab_summary} provides a summary of key characteristics of the different 
samples of stars analyzed in this Section. The columns identify the UF designation of the sample, or for the case of \Gaia stars that do not match to POLCAT stars, a ``\gaia, No POLCAT" column. 
The rows list the source of the properties in the preceding text, mostly derived from the
Figure~\ref{fig_G_para}, Figure~\ref{fig_G2_hk_mh}, and Figure~\ref{fig_G3_hk_para} 
plots of the two-dimensional distributions of stars that match among
the POLCATs, \gaia, and 2MASS. 
The locations of the peak stellar
densities in Figure~\ref{fig_G_para} are listed, but have not been corrected for extinction effects.
In the Figure~\ref{fig_G2_hk_mh} - based rows, the UF1+\gaia$\pi$ subsample is seen to account for 
about 46\% of all UF1 stars, while the fainter 
UF2 and UF3 stars are less well-represented by matches to the 2MASS and \gaia$\pi$ archival 
catalog data and so 
the propertied derived might not be as accurately characterized for these stars. 
In the Figure~\ref{fig_G3_hk_para} - based rows, the \gaia$\pi$ match subset estimates for near, far, and mean distances for each sample are listed for multiple population percentage steps in the cumulative probability distributions.

To estimate the spectral types of the POLCAT stars, the \gaia HR Diagram for 
low-extinction stars presented as Figure~5 in \citet{Gaia_HRD} was employed. For each of the 
A$_V$-corrected $M_g$ values noted above, 
the range of \gaia colors ($G_{BP} - G_{RP}$) spanned by the majority of stars 
were identified for the three main luminosity classes (III, V, VII) and these values are
reported near the bottom of Table~\ref{tab_summary}. These colors were converted to Johnson-Cousins $(V - I)$
using the analysis of \gaia colors reported in \citet{Evans18}.
The $(V - I)$ colors were used to retrieve spectral types for the associated luminosity classes
using the tables published in \citet{Ducati01}, except the white dwarfs. In the
absence of characterizing spectra, these were assigned DA(?) types.

\begin{deluxetable}{lcccc}
\tablecaption{Summary of UF Sample Stellar Distributions Properties\label{tab_summary}}
\tablewidth{0pt}
\tablehead{
&\multicolumn{4}{c}{Sample}\\
\cline{2-5}
\colhead{Property \hfil} & UF1 & UF2 & UF3 & \gaia, No POLCAT\\
& \colhead{(1)} & \colhead{(2)} & \colhead{(3)} & \colhead{(4)}
}
\startdata
\multicolumn{5}{l}{Figure~\ref{fig_G_para} Based: Locations of distribution peaks, no A$_V$ corrections applied}\\
\ \ \ \ \ \ $m_g$ (mag) & 15.6 & 16.8 & 18.2 & 20.4 \\
\ \ \ \ \ \ log($\pi$) (mas) & $-$0.50 & $-$0.36 & $-$0.30& +0.11 \\
\ \ \ \ \ \ Distance (kpc)& 3.2 & 2.3 & 2.0 & 0.78 \\
\ \ \ \ \ \ $M_g$ (mag) & 3.1 & 5.0 & 6.7 & 10.9 \\[6pt]
\multicolumn{5}{l}{Figure~\ref{fig_G2_hk_mh} Based: Sample Fractions of GPIPS POLCAT stars}\\
\ \ \ \ \ \ No 2MASS ($\sigma_{(H-K)} > 0.5$~mag) (\%)&  8.2 & 27.3 & 69.4 & ... \\
\ \ \ \ \ \ No \Gaia Match (\%)					& 14.0 & 24.2 & 17.5 & ... \\
\ \ \ \ \ \ \Gaia Match, no $\pi$ (\%)				& 31.9 & 22.5 & 5.4 &...\\
\ \ \ \ \ \ \Gaia Match, with $\pi$ (\%)				& 45.9 & 26.0 & 7.6 & ...\\[6pt]
\multicolumn{5}{l}{Figure~\ref{fig_G3_hk_para} Based: Distribution Properties for \Gaia$\pi$ Subsamples}\\
\multicolumn{5}{l}{\ \ \ \ \ \ Distance, in kpc, to cumulative percentage of subsample:}\\
\ \ \ \ \ \ \ \ 0.5\% &0.35 & 0.35 & 0.33 & ... \\
\ \ \ \ \ \ \ \ 10\% &0.91 & 0.72 & 0.71 & ... \\
\ \ \ \ \ \ \ \ 50\%&2.63 & 1.95 & 1.78 &... \\
\ \ \ \ \ \ \ \ 90\%&6.17 & 4.90 & 4.22 & ...\\
\ \ \ \ \ \ \ \ 99.5\% &11.89 & 10.59 & 9.44 & ... \\
\ \ \ \ \ \ Av at distrib. peak (mag)&3.2&1.6&2.2&$\sim$0\\[6pt]
\multicolumn{5}{l}{Derived properties, with A$_V$ corrections applied}\\
\ \ \ \ \ \ $M_g$ (mag)&$-$0.1&3.4&4.5&$\sim$11\\
\multicolumn{5}{l}{\ \ \ \ \ \ \Gaia Colors (G$_{BP}$ - G$_{RP}$) (mag) and Approx. Spectral Types}\\
\ \ \ \ \ \ \ Lum. Class III&+1.04 -- +1.46&+1.07 -- +1.20&...&...\\
&G7--K2.5\,III &G7--K0\,III&...&...\\[3pt]
\ \ \ \ \ \ \ Lum. Class V&$-$0.04 -- +0.19& +0.63 -- +0.98& +0.74 -- +0.96&+2.63 -- +2.91\\
&A2--A7\,V     &F4--G8\,V&F8--G7\,V&M1.5--M3\,V\\[3pt]
\ \ \ \ \ \ \ Lum. Class VII&...&...&...&$-$0.27 -- $-$0.11\\
&...&...&...&DA(?)\,VII\\
%
\hline
\enddata
\end{deluxetable}

These spectral type ranges
are based on the $g$-band apparent magnitudes, \gaia parallaxes, 2MASS colors,
and standard NICE conversions to A$_V$ drawn from values at the {\it peaks} of each distribution
in Figures~\ref{fig_G_para}, \ref{fig_G2_hk_mh}, and \ref{fig_G3_hk_para}.
The contoured distributions shown in these Figures also span ranges of all those quantities, so the 
derived spectral types should be viewed as notional characterizations, not quantitatively limited
ones.

In Table~\ref{tab_summary}, the UF1+\gaia$\pi$ matched stars span spectral types
of G7--K2.5 in the giant luminosity class and A2--A7 in the dwarf class. However, the former
is more likely to represent the stars in the UF1 subset, as A-type stars are rare in 
Figure~5 of \citet{Gaia_HRD}, whereas that giant range takes in much of the 
highly populated Red Clump
\citep[e.g.,][]{Pavel14}. For the UF2 stars, both giant and dwarf branches are likely, with
a slight dominance by the dwarfs. This luminosity class bifurcation might be a partial cause
of the A$_V$ bifurcation of the UF2+\gaia$\pi$ distribution (blue contours) in 
Figure~\ref{fig_G2_hk_mh}. The UF3 stars show distribution peak $M_g$ values that only
slice through the dwarf sequence of Figure~5 in \citet{Gaia_HRD} and only do so for
spectral types somewhat later than those of UF2 and quite distinct from the ones for UF1.

The \gaia stars not matched to POLCAT stars were already shown to be much closer than
the UF1, 2, or 3 stars. Given the proximity of these \gaia-only stars, extinctions are likely 
much less than one magnitude of A$_V$. As such, their inferred $M_g$ values slice through
both the white dwarf and dwarf sequences in Figure~5 of \citet{Gaia_HRD}. Given the high
space density of red dwarf stars, it is likely they dominate this POLCAT-unmatched subset of stars.
 
It is more difficult to assign distances and spectral types to POLCAT stars in the
no-\Gaia and \gaia${\rm no}$~$\pi$ subsamples. Figure~\ref{fig_G2_hk_mh} includes a reddening
line that can aid in interpreting the nature of the stars in these subsamples. For example, the
slope of the reddening line is such that the UF1+\gaia${\rm no}$~$\pi$ (orange contours) subset
could be deextincted by about 4~A$_V$ mag to fall closely over the UF1+\gaia~$\pi$ (red) 
subset. If this is the case, then both subsets would have nearly identical spectral types and
distances, with the \gaia${\rm no}$~$\pi$ subset merely suffering additional extinction, likely 
associated with denser molecular cloud directions. Interestingly, the UF1~no-\gaia (yellow contours)
subset, if deextincted by 10~mag to the UF1+\gaia$\pi$ value of about 4~A$_V$~mag would have 
brighter apparent magnitudes than the UF1+\gaia$\pi$ subsample. This could be caused
by the UF1~no-\gaia stars being closer or by having greater luminosities, perhaps caused by
supergiants located at much greater distances. The very red $(H - K)$ colors
for the UF1,~no~\Gaia stars, of 1.0~mag and beyond, 
cannot be due to nearby red dwarfs, however, as shown in the relative color-color diagram
of Figure~19 in \citetalias{Clemens12c}. Simulations of stellar types, distance,
and extinctions with GPIPS and \gaia detection limits imposed as priors are beyond this present
treatment. Yet, it does appear likely that all of the UF1 stars, across all \gaia-based subsets, 
are similar enough as to be recognized as red giants, many in the Red Clump, with various
degrees of foreground reddening and extinction.
The extinctions could be associated with individual dark clouds, with spiral arms that lie between 1 -- 7~ kpc such as Sagattarius and Scutum, or even with the central Galactic Bar.

Figures~\ref{fig_G_para}, \ref{fig_G2_hk_mh}, and 
\ref{fig_G3_hk_para}, taken together, provide the information necessary for identifying the near- and far-side GPIPS ``horizons."  As listed in Table~\ref{tab_summary}, 99\% of UF1 stars are located between distances of 0.35 and 11.9~kpc, with 90\% between 0.63 and 7.5~kpc. Thus, there
are sufficient UF1 and UF2 stars to conduct limited angular resolution polarization probes of
molecular clouds as close as 350-400~pc as well as to characterize magnetic fields in spiral arms in the Galactic midplane.

Having established the bases and representative values for the GPIPS horizons, an exploration of
the NIR GPIPS linear polarization properties for one
FOV is described in the following.

\subsection{Methods and Characterizations for One Field, GPIPS-1619}\label{one_field}

The data from  the GPIPS-1619 field, shown in Figure~\ref{fig_mosaic}.A, were used to explore methods of analyzing and characterizing the polarization properties of the stars for one FOV.
This began with plots of basic data properties and proceeded to analyzes of distributions of those properties. 

\begin{figure}
\includegraphics[width=7.25in]{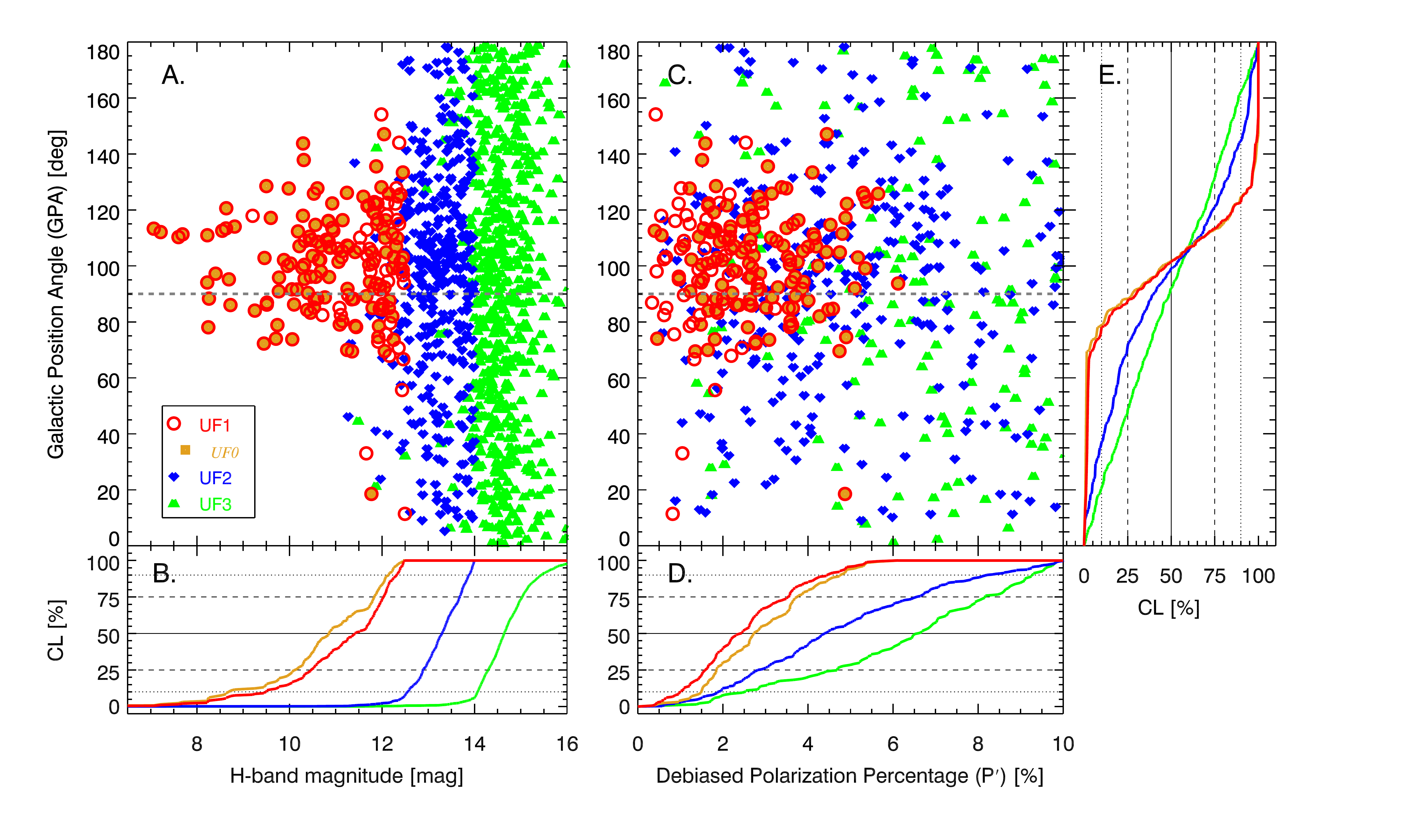}
\caption{Polarization position angle in Galactic coordinates (GPA) versus $H$-band stellar magnitude (Left, top A-panel) and versus debiased polarization percentage $P^\prime$ 
(Center, top C-panel) for GPIPS FOV number 1619.
Symbol colors identify stellar classification by UF number, as shown in the legend in the lower 
left corner of the A-panel. The dashed horizontal gray lines at  90$\degr$ in the A- and C-panels indicate the 
Milky Way disk-parallel GPA value. The brighter, UF1 stars (red open circles), most of which are also members of the UF0 subset (light-brown filled squares), exhibit lower 
polarization percentages and a smaller spread in GPA values, compared to the moderately fainter UF2 (blue) and much fainter UF3 (green) stars, though the non-UF1 stars retain some information concerning the B-fields they trace. The B-, D-, and E-panels present cumulative
likelihood (CL) distributions, after marginalizing over the other dimension, for each UF stellar subset. The B-panel (Left, bottom) reveals 
median $H$-band magnitudes of 10.8, 11.5, 13.3, and 14.6 for the UF0, UF1, UF2, and UF3 subsets. The D-panel (Center, bottom) cumulative likelihood distributions do not include polarization upper limits or $P^\prime$ values in excess of 10\%. The UF2 and UF3 distributions are shifted to higher $P^\prime$ values due to bias from their uncertainty contributions. The E-panel (Rightmost) shows CL distributions with a strong concentration near 100\degr\ (UF0, UF1) or much weaker concentrations closer to 90\degr\ (UF2, UF3).
\label{fig_GPA_vs_P_Hmag}}
\end{figure}

Figure \ref{fig_GPA_vs_P_Hmag} displays, for the stars in the GPIPS-1619 FOV, the dependence of polarization position angle relative to the Galactic frame of reference (Galactic Position Angle; GPA) on stellar $H$-band magnitude (upper-left, A-panel) and on debiased polarization percentage (upper-center, C-panel). The 197~UF1, 557~UF2, and 1172~UF3 stars in the POLCAT for GP1619
were trimmed to remove polarization upper limits ($\sigma_P \ge P_{RAW}$), as they contribute no meaningful GPA values. The remaining 
179 UF1, 352 UF2, and 563 UF3 stars are plotted as the red, blue, and green
symbols, respectively. 
The 129 member subset of UF1 stars that exhibit P$^\prime$SNR ($P^\prime / \sigma_P$)~$\ge 3$ (equivalent to $\sigma_{PA} < 9.6$\degr), and designated UF0, are plotted as filled light-brown squares, which appear within most of the red UF1 star location circles. 

The UF1 (and UF0) stars, and many UF2
stars, exhibit GPA values in the 70 to 130\degr\  range (A-panel) and $P^\prime$ values in the 
0.5 to 5\% range (C-panel).
The UF2 stars tend to exhibit higher $P^\prime$ values than do the UF1 stars and this
tendency is even stronger for the fainter, UF3 stars. 
Some of these higher polarization percentages could be real
and could correlate with higher dust column densities and coherent magnetic field orientations, traceable through extinction. 
However, these fainter stars are more likely to exhibit higher $P^\prime$ values due to noise bias 
and those values should be treated with caution \citepalias{Clemens12c}. 
Lower limits on P$^\prime$SNR alone are not sufficient to select high-confidence stellar polarizations, 
unless the limit values are quite high \citep[e.g., greater than 3-5;][]{Simmons85}.

In Figure~\ref{fig_GPA_vs_P_Hmag}, the upper-left, A-panel exhibits the expected  delineations into the UF designations of stars as a function of  
$H$-band mag. These boundaries are sharpest at the faint limits of UF1 ($m_H = 12.5$~mag) and UF2 ($m_H = 14$~mag).
However, since the UF definitions are based on both magnitude {\it and} polarization uncertainty, the minority presence
of some UF2 and UF3 stars across the magnitude boundaries is to be expected. The main conclusion
from the distribution of stars in the A-panel is that the UF1 (and UF0 subset) stars appear near, but not
completely on, the GPA~=~90$\degr$ disk-parallel line. 
The UF2 stars have a similar, though much
weaker correlation with the same orientation, and the UF3 stars are only weakly constrained in their
GPA values. 

In Figure~\ref{fig_GPA_vs_P_Hmag}, the B-, D-, and E-panels present cumulative likelihoods of the marginalized distributions, color-coded by UF designation in the same fashion as for the plotted symbols in the A- and C-panels. The B-panel (lower-left) shows how the UF subsets are distributed with 
$H$-band magnitude. Median magnitudes are 10.8, 11.5, 13.3, and 14.6~mag for the UF0, UF1, UF2, and UF3 samples, respectively. The D-panel (center-bottom) shows the CL curves of $P^\prime$ after being marginalized over GPA. However, the likelihood accumulation window was truncated at 10\%\ of $P^\prime$, which affects the UF2 and UF3 CLs, as some of those stars have $P^\prime$ values in excess of that limit. In that D-panel, the $P^\prime$ medians of 2.4 and 2.7\%\ for UF1 and UF0 are accurate, but the UF2 and UF3 medians of 4.4 and 6.5\%\ represent only lower limits because of 
the truncation.

The GPA CLs, marginalized over $P^\prime$, but including the truncation at 10\%, are shown in the E-panel (rightmost) of Figure~\ref{fig_GPA_vs_P_Hmag}. The medians for UF0 and UF1 are both 102\degr, while UF2 shows 98\degr\ and UF3 shows 94\degr. The UF0 and UF1 CL curves show the least separations between their first and third quartiles at 25\degr, UF2 is intermediate at 52\degr, while UF3 is the widest at 80\degr, a value close to the 90\degr\ expected for a random distribution.

The GPA distributions of Figure~\ref{fig_GPA_vs_P_Hmag} were used to compute a mean value of GPA and its standard deviation ($\Delta_{GPA}$) for each UF subset of stars. These values were computed both in an unweighted fashion, without respect to individual GPA uncertainties, and using variance weighting by those uncertainties. (Note that Stokes $U$ and $Q$ averaging was not performed - the analyses used GPA values only.) The resulting values appear in Table~\ref{tab_GP1619}. The first column in the Table indicates the UF star subset, with UF0 entries shown in italics to highlight that it is a subset of UF1. The numbers of GPIPS-1619 stars with detected debiased polarizations used in each UF sample is in the second column. The remaining columns list the unweighted GPA means, the unweighted GPA standard deviations, the weighted GPA means (with propagated uncertainties in parentheses), and the weighted GPA standard deviations (and uncertainties).

The UF1 (and UF0) stars show the greatest mean GPA departure from the disk-parallel value of 90\degr, while the UF3 stars show the least such departure. The standard deviation progression with UF type is stronger, with the UF1 (and UF0) stars having weighted $\Delta_{GPA}$ of under 15\degr. The UF2 and UF3 samples, with $\Delta_{GPA}$ values of 41\degr\ and 51\degr, respectively, are close to being completely uniform in their GPA spreads (for which $\Delta_{GPA}$ would be 52\degr). There are no major differences in mean GPA or $\Delta_{GPA}$ between the unweighted and weighted values, though the latter provide uncertainties that give context to the values.

\begin{deluxetable}{cccccc}
\tablecaption{GPIPS-1619 Average Galactic Position Angles \label{tab_GP1619}}
\tablewidth{7truein}
\tablehead{
\colhead{UF} & \colhead{Num.} & \multicolumn{2}{c}{Unweighted} & \multicolumn{2}{c}{Weighted}  \\
\colhead{Subset} & \colhead{$P^\prime > 0$\%} & \colhead{$\langle $GPA$ \rangle$} & \colhead{$\Delta_{GPA}$} & \colhead{$\langle $GPA$ \rangle$} & \colhead{$\Delta_{GPA}$}  \\
&&\colhead{(deg.)}&\colhead{(deg.)}&\colhead{(deg.)}&\colhead{(deg.)}\\
\colhead{(1)}&\colhead{(2)}&\colhead{(3)}&\colhead{(4)}&\colhead{(5)}&\colhead{(6)}
}
\startdata
UF1 & 179 & 100.8 & 16.1 & 102.88 (0.32) & 14.69 (0.32) \\
\ \ \ \ {\it UF0}& {\it 129} & {\it 101.4} & {\it 15.9} & {\it 102.94 (0.33)} & {\it 14.62 (0.33)} \\
UF2 & 352 & 95.2 & 39.7 & 97.5 (1.0) & 40.6 (1.0) \\
UF3 & 563 & 90.9 & 50.2 & 88.5 (0.9) & 51.3 (0.9) \\
\enddata
\end{deluxetable}

\subsubsection{Analyses Based on UF1 Stars}

Characterizing the properties of the Galactic disk magnetic field using GPIPS data products can proceed along many different paths, invoking a wide variety of data selection criteria. This could include, for example, utilizing all of the GPIPS stellar Stokes $U$ and $Q$ measurements, or alternatively, selecting only the data corresponding to high P$^\prime$SNR (e.g., $\ge$ 5). All choices of data selection schemes bring some type of bias, or focus, on a particular data range or ISM characteristic.
High P$^\prime$SNR cutoffs select the highest quality polarization measurements, but reduce the number and
spatial sampling that could reveal magnetic field properties with adequate confidence levels.
Low P$^\prime$SNR cutoffs introduce excessive noise and false positives. 
In Appendix~\ref{Appendix_PSNR}, the UF1 stellar subset is shown to reveal similar properties to those found in the high P$^\prime$SNR UF0 subset. This reduces the need to restrict further analyses to the smaller UF0 subset, as the larger UF1 set is adequate for establishing overall magnetic field properties and correlations. The UF1 stars suffer less noise-biasing effects
than fainter stars and so enable higher-significance differential comparisons of polarization properties between GPIPS FOVs than would be possible using UF2 and/or UF3 stars.

For those reasons, the analyses presented in the remainder of this Section and in all of the 
following Sections were
performed by focusing on the properties of the stars classified as UF1 in each of the 3,237 GPIPS FOVs. Future studies utilizing other subsets of the GPIPS stars, with other selection criteria
and with different weighting schemes, could reveal additional magnetic field behavior that 
may be missed in the current analyses. Indeed, Bayesian analyses \citep[e.g.,][]{Clemens18}
have already shown great promise for combining a wide range of P$^\prime$SNR stellar polarization values with \Gaia distance information. 

\subsubsection{Polarization Properties of the UF1 Stars}\label{sec_uf1}

\begin{figure}
\includegraphics[width=7.25in]{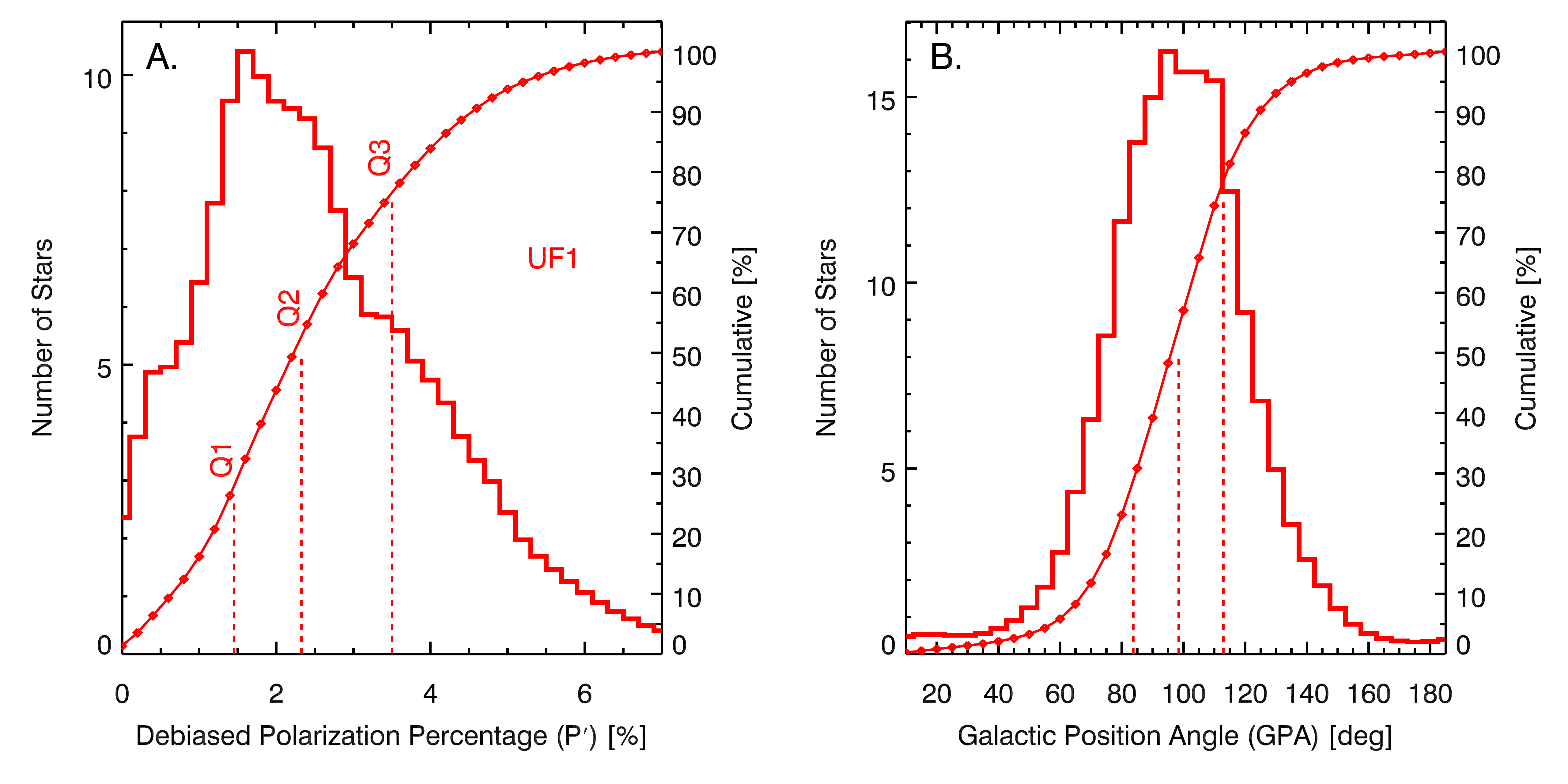}
\caption{Histograms of debiased polarization percentage $P^\prime$ (Left, A-panel) and Galactic position angle GPA
(Right, B-panel) for UF1 stars in GPIPS-1619. The histograms were
constructed using accumulated gaussian distribution representations for the properties 
of each star, as noted in the
text. Cumulative probability distributions are shown as the smooth curves that connect points
centered in each bin and are referenced to the right axis labels. Vertical red dashed lines and labels identify
the locations of the three quartile boundaries of the cumulative distributions. 
\label{fig_UF1_histos_GPA_and_P}}
\end{figure}

Figure \ref{fig_UF1_histos_GPA_and_P} shows histograms of debiased polarization percentage
$P^\prime$ (left, A-panel) and Galactic position angle GPA (right, B-panel) for the 
179 UF1 stars that have $P^\prime > 0$\% in the GPIPS-1619 FOV.
These histograms were constructed in a manner designed to account for the uncertainties in the individual measured 
$P^\prime$ and GPA values via accumulation of representative gaussian probability distributions \citep[described in the Appendix of][]{Clemens13}. This process results in smoother histograms than if the 
$P^\prime$ and GPA values were directly binned, and more accurately reflects
the likelihood functions for these quantities. The $P^\prime$ histogram (A-panel) peaks at just 
under 2\% and the GPA histogram\footnote{Angle-aliasing in the GPA histogram was reduced by accumulating the gaussian probability distributions across an angle range eleven times wider than the usual 0 to 180\degr\ range, instead of truncating at the range boundaries, and then shifting and adding the probabilities outside the usual range into the bins within that range.} is centered near 100$\degr$. 

Across the full set of GPIPS FOVs, many of the resulting histogram probability functions 
were asymmetric and sometimes double-peaked. Hence, simple representative functions or
fits, such as gaussians, cannot accurately characterize the actual distributions.

Accordingly, cumulative probability distributions were computed for the UF1 subset for each histogram 
and are overlaid as thin red lines in Figure~\ref{fig_UF1_histos_GPA_and_P}. These were analyzed to find the quartile boundaries 
(Q1=25\%, Q2=median, and Q3=75\% cumulative probability). The resulting median $P^\prime$ 
(hereafter P50) for this field is 2.32\%, while the median GPA (hereafter GPA50) 
is 98.5$\degr$. The combination of GPA50 in Figure~\ref{fig_UF1_histos_GPA_and_P}.B being 98.5\degr\ for UF1 stars while the weighted mean for the same stars in Table~\ref{tab_GP1619} is $102.9\degr \pm 0.3$\degr\ is evidence that the Figure~\ref{fig_UF1_histos_GPA_and_P}.B UF1 distribution has some non-gaussian nature, despite it gaussian appearance.
The values for the first and
third UF1 quartile boundaries are 1.45 and 3.50\% for $P^\prime$ and 83.7 and 112.9\degr\ 
for GPA, as indicated on Figure~\ref{fig_UF1_histos_GPA_and_P}. 

In order to obtain robust measures of the widths of the GPA distributions, the interquartile ranges, computed from the differences between the $x$-axis locations of the first and third quartiles (Q3 - Q1), were adopted. For the GPIPS-1619 GPA histogram, this width difference, designated 
WGPA, is 29.2\degr\ for the UF1 stars. If the GPA distributions were perfectly gaussian, the gaussian
width parameters would be 42.5\% of the WGPA values, or about 12.4\degr\ for the GPIPS-1619 FOV. This value is somewhat less than the standard deviations reported in Table~\ref{tab_GP1619}, again indicating some non-gaussian nature characterizes the GPIPS-1619 UF1 GPA values.

Similar histogram analyses were performed using the UF1 stars contained in the POLCATs 
for each of the GPIPS FOVs. The four key characterizing values extracted for each FOV included 
the numbers of UF1 stars in each field, their median $P^\prime$ values (P50), 
their median GPA (GPA50), and their interquartile range of 
the GPA histograms (WGPA), all of which are evaluated in the following.

\subsection{FOV-based GPIPS Characterization of Polarization and Magnetic Field Properties}\label{all fields}

In this Section, histograms of these four key characterizing quantities are presented, as are 
their distributions as functions of Galactic longitude and latitude, at the 10~arcmin 
angular resolution corresponding to the Mimir FOV size. This selection enables generation of high-significance values by quantifying the behavior of the properties of the many UF1
stars in each FOV, and doing so reveals both large-scale trends and moderate-scale
departures from those trends. Detailed examination of properties on finer angular scales
and across overlapping FOVs is needed, for example to establish the magnetic field
properties associated with resolved molecular clouds \citep[e.g.,][]{Marchwinski12, Hoq17}, but is beyond the scope of this paper.

The large-scale, FOV-based characterizations begin with histogram analyses of 
the numbers of UF1 stars measured for polarization in each FOV,
the median polarizations, the median Galactic position angles, and GPA widths, followed
by representations of the sky distributions of these same quantities.


\subsubsection{Histograms of FOV-based Properties}\label{sec_FOV_histos}

\begin{figure}
\includegraphics[width=7.2in]{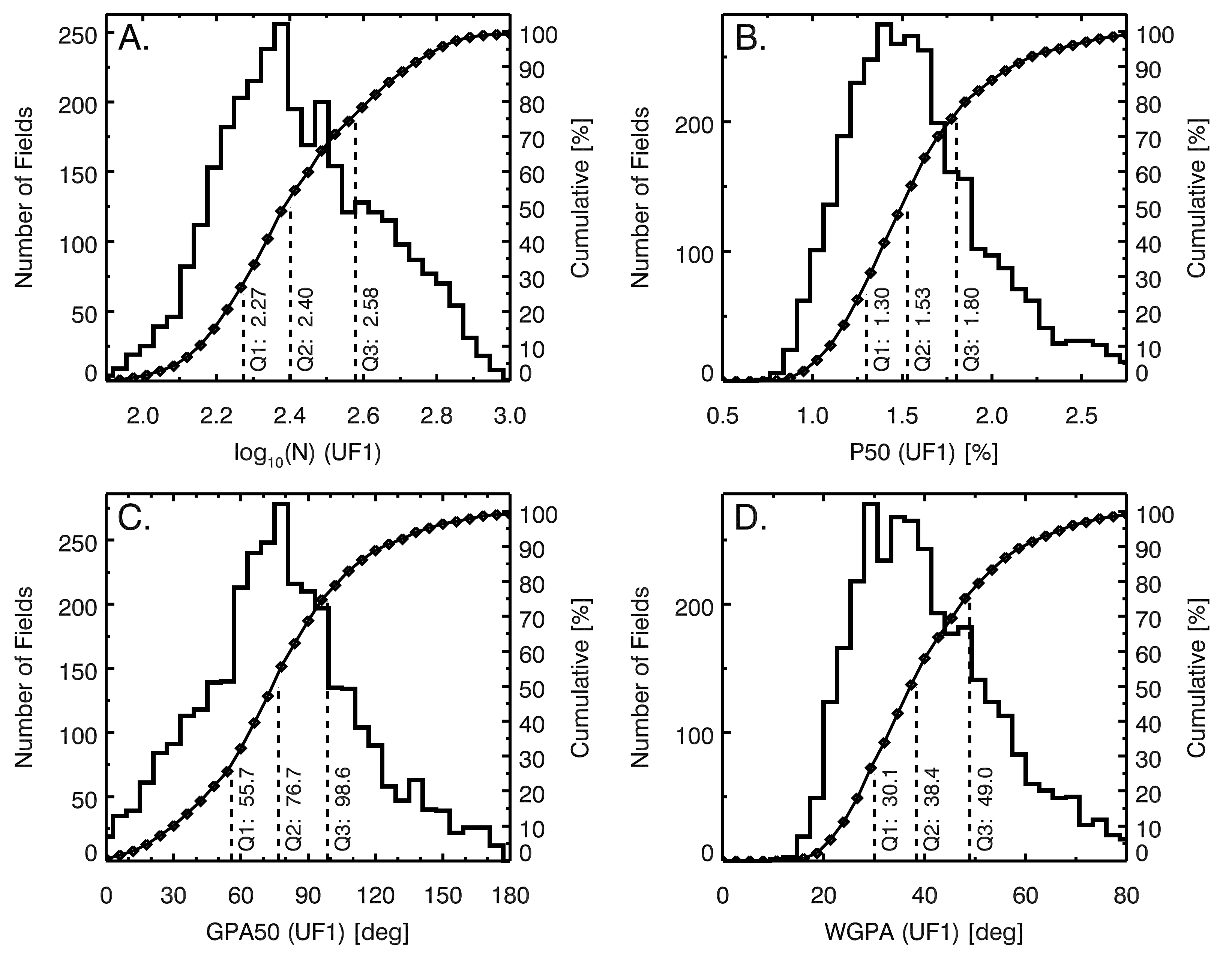}
\caption{Histograms of the four key quantities measured for the UF1 stars in each 
of the GPIPS FOVs. Curves connecting diamonds are the 
cumulative probability distributions, with downward dashed lines locating the distribution 
quartile boundaries.
(Top left, A-panel): Histogram of the base-10 logarithm of the number of UF1 stars in 
each GPIPS FOV.  The median number of UF1 stars per 
Mimir FOV is 252. 
(Top right, B-panel): Distribution of median percentage polarizations (P50) for 
UF1 stars in each FOV. The median of this distribution is 1.53\% at $H$-band.
(Bottom left, C-panel): Distribution of median Galactic polarization position angle (GPA50) 
of the UF1 stars in each GPIPS FOV, where 90\degr\ represents a disk-parallel orientation.
(Bottom right, D-panel): Distribution of interquartile ranges (WGPA) of the GPA distributions of the UF1 stars in each GPIPS FOV. 
\label{fig_field_histos}}
\end{figure}

Figure \ref{fig_field_histos} presents histograms of values for the four FOV-based properties,
as determined from the distributions of properties for the UF1 stars in each of the 3,237 GPIPS FOVs. 
Table~\ref{tab_quartiles} lists the quartile boundaries for the distribution functions for each
quantity.

The upper-left, A-panel in Figure~\ref{fig_field_histos} shows the distribution 
of numbers of UF1 stars in each 
FOV, binned by the base-10 logarithm of the number of star counts. The cumulative probability distribution is overplotted. As was done for the distributions
in the GPIPS-1619 FOV, this cumulative distribution was analyzed to identify the distribution quartile
boundaries. 
The median number of UF1 stars
in a GPIPS FOVs is 252, while less than 25\% of the FOVs have fewer than 188 
stars and less than 25\% have more than 378 stars. 

The upper-right, B-panel of Figure~\ref{fig_field_histos} displays the histogram of the P50 (median debiased polarization percentage
$P^\prime$) for
UF1 stars in each of the GPIPS FOVs. The median of this distribution is 1.53\%, with quartile
boundaries at 1.30 and 1.80\%. This median is only 1.06 times greater than the value found in an analysis of the first 18\% of the GPIPS FOVs (DR1---\citetalias{Clemens12c}).
These polarization percentages are smaller than optical wavelength values, which average
closer to 5---7\%\  in the interstellar medium \citep{Hall49, Hiltner49a, Hiltner49b}.
This difference is expected due to the wavelength dependence of starlight polarization \citep{Serkowski73, Serkowski75, Wilking80}.

The lower-left, C-panel of Figure~\ref{fig_field_histos} displays the distribution of
median Galactic polarization position angles (GPA50) for the UF1 stars in each GPIPS FOV.
As found in \citetalias{Clemens12c}, the distribution is centrally peaked, though offset by 
$13.\degr3$ from the expected, disk-parallel value
of GPA~=~90$\degr$ that characterizes most magnetic field models \citep[e.g.,][]{Ferriere00}. 
Over the region of the
Galactic mid-plane surveyed by GPIPS in the first Galactic quadrant, the median magnetic field orientation
is not purely parallel to the Galactic disk. It also exhibits 
broad orientation deviation wings that extend to both Galactic pole directions.

\begin{deluxetable}{cccc}
\tablecaption{GPIPS FOV-Based UF1 Polarization Distributions Properties\label{tab_quartiles}}
\tablewidth{0pt}
\tablehead{
\colhead{Quantity} & \multicolumn{3}{c}{Quartile Boundary}\\
&\colhead{First} & \colhead{Second} & \colhead{Third} \\
\colhead{(1)}&\colhead{(2)}&\colhead{(3)}&\colhead{(4)}
}
\startdata
log$_{10}$(N)	&2.27& 2.40&2.58\\
P50 [\%]&1.30& 1.53&1.80\\
GPA50 [\degr]&55.7&76.7&98.6\\
WGPA [\degr]&30.1&38.4&49.0\\
\enddata
\end{deluxetable}

The lower-right, D-panel of Figure~\ref{fig_field_histos} shows the distribution of interquartile ranges 
(WGPA) of the UF1 GPAs of each individual GPIPS FOV. The distribution 
peaks near 30-40$\degr$, with quartiles at 
30.1, 38.4, and 49.0$\degr$. These WGPA distributions in the 
GPIPS FOVs indicate that uniformly parallel magnetic fields are rare across 10~arcmin FOVs in the Galactic disk. This result has implications for assessments of the degree of
magnetic or hydrodynamic turbulence and the ratios of energy density in the random
and uniform magnetic field components \citep[e.g.,][]{Jones89}.

The offset of the median GPA from being purely disk-parallel and the wide range of position 
angles present in each GPIPS FOV indicate that magnetic field models dominated by strongly 
uniform, disk-parallel behavior may not be adequate to describe these characterizations. In order to ascertain how these key properties vary with location in the
Galactic disk, their one-dimensional (1-D) and two-dimensional (2D) distributions were 
examined next.

\subsubsection{Galactic Latitude and Longitude 1-D Behavior of FOV-based Properties}\label{1d}

\begin{figure}
\includegraphics[width=7.2in]{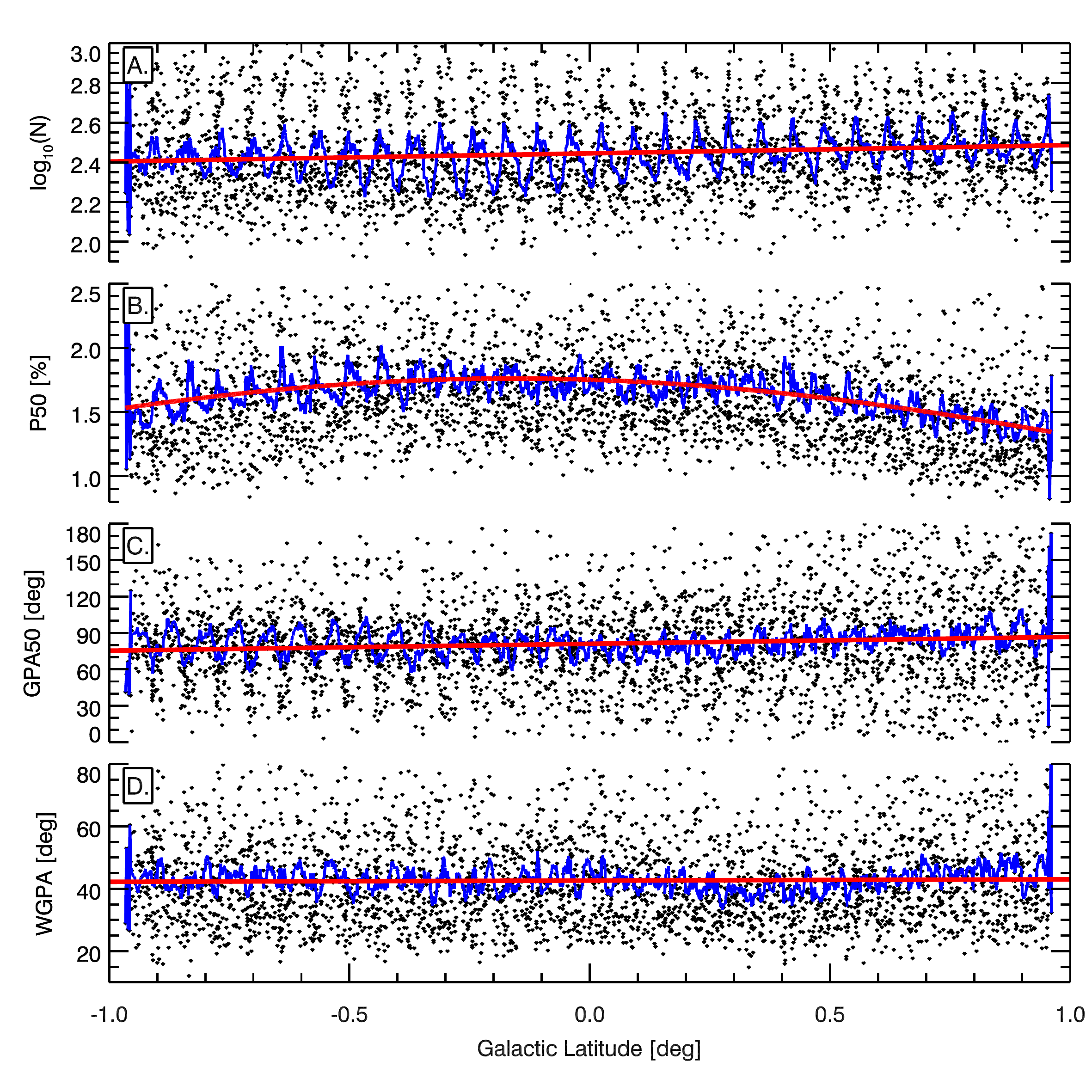}
\caption{Galactic latitude variations of the four characterizing quantities measured for the UF1 stars in each of the
GPIPS FOVs (black dots), running averages (10-point; blue lines), and best-fit lines or 
gaussian (red lines - see text). 
(Top, A-panel) Base-ten log of the number of UF1 stars in each FOV, log(N).
(Middle-top, B-panel) P50.
(Middle-bottom, C-panel) GPA50. 
(Bottom, D-panel) WGPA.
\label{fig_Quad_B}}
\end{figure}

\begin{figure}
\includegraphics[width=7.2in]{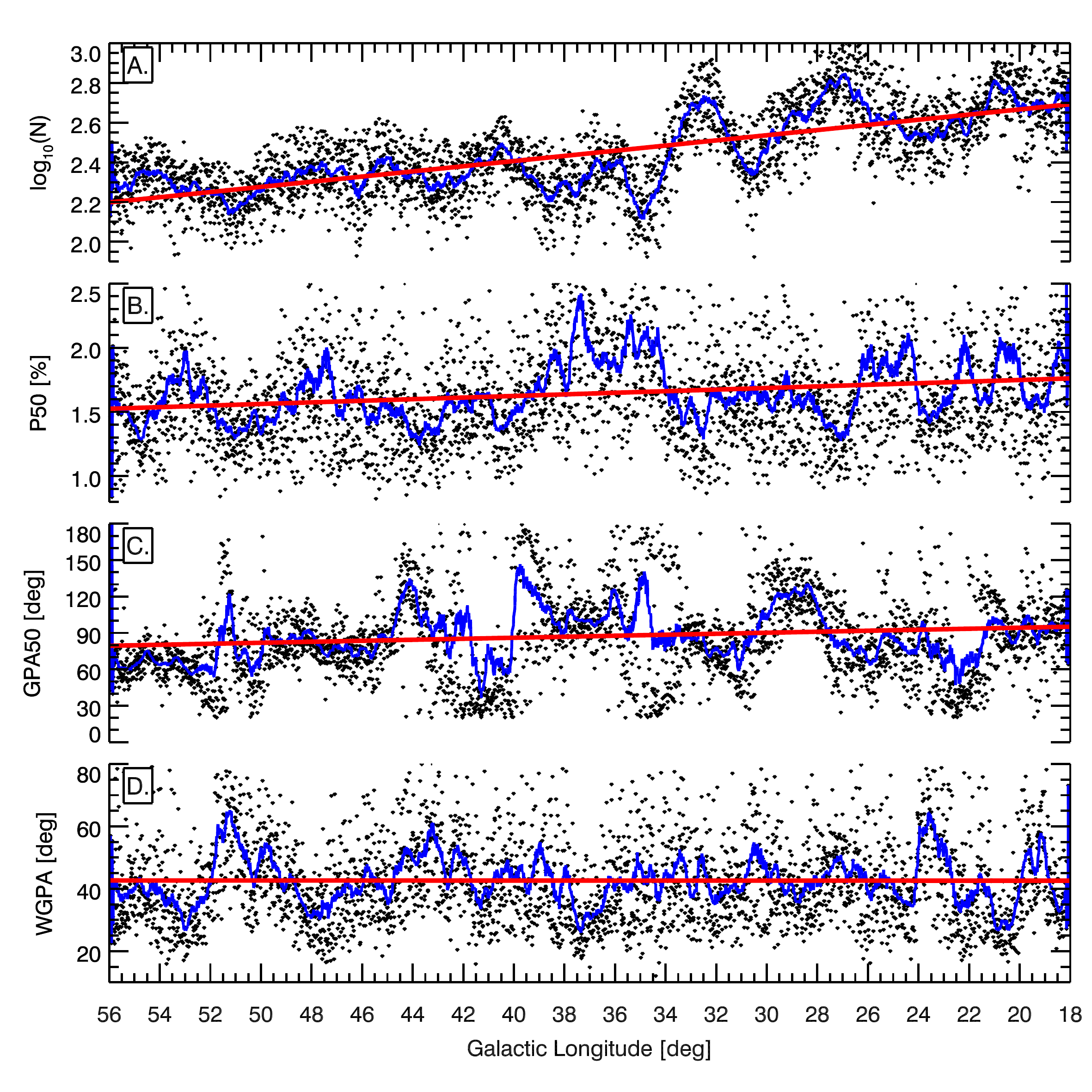}
\caption{Galactic longitude variations of the four characterizing quantities measured in each 
GPIPS FOV (black dots), running averages (10 point; blue lines), and best-fit lines (red). 
(Top, A-panel) Base-ten logarithm of the number of UF1 stars in each FOV---the Galactic bulge may contribute equal numbers of stars as does the Galactic disk from longitude 18\degr\ to about 34\degr, 
but there are strong variations on few-degree size scales.
(Middle-top, B-panel) P50. There are regions of coherent,
high-percentage polarizations and a slow trend toward weaker percentage polarization as longitude increases (to the left).
(Middle-bottom, C-panel) GPA50.  Regions of coherent, significant departures 
from the best fit line, which itself swings from 95$\degr$ to 80$\degr$ for low to high
longitudes, are present.
(Bottom, D-Panel) WGPA shows no significant
trend with longitude, though it has zonal departures.
\label{fig_Quad_L}}
\end{figure}

The same four, FOV-based quantities for each GPIPS FOV, derived from the UF1 stars, 
are plotted versus Galactic latitude in Figure~\ref{fig_Quad_B} and versus
Galactic longitude in Figure~\ref{fig_Quad_L}. In both Figures, stacked 
plots are shown that display log(N), P50, GPA50, and WGPA, respectively, from top to bottom. Black dots mark the 
values of each quantity measured for each of the GPIPS FOVs. 

\paragraph{Galactic Latitude 1-D Behaviors}\label{b_1d}
In Figure~\ref{fig_Quad_B}, in addition to showing the black points for the FOV-based 
values, a 10-point running average is shown as the blue curve, a linear fit with latitude 
is shown in red for three of the four 
quantities, and a gaussian fit with latitude is shown in red for the P50 plot.
The top, A-panel shows that log(N) displays a strong, high-angular frequency modulation 
(30 cycles over 2\degr) that is
most easily seen in the blue running average curve. This modulation is likely due to the slight
variation of FOV Galactic longitude with latitude, related to the equatorial grid used to conduct the GPIPS observations (e.g., Figures~\ref{fig_N_LB}  to \ref{fig_WGPA_LB}, below).
The modulation suggests that the log(N) dependence on longitude will be more pronounced than it is on latitude. Ignoring this modulation, the overall log(N) dependence
on latitude appears to show a slight depression in star counts from about B$= -0.6$\degr\ 
to $+0.1$\degr\ 
(depicted by the blue, running-average means falling below the red linear fit), 
coupled with a weak rise in counts for positive latitude values (or a decrease in star counts for negative
latitude values). The mid-plane decrease in counts is likely due to extinction by distant dust cloud complexes in the disk, 
which are expected to 
have small angular scale heights. The decrease in counts for negative latitude may be caused by
extinction from more nearby clouds with larger angular offsets. The red line shows the linear fit, namely:
$log_{10}(N) = (2.445 \pm 0.004) + (0.041 \pm 0.007) \times {\rm GB}$, where GB represents Galactic latitude, in degrees.

The middle-top, B-panel of Figure~\ref{fig_Quad_B} that presents P50 also exhibits some of the same effects of the longitude beating. 
In addition, it
also shows a weak rise in value, by about 0.2-0.3\%, centered near the mid-plane and 
falling toward the survey latitude limits. The data were fit with a gaussian, 
returning an amplitude of $1.760 \pm 0.011$\%, a gaussian
width of $1.53 \pm 0.06\degr$, and center at GB=$-0.160 \pm 0.022\degr$.
This polarization rise near the Galactic equator may be due to polarization contributed by 
more distant dust cloud complexes or ones with higher dust column densities.

The fitted gaussian latitude width corresponds to a 3.6\degr\ FWHM, which, if associated with the 2.6~kpc median
distance inferred for the UF1 stellar count density in the Figure~\ref{fig_G3_hk_para}
\Gaia-GPIPS 
comparisons described earlier,  implies an NIR polarization Galactic disk thickness FWHM of 
160~pc, only $\sim$30\%\  thicker than the 120~pc FWHM for the CO-traced gas layer 
\citep{Clemens88}. 
A more realistic estimate might use half the distance to the UF1 star distribution as better representing the effective mean polarizing dust distance, resulting in an even thinner dust layer thickness, and one in agreement with the CO layer thickness.
Both estimates are thinner than the $400 \pm 30$~pc exponential scale
height found for the Galactic magnetic
field model developed by \citet{JF12}, based primarily on synchrotron Rotation Measures (RMs)
of extragalactic radio sources. For this GPIPS analysis, the apparently smaller NIR polarization scale height likely results from the combined effects of a larger magnetic field scale height and a smaller dust layer scale height. 

The middle-bottom, C-panel of Figure~\ref{fig_Quad_B}, showing GPA50, also exhibits longitude beating, though with beat frequencies and amplitudes that change
with latitude, suggesting complex longitude and latitude dependencies. The blue, running-average line segments
and the red linear fit both exhibit means similar to the median in the histogram
analysis described in Section~\ref{sec_FOV_histos}. The linear fit gave a GPA50 value at GB=$-1$\degr\ of $75.4 \pm 1.3\degr$ which rises to $86.7 \pm 1.3\degr$ by GB=$+1$\degr. Hence, the median  magnetic field orientation exhibits a weak mean dependence ($5.7 \pm 1.1$\% per deg) on
latitude over the 2\degr\ sampled by GPIPS. 

Finally, the bottom, D-panel of Figure~\ref{fig_Quad_B} shows that WGPA exhibits weak longitude beating
and almost no systematic variation with latitude. A linear fit returned a mean WPGA of 
$42.63 \pm 0.26$\degr\ and a slope of $0.41 \pm 0.47$ degrees of WGPA per degree of latitude. This result indicates that the significantly dispersed 
magnetic field orientation angles associated with WGPA values of 40$\degr$ or more are common and do not disappear at any latitude within the GPIPS region. The black points
at all latitudes do distribute along WGPA with the same pile-up at low values and greater 
spread at high values characterizing the D-panel WGPA histogram of Figure~\ref{fig_field_histos}.
Thus, to first order, the relative incidences of coherent polarizations (low WGPA) and 
disordered polarizations are independent of Galactic latitude (but see Section~\ref{sec_b-zones}, below).

\paragraph{Galactic Longitude 1-D Behaviors}\label{l_1d}

Figure~\ref{fig_Quad_L} presents the same four, FOV-based quantities versus Galactic
longitude L, along with the same 10-point running averages as blue lines and linear fits 
as red lines.  In all four of the panels in this Figure, the quantities vary strongly and 
more coherently than they do with latitude B (Figure~\ref{fig_Quad_B}).

The top, A-panel of log(N) in Figure~\ref{fig_Quad_L} 
shows a rise in average numbers of UF1 stars
per FOV toward lower Galactic longitudes, mostly for GL~$< 34$\degr. 
This result was attributed in \citetalias{Clemens12c} to the appearance of the Galactic bulge stars for these longitudes. 
Here, the $\sim$0.3 dex offset, between FOVs with GL~$> 34$\degr\ and FOVs with GL~$< 34$\degr, may indicate that the Galactic bulge
contributes about the same number of stars to the GPIPS FOVs as are contributed by the Galactic disk, for 
FOVs with GL~$< 34$\degr. The predominance 
of red giants in the bulge makes many of them detectable at the UF1 GPIPS sensitivity level,
as predicted in \citetalias{Clemens12a} and shown earlier in Section~\ref{gaia_analysis}. This finding is another indication that UF1 stars probe to distances
well into the bulge along these longitudes, likely in the 5-7~kpc range. In Figure~\ref{fig_Quad_L}.A, there are also some significant decreases in UF1 star counts, especially near
longitudes 24, 30, 35, 38, and 51\degr\ and some apparent increases in star counts (e.g., near 21, 27, and 32\degr). The decreases span 0.5 to 2\degr\ of longitude and
may be due to large complexes of dust (and gas) foreground to the bulk of the background
stars seen in the neighboring fields. 
They could also represent interarm regions with fewer stars between richer spiral arms. A more detailed comparison of GPIPS star counts to star counts for \Gaia and GLIMPSE as well as to 2MASS (H-K) colors is described in Section~\ref{star_counts} below.

The middle-top, B-panel in Figure~\ref{fig_Quad_L} shows a linear rise of P50 with decreasing longitude, though only by $0.24 \pm 0.05$\% over the mean value of 
$1.53 \pm 0.05$\%. Individual departures to higher and lower values
far exceed this weak gradient. Some longitudes exhibit P50 decreases (e.g., 19, 22, 23.5, 27, 33, 
40.5, 44, 51, and 55\degr) while others appear to have P50 increases (18.5, 20.5, 22.7, 25, 35, 38, 48, 53\degr).
Whether P50 departures correlate with log(N) departures is examined in Section~\ref{2d}.

The middle-bottom, C-panel in Figure~\ref{fig_Quad_L} shows that GPA50 exhibits the most dramatic variations with longitude. 
Defining any low-order polynomial trend was difficult. The red line, representing a linear fit, 
traces changes in the mean GPA50, from $95.1 \pm 2.6$\degr\ at longitude 18\degr, to 
$79.4 \pm 4.1$\degr\ at longitude 56\degr. However, the GPA50 values swing by 90\degr\ 
or more over longitude intervals as short as 2-4\degr. The blue, 10-point moving averages
also show that some GPA50 swings span nearly 180\degr\ over a few degrees of 
longitude, especially near longitudes 35, 40, and 52\degr. In contrast to the generally
uniform behavior of GPA50 with Galactic latitude (Figure~\ref{fig_Quad_B}), the GPA50 longitude behavior is strongly
non-uniform. Quantitatively, the 10-point moving averages deviate by more than 30\degr\ 
from disk-parallel orientations for over 30\% of the longitude range surveyed (when 
averaged over latitude), compared to no such deviations for the latitude range survey 
(when averaged over all longitudes). Magnetic field orientation deviations in the disk
midplane zone are mostly dominated by longitude effects and are nearly unaffected by purely
latitude effects.

The bottom, D-panel in Figure~\ref{fig_Quad_L} shows that WGPA exhibits the same lack of overall change with longitude that it exhibited with latitude in Figure~\ref{fig_Quad_B}.D. However, as for the previous
three quantities, there are short intervals of longitude over which WGPA strongly deviates
from the mean value, especially near longitudes 21, 24, 37, 43, and 52\degr. Two of
these positive WGPA deviations, at 24 and 52\degr, correspond to longitude zones where
large changes in GPA50 take place. Such rapid GPA changes could help drive WGPA to larger than average values.
There also appear to be additional two- or three-way correlations, especially for longitude ranges
20-21, 36-38, 47-49, and 52.5-54\degr, where GPA50s are near disk-parallel, P50 values
are greater than average, and WGPA values are lower than average. These behaviors
could be signaling the existence of regions of uniform, strong magnetic fields that are oriented
parallel to the Galactic disk, an aspect explored further in Section~\ref{sec_b-zones}.

These variations and correlations are intriguing and likely can help constrain models
of the Galactic disk magnetic field. However, the lines
of sight to individual GPIPS stars in each field can span quite
different path lengths and encounter a variety of distinct dusty molecular clouds and/or
spans of diffuse ISM. Interpretation of the variations must account for these line-of-sight 
differences.

\subsubsection{Comparisons of Star Counts and Colors versus Longitude}\label{star_counts}

\begin{figure}
\includegraphics[width=6.5in]{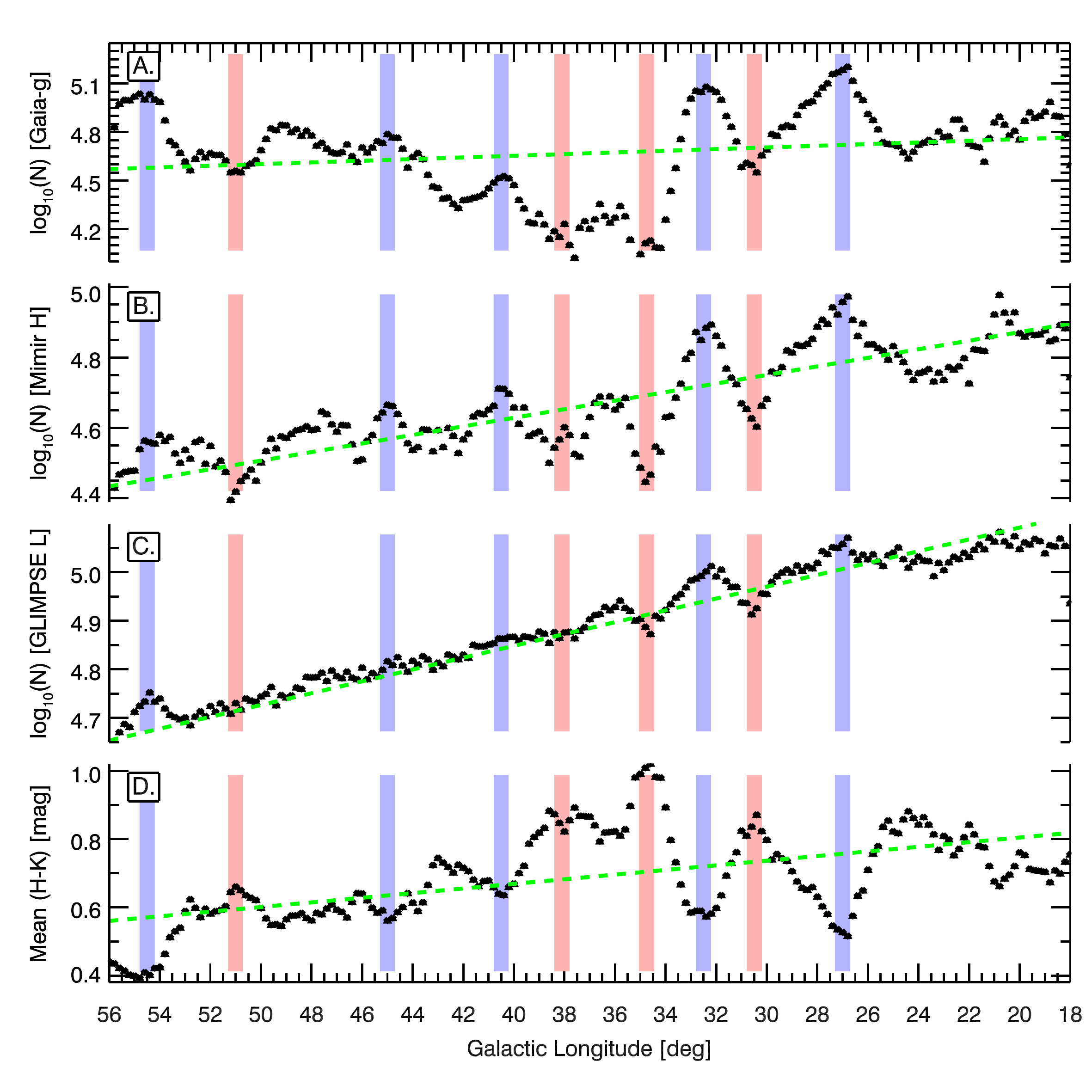}
\caption{Comparison of star counts and colors for \Gaia, Mimir, \glimpse, and 2MASS.
In all panels, binned data points are shown as black symbols, unweighted fits with longitude 
are shown as dashed green lines, and local regions with changes in properties are identified
by vertical light-red or light-blue rectangles. The top, A-panel, presents the base-ten logarithm of
the numbers of \gaia stars found in the GPIPS FOVs and having $g$-band magnitude uncertainties under 0.33~mag. The middle-top, B-panel, presents similar star counts for Mimir $H$-band DR4 POLCAT stars.
The middle-bottom, C-panel, shows the star counts for \glimpse 3.6~\micron\ (``$L$''-band) stars.
The bottom, D-panel, shows the run of unweighted mean (H-K) stellar colors, from 2MASS.
The photometric tracers show decreases in star counts with increasing longitude, while
the colors become less reddened with increasing longitude. The light-red vertical rectangles show
the correlation of localized redder star colors with star count decrements. The light-blue
rectangles show where less reddened stellar colors correspond to local increases in star counts.  
\label{fig_count_compare}}
\end{figure}

Figure~\ref{fig_count_compare} shows the Galactic longitude behavior of the base-ten
logarithm of the star counts in the
\Gaia $g$-band (top, A-panel), Mimir POLCAT $H$-band (middle-top, B-panel), and \glimpse 3.6~\micron\ band (``$L$-band''; middle-bottom, C-panel) as well as mean (H-K) colors from 2MASS (bottom, D-panel) in
0.2~deg wide longitude bins.
For the star count panels, stars were included if they appeared in the GPIPS FOVs and had
photometric uncertainties under 0.33~mag. For the color panel, stars from 2MASS were included if their $H$ and $K$ magnitudes had uncertainties under 0.5~mag and the propagated (H-K) uncertainty was under 0.33 mag. The colors of the 2MASS stars meeting these criteria in each longitude bin were averaged without weighting to produce mean color values. The green dashed line in each panel indicates an unweighted fit to the values. The resulting slopes 
suggest e-folding (decrements by 0.43~dex) longitude spans of 84\degr\ for $g$-band and 36\degr\ for $H$- and $L$-bands. These imply that \gaia star counts do not 
include many stars that are as distant as the Galactic bulge, while the \glimpse $L$-band counts do. 
The Mimir $H$-band POLCAT star counts exhibit the same slope as does the \glimpse $L$-band,
indicating that POLCATs include bulge stars.

Vertical light-red and light-blue rectangles identify longitudes where color and star count changes are
correlated. Light-red rectangles at longitudes 30.5, 34.75, 38, and 51\degr\ are where the (H-K) 
colors become redder and the star counts in the photometric bands strongly decrease. 
Light-blue rectangles at longitudes 27, 32.5, 40.5, 47, and 54.5\degr\ show where the (H-K) colors become
less red and the star counts correspondingly increase. Strong color changes due to changes
in spatially-distributed stellar populations (e.g., disk versus bulge stars) are not expected. Instead, where the colors become
redder, more dust extinction must be present along the line of sight to the stars, likely associated
with molecular cloud complexes and/or Galactic spiral arms. Interestingly, where the colors
are less red, there must be a corresponding relative deficit of extinction along these sightlines, resulting
in more stars being revealed and perhaps seen to greater distances. These modulations are strongest
for longitudes between 25 and 40\degr\ and beyond 54\degr, with all tracers participating. Outside of these regions, the \glimpse $L$-band star counts, in particular, show only smooth
bin-to-bin behavior.

\subsection{Correlated Behaviors of FOV-based Properties}\label{corner_analysis}

\begin{figure}
\includegraphics[width=7.2in]{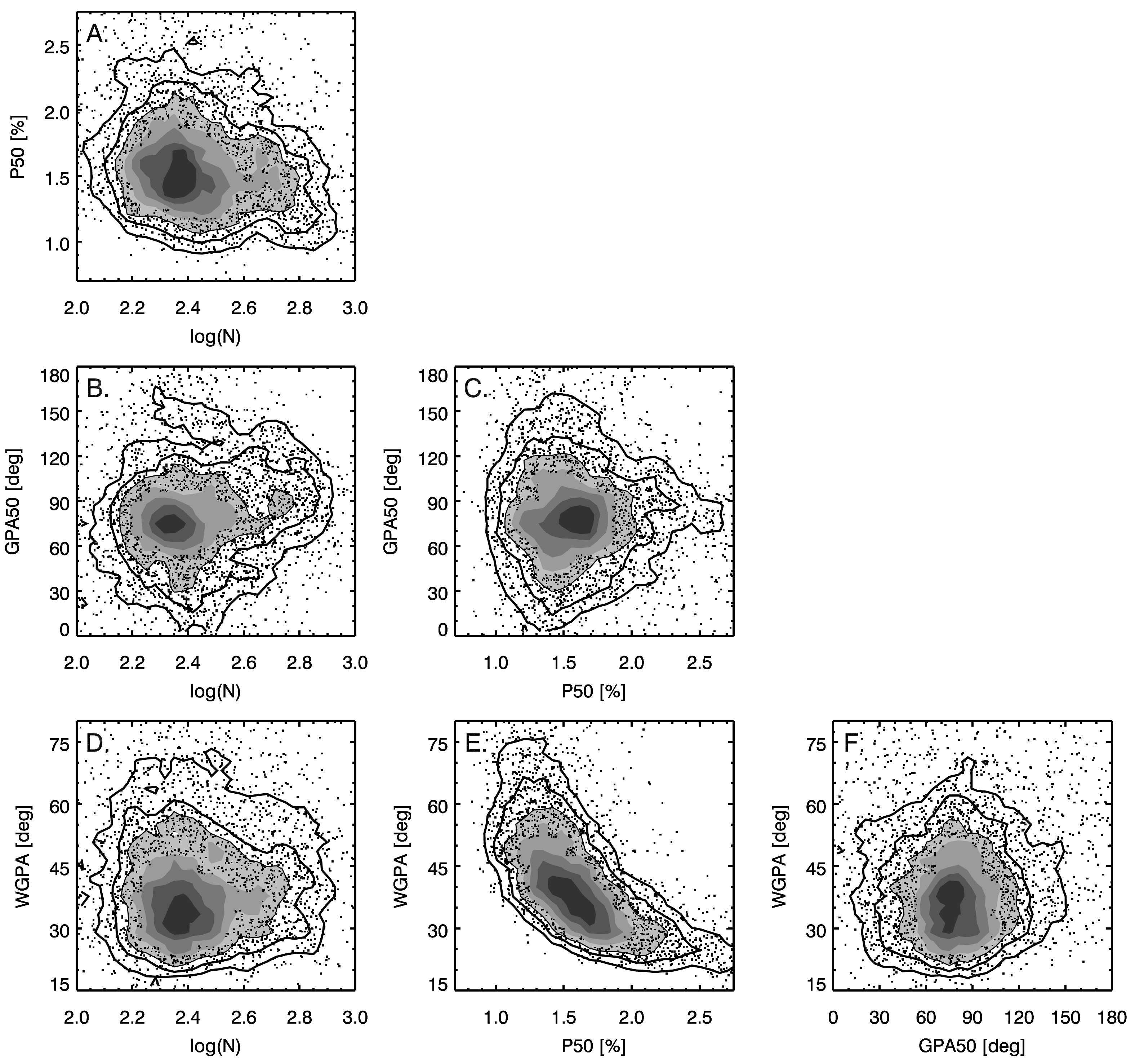}
\caption{Corner plot comparisons of key polarization properties for the FOV-based, 
median-value quantities in the GPIPS FOVs. 
In each panel, individual FOV values are shown as black dots
until the density of dots becomes high. Thereafter, open and filled contours 
continue the density representation, stepped linearly with point density by 12.5\% of the peak value, starting at that 12.5\%  value.
(Upper-left, A-panel): P50 versus log(N). (Middle-left, B-panel): GPA50 versus log(N).
(Middle-center, C-panel): GPA50 versus P50. (Lower-left, D-panel): WGPA versus log(N).
(Lower-center, E-Panel): WGPA versus P50. (Lower-right, F-panel): WGPA versus GPA50.
\label{fig_corner}}
\end{figure}

To explore the nature of possible correlations among the four key characterizing quantities, 
a corner-plot representation was developed, using the median-value characterizations for
each of the GPIPS FOVs, and is shown in Figure~\ref{fig_corner}. For each
of the six panels in the Figure, their 2-D spans were gridded uniformly into 
$31 \times 31$ bins and the number density of GPIPS FOV values in each bin was found. Where the FOV counts in bins were high, open and filled contours were added to convey isodensity locations and changes.

The upper-left, A-panel in Figure~\ref{fig_corner} shows how P50 
and the log of the number of UF1 stars in each field are related. There is a peak
likelihood at about 1.5\% for P50 and 240 stars per field (log(N)$\sim$2.37) with
gaussian-like decays away from the peak. There is a weaker, secondary
peak at 1.5\% in P50 and log(N)$\sim$2.7, which comes from the longitude range less
than 34\degr, as it is only in that range that such high UF1 star counts are seen
in Figure~\ref{fig_Quad_L}. This local increase could be due to distant bulge stars.
There is also a minor tertiary peak at 1.25\% for P50 and 2.8 for log(N), which
corresponds to the L=27\degr\ count increase and polarization dip seen in 
Figure~\ref{fig_Quad_L}.

The middle-left, B-panel in Figure~\ref{fig_corner} compares GPA50 with log(N), revealing the likelihood
peak at 75\degr\ of GPA50 and 210 stars per field.
The secondary and tertiary peaks noted appear to blend into one, weaker peak near 
GPA50 = 90\degr\ and log(N) = 2.7.

The middle-right, C-panel in Figure~\ref{fig_corner} compares GPA50 with P50, showing a likelihood
peak at about 1.7\% and 77\degr, respectively. Near 1.4\% of P50, the GPA50 values extend from 
zero to 180\degr, suggesting that angle aliasing may be present. As the P50 values increase beyond
1.5\%, the GPA50 range decreases and GPA50 values appear to move closer to a 
disk-parallel (GPA=90\degr) orientation. There are few fields with P50 values below 1\%, as was already
shown in Figure~\ref{fig_field_histos}.B., and this could be a consequence of the UF1 selection criteria.

The lower-left, D-panel in Figure~\ref{fig_corner} examines how the widths of the GPA histograms in each FOV,
WGPA, depend on the numbers of UF1 stars in those FOVs. The likelihood maximum
is at about 240 stars per FOV and 33\degr\ of WGPA. The number of UF1 stars in a FOV does not 
appear to constrain WGPA, as the latter shows values extending to 75\degr\ or beyond
for log(N) values similar to those at the peak. The weak secondary and 
tertiary log(N) peaks from the A-panel appear here with slightly higher WGPA values than
for the main peak, perhaps as high as 40-45\degr.

The lower-center, E-panel in Figure~\ref{fig_corner} shows a strong anti-correlation of WGPA and P50, which was
suggested in the discussion of the previous section. While there does appear to be a
likelihood peak near 1.7\% and 35\degr, the overall curved shape of the contours indicate some degree of correlation.
For GPIPS FOVs exhibiting high median polarization fractions, the widths of their 
polarization position angle distributions are narrow, as small as 15-20\degr.
On the other hand, the fields showing the widest GPA distributions exhibit the weakest
P50 values, as small as 1-1.25\%. 
Such anti-correlations of polarization with dispersion
in position angle have been seen elsewhere \citep{Planck_XIX_2015, Planck_XII_2018}, and 
are examined in more detail in Section~\ref{SxP}.

The lower-right, F-panel in Figure~\ref{fig_corner} compares WGPA to GPA50. The likelihood maximum has a center near 76\degr\ of GPA50 and 36\degr\ of WGPA. The contours appear mostly symmetric across GPA50 and non-symmetric along WGPA.
The lack of correlation between WGPA and GPA50 here and the lack of 
correlation between GPA50 and P50 (C-panel) indicate that the anti-correlation 
of WGPA and P50 (E-panel) is the fundamental dependence. The mean orientation
of the magnetic field in the plane of the sky has no bearing on the anti-correlation of
polarization percentage with width of the GPA distribution (or, equivalently, 
dispersion of PAs).

\subsubsection{Latitude and Longitude 2-D Behavior of FOV-based Properties}\label{2d}

This Section presents the 2-D distributions of the FOV-based polarization properties 
of log(N), P50, GPA50, and WGPA for the GPIPS UF1 stars.
Due to the large aspect
ratio of longitude coverage to latitude coverage of GPIPS, the 2-D representations are shown
as stacked longitude slices with aspect ratios chosen to retain sky-true shapes and
orientations. In all cases, color look-up rectangles are shown, labeled with limiting values, and have gray bars indicating the corresponding values of the contours that are overlaid
on the images. Each of the GPIPS FOVs is shown as a rotated square to
emphasize the equatorial orientation of the Mimir instrument. Each is placed
at the equatorial grid center used for the GPIPS observations. Appendix~\ref{Appendix_CBLUT} presents
alternative representations of the same Figures, using color look-up tables better suited to individuals 
with color vision deficiency (CVD).

\begin{figure}
\includegraphics[width=7.2in]{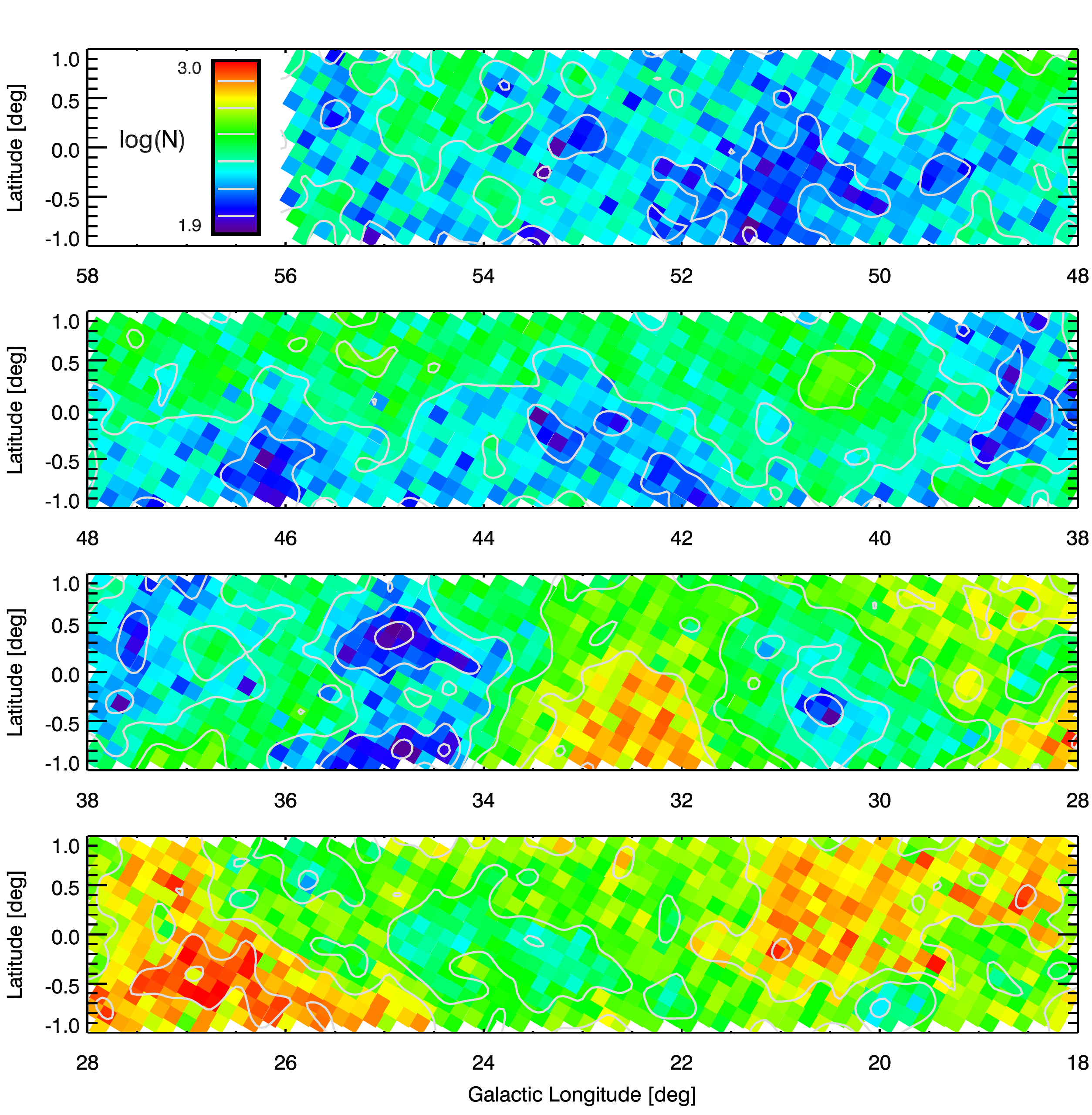}
\caption{False-color representation of the base-10 log of the number of UF1 stars in each
of the GPIPS FOVs, versus Galactic longitude and latitude, as four longitude 
slices. 
Color look-up rectangle, with six nearly equally spaced gray contours, corresponding to 100, 150, 225, 340, 500, and 750 UF1 stars
per FOV, is shown in the top slice. 
The smallest number of UF1 stars in any FOV is 64, the largest is 1,112. 
Note that extinction and the Galactic bulge both likely play roles in the variations in the numbers of 
detected stars with direction. (Also shown as Figure~\ref{fig_D5} for CVD individuals.)
\label{fig_N_LB}}
\end{figure}

\paragraph{Log(N) versus Galactic longitude and latitude}

Figure~\ref{fig_N_LB} presents a false-color representation of the log(N) counts
of UF1 stars in each GPIPS FOV as functions of Galactic latitude and longitude,
for four stacked longitude slices. Each longitude slice spans 10\degr, starting with
the lowest longitude surveyed (18\degr) at the rightmost limit of the lowest slice in 
Figure~\ref{fig_N_LB}, and proceeding from right to left and then up to the next slice until the
final survey longitude of 56\degr\ is reached. Each longitude slice presents the full
2\degr\ latitude extent of the survey. The color-table look-up rectangle in the uppermost
slice indicates that the least populous FOVs, shown in the darkest blue colors, contain
about 80 UF1 stars on average (log(N) = 1.9). The most populous FOVs, shown in
red colors, contain 1,000 or more UF1 stars. The general rise in number of UF1 stars for GL~$< 34$\degr\ is
likely due to the contribution of bulge red giant stars in addition to the disk 
stars sampled at all longitudes, as noted earlier.

In Figure~\ref{fig_N_LB}, three extended zones of enhanced numbers of UF1 stars
in the FOVs are apparent. These span Galactic longitude 18-21.5\degr\ for positive latitudes
down to midplane ones, longitudes 24-29\degr\ for mostly negative latitudes (though with a small
secondary peak to positive latitudes at longitude 27.5\degr), and from longitudes 31.9-33.2\degr\  for
only negative latitudes. These three zones appear to account for the three peaks in
the run of log(N) versus longitude in the lowest panel of Figure~\ref{fig_Quad_L}
and in the photometric panels (A, B, C) of Figure~\ref{fig_count_compare}.
Similarly, reductions in the numbers of stars per FOV for large, resolved regions near  longitudes 30.5, 35, 
37.5-39, 42-44, 46, and 51\degr\ correspond to specific dips in the log(N) versus longitude
curve in Figures~\ref{fig_Quad_L} and \ref{fig_count_compare}. 
The patterns
of high and low numbers of UF1 stars seen in the different GPIPS FOVs are likely
due to the extinction effects described earlier regarding Figure~\ref{fig_count_compare}, but here they are
resolved in longitude and latitude.

\begin{figure}
\includegraphics[width=7.2in]{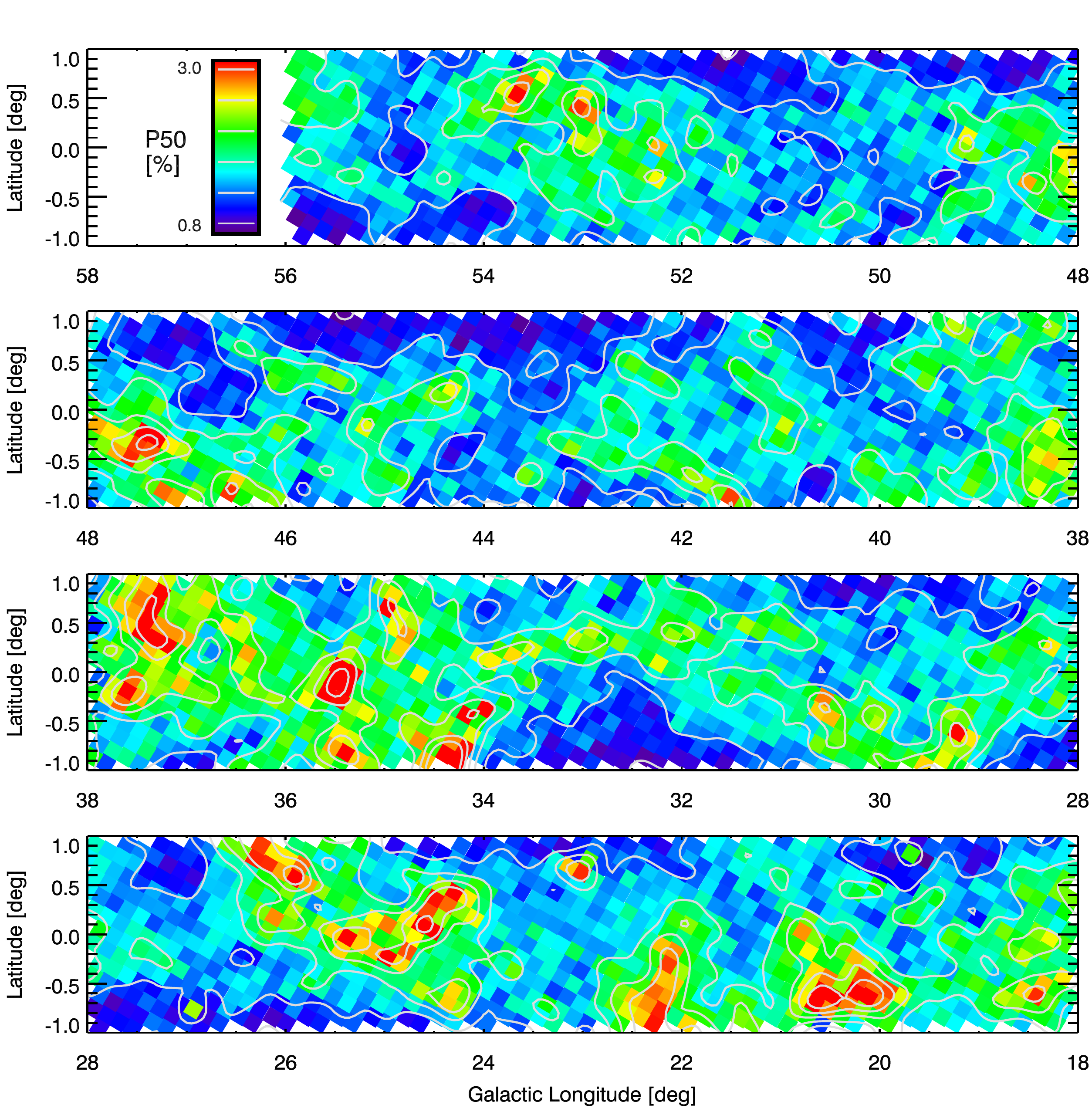}
\caption{Galactic longitude and latitude distribution of the median percentage polarization P50 in each GPIPS FOV. 
Polarization medians range from around 0.8\% to over 3\% in the $H$-band,
with a weak tendency to be higher along the disk midplane. Six gray contour 
levels corresponding to P50 values of 0.9, 1.3, 1.7, 2.1, 2.5, 
and 2.9\% are drawn over the false color panels and inside the color look-up rectangle in the top panel.
(Also shown as Figure~\ref{fig_D6} for CVD individuals.)
\label{fig_P_LB}}
\end{figure}

\paragraph{P50 versus Galactic longitude and latitude}

The Galactic longitude and latitude distribution of UF1 median percentage polarization
P50 is shown for the GPIPS FOVs in Figure~\ref{fig_P_LB}. The false-color mapping scheme
spans from about 0.8 to 3.0\%, though the maximum value across the map is closer to 5\%
and the average is about 1.5\% (Figure~\ref{fig_field_histos}.B).
Regions of polarization percentage greater than that average 
(light blue to green colors) tend to be found closer to the 
mid-plane. Comparing Figure~\ref{fig_P_LB} to 
Figure~\ref{fig_N_LB}, the zones of higher star counts in Figure~\ref{fig_N_LB} are seen 
in Figure~\ref{fig_P_LB} to 
exhibit lower median polarization percentages (e.g., longitudes $32-33$\degr, latitudes $\le 0$\degr). 
The reverse is also true---lower star count 
regions tend to exhibit higher polarization percentages (e.g., longitude $30.6$\degr, latitude $-0.4$\degr), likely the signature of dust 
along those latter sight lines both extincting and polarizing the background starlight. 
This trend is more pronounced in Figure~\ref{fig_P_LB} than in 
Figure~\ref{fig_corner}.A, so seeing
the trend in Figure~\ref{fig_P_LB} helps reveal the weak  
anti-correlation of log(N) with P50 that is present in Figure~\ref{fig_corner}.A. 

The roughly 25-30 regions showing the greatest 
polarization percentages (red colors) in Figure~\ref{fig_P_LB} have small angular extents. They also tend to be surrounded by regions of higher than average
polarization (green colors), which are themselves more extended. The natures of these high-P50 regions are explored further in Section~\ref{SxP}.

\begin{figure}
\includegraphics[width=7.2in]{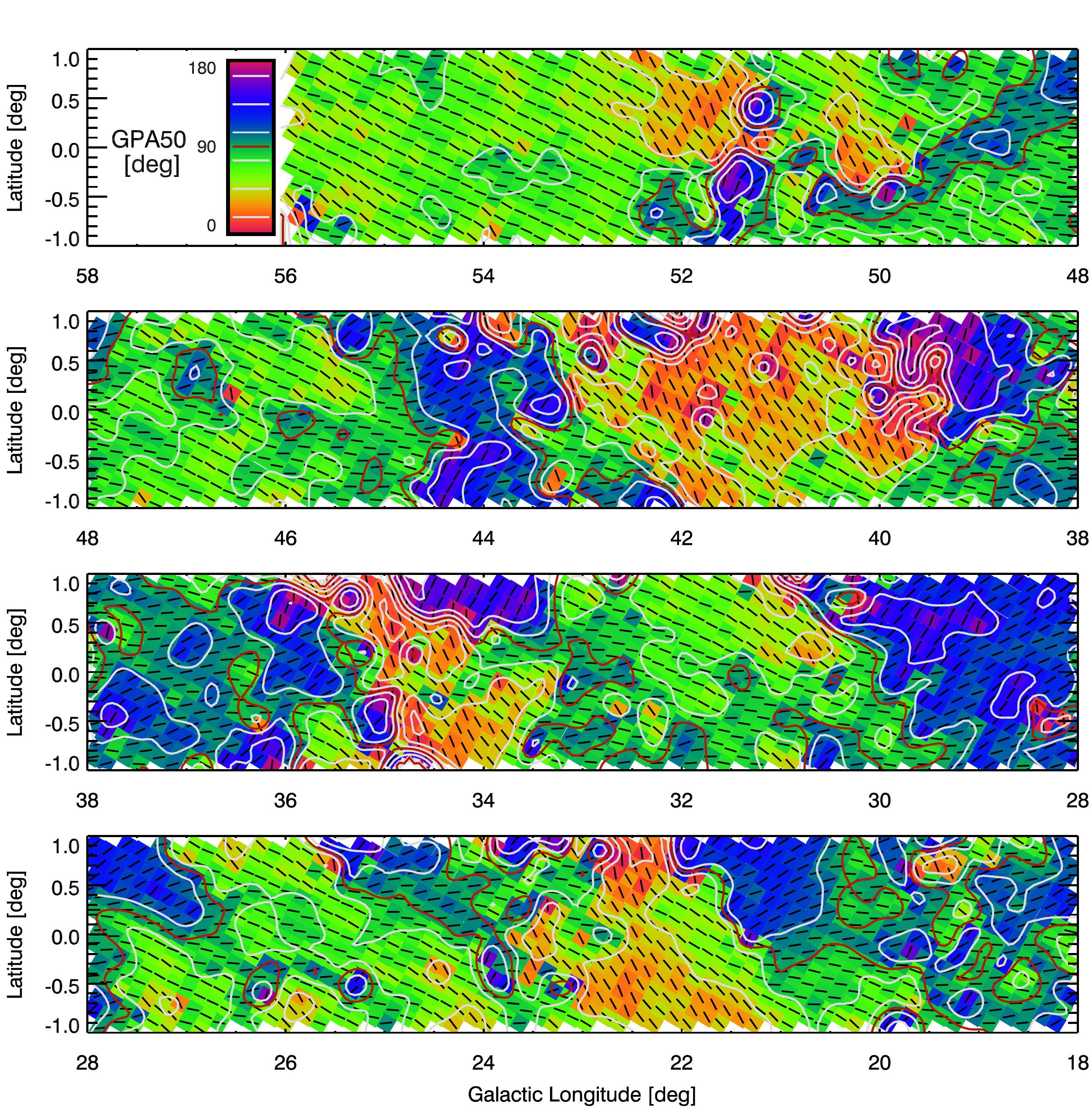}
\caption{Galactic longitude and latitude distribution of GPA50 in each GPIPS FOV
in the same four-slice presentation as in the previous Figures. 
Note that the
look-up table colors wrap around 0\degr\ and 180\degr\  to reduce the appearance of aliasing. 
Black lines within each FOV box indicate the GPA orientation. Six gray contour 
levels, corresponding to GPA50 values of 15, 45, 75, 105, 135, 
and 165\degr, and one red contour, corresponding to the disk-parallel GPA50 value of 90\degr,
are drawn over the false color panels and inside the color look-up rectangle in the top panel.
(Also shown as Figure~\ref{fig_D7} for CVD individuals.)
\label{fig_GPA_LB}}
\end{figure}

\paragraph{GPA50 versus Galactic longitude and latitude}

The Galactic midplane distribution of GPA50 values at 10~arcmin resolution is shown
as Figure~\ref{fig_GPA_LB}. There, the color look-up conversion has been modified so that GPA50
values of 0\degr\ and 180\degr\ are represented by the same false color in order to reduce
display confusion caused by angle aliasing. The colors representing Galactic disk-parallel magnetic 
field orientations are close to the green-blue false-color boundary. A red contour is shown which corresponds to GPA 90\degr. Some large regions of the
stack of longitude slices do appear dominated by colors that are mostly green to blue and so have 
GPA50 values similar to being disk-parallel.

However, there are also several regions that exhibit GPA50 values that are
far from being disk-parallel. The larger of these regions span longitudes of 21-23,
33.5-36, 39-43, and 50-52.5\degr. The first two of these regions correspond to the two dips in
GPA50 seen in Figure~\ref{fig_Quad_L}. 
The region near longitude 22\degr\  appears
to be a fairly uniform zone showing GPA50 values of 40-45\degr\ and could be due to 
the magnetic field characterizing a single
molecular cloud or complex. The feature near longitude 34\degr\  is more complex, seemingly consisting
of extended zones above and below the Galactic equator. The negative latitude zone is similar
in characteristics to the longitude 22\degr\  zone, but the positive latitude zone is different in that it shows
strong GPA50 variations across a ridge of nearly constant R.A. 

\begin{figure}
\includegraphics[width=7.2in]{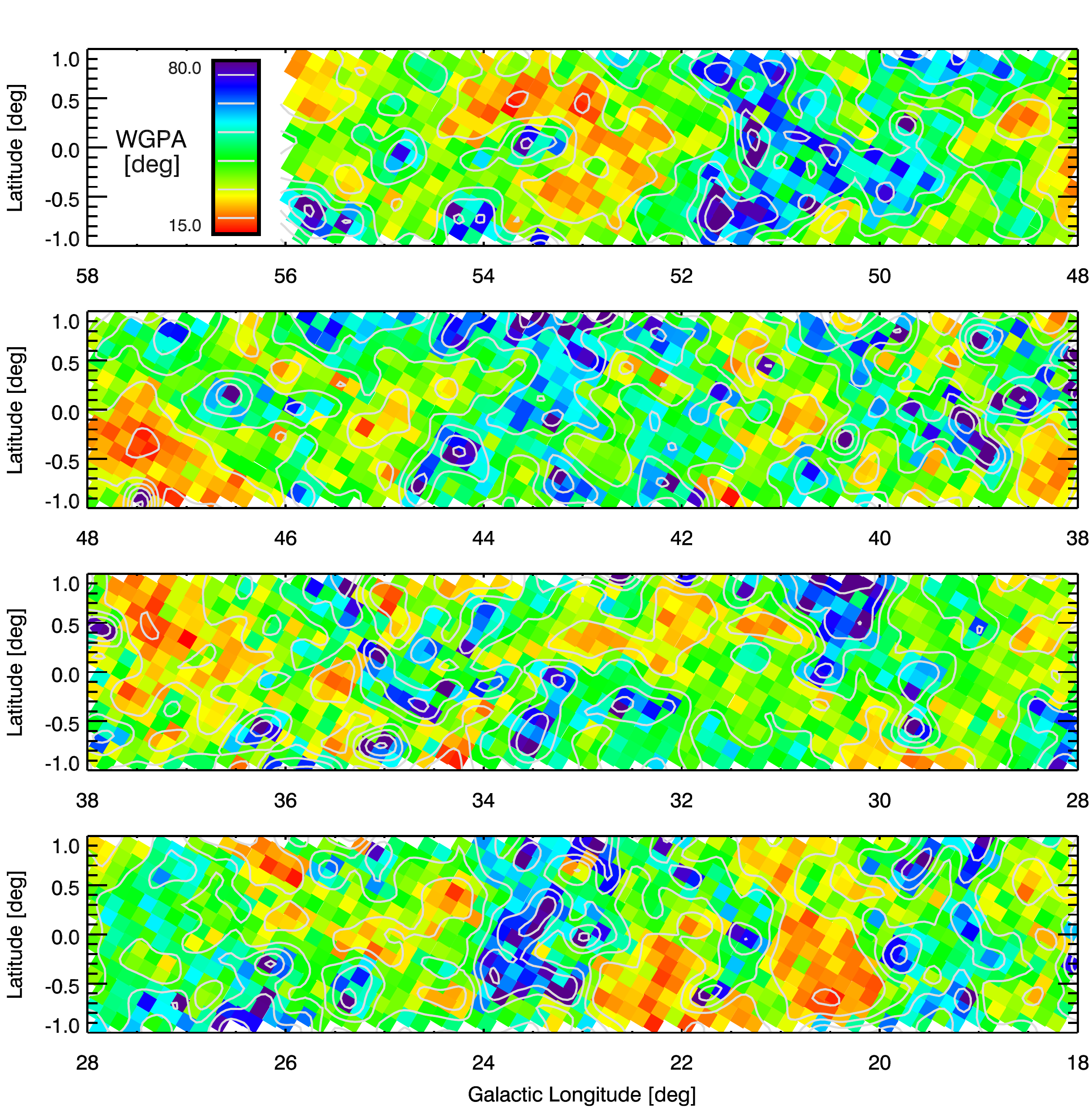}
\caption{Galactic longitude and latitude distribution of WGPA, the interquartile ranges of the
GPA distributions in each of the GPIPS FOVs. The color look-up rectangle in
the uppermost panel indicates that WGPA values range from a low of about
15\degr\ (red colors) to a high of about 80\degr\ (dark blue colors). Six gray contour 
levels corresponding to WGPA values of 20, 31, 42, 54, 64, 
and 75\degr\ are drawn over the false color panels and inside the color look-up rectangle in the top panel. Note that the value to color mapping in this Figure is inverted from previous Figures to highlight 
the low-WGPA regions in red. 
(Also shown as Figure~\ref{fig_D8} for CVD individuals.)
\label{fig_WGPA_LB}}
\end{figure}

\paragraph{WGPA versus Galactic longitude and latitude}

Figure~\ref{fig_WGPA_LB} presents the Galactic midplane distribution of the final key
quantity, WGPA, the interquartile ranges of the UF1 star GPA distributions in each
GPIPS FOV. In this Figure, the color table is reversed to associate
lesser WGPA values (those exhibiting greater GPA coherence) with red colors and greater WGPA values with darker, bluer colors. The Davis-Chandrasekhar-Fermi \citep[DCF:][]{Davis51,CF53} method for estimating magnetic field strengths from linear polarization data has an inverse dependence on polarization position angle dispersion, {a direct linear dependence on gas velocity dispersion, and a square root
dependence on gas density. Assuming the latter two quantities exhibit roughly constant mean values
across all FOVs (a weak, but useful first step),} the red highlighted zones in Figure~\ref{fig_WGPA_LB}
could represent regions of locally stronger magnetic field strength.
In Figure~\ref{fig_WGPA_LB}, the overall impression is one of a relatively uniform background
of WGPA values near 31\degr\ 
(green colors; $\Delta_{GPA} \sim 13$\degr), with many small regions
and 9-10 larger regions of low-WGPA values, seen as red-colored zones. 
High-WGPA values are present in many small
zones and a few larger zones, but in fewer numbers of regions, both small and large, 
than for the low-WGPA zones. Low-WGPA FOVs correlate with high-P50 FOVs,
as already noted in the discussion related to Figure~\ref{fig_corner}.E and as examined in Section~\ref{SxP}

\section{Discussion and Applications}\label{discussion}

The types of studies made possible by GPIPS data will be different
in nature and scope than those performed using other existing magnetic field probes.
The data contained in GPIPS DR4 include more than 1~million measured stellar polarizations
of high quality that have the potential for being used individually \citepalias[UF1 type;][]{Clemens12c}. 
The UF2 and UF3 stars account for another nearly 9~million stars, but need careful {distance} selection and averaging to return magnetic field information. \planck polarization
at 353~GHz, with its 5~arcmin resolution, yields about 6~million 
independent samples
across the entire sky, but only 11~thousand across the GPIPS zone. The pulsar and
extra-galactic radio source list used by \citet{Han18} for characterizing the magnetic
field of the Galactic disk via Rotation Measures {(RM)} includes 4,700 sources, but spans
eight times the latitude range and 9.5 times the longitude range of GPIPS. 
If those sources were uniformly distributed 
in latitude and longitude \citep[a poor, but serviceable assumption - 
see][]{Han18}, then only about 60 of these sources probe lines of sight through the GPIPS region. 
Hence, whether compared to
the sampling density of the radio continuum magnetic field RM studies (0.8 samples deg$^{-2}$) or to \planck thermal dust emission
from the Galactic plane (183 beams deg$^{-2}$), GPIPS data achieves much finer
angular sampling (14,400 UF1 stars deg$^{-2}$) than any other technique. 

The science applications of the GPIPS data
will likely be many, and are well beyond the scope of this introductory presentation. 
GPIPS data from earlier data releases have been used to create the first resolved map
of the plane-of-sky magnetic field strength for a molecular cloud \citep{Marchwinski12}
and used with deeper $K$-band (2.2~$\mu$m) Mimir polarimetry to map the magnetic field strength across an
Infrared Dark Cloud \citep{Hoq17}. 

In the following, example studies based on comparisons of GPIPS to \planck 
yielded some new findings and insights. The first comparison involved the plane-of-sky
polarization position angle orientation distributions of \planck and GPIPS. The second was a deeper look at the
anti-correlation of WGPA and P50.


\subsection{Plane-of-Sky Polarization Orientation Distributions Comparison}

The histograms of Galactic polarization position angle using GPIPS NIR
background starlight polarimetry 
were shown in Figure~24 of \citetalias{Clemens12c} and here in Figure~\ref{fig_field_histos}.C. 
The first was constructed from values measured in the DR1 GPIPS release
of 50,000 stars of UF1 type with the additional criterion of requiring 
$P^\prime / \sigma_P \ge 2.5$, to select higher quality measurements. 
The resulting histogram of the numbers of such stars versus GPA revealed a single
peak at about GPA=75\degr\ with a FWHM width of about 50\degr. 

Here, using GPIPS DR4, the distribution of the FOV-based median values derived from the GPA
distributions for the UF1 stars in each of the GPIPS FOVs was presented in Figure~\ref{fig_field_histos}.C. There, 
the median of the
GPA50 distribution is 76.7\degr, with an interquartile range
of about 43\degr. So, while the distribution of GPA50 values appears to be somewhat narrower
than the corresponding distribution of high quality DR1 GPIPS star GPA values, the
distribution centers are similar.

For \planck 353~GHz polarization, \citet{Planck_XIX_2015} showed, in their
Figure~3, the distribution of GPAs with, and without, bandpass correction terms. When
corrected using their favored approach, the GPA distribution for the first Galactic 
quadrant (longitudes 0 to 90\degr, encompassing the GPIPS region) shows a peak 
at a GPA equivalent\footnote{\citet{Planck_XIX_2015} reported position angles $\psi$ for the
peak of the electric field vector, which is perpendicular to the magnetic field orientation
for thermal dust emission. Those angles are rotated by 90\degr\ here to become GPA values.}  of about 95\degr\ and a FWHM of about 40\degr. Their PA distribution shows the same asymmetry 
of lower likelihoods for GPAs greater than 90\degr\ and higher likelihoods for GPAs less
than 90\degr, as seen in the GPIPS distributions. Thus, the \planck results disagree in 
mean or median GPA values with respect to GPIPS by about one half of the distribution
widths, although the distribution shapes appear to be similar. 

More recently, \citet{Planck_XII_2018}, in their Figure~6, reported a 353~GHz GPA histogram
for the entire sky, computed for 80~arcmin resolution, that peaks at about $86$\degr\ and has a 
FWHM of about 40\degr. This is closer to the $\sim$77\degr\ value found in GPIPS data, but 
still differs by 9\degr.

The exact reason for the difference in GPA distributions between GPIPS and \planck remains unknown. It could arise from differences in the effects of angular resolution
and/or the line-of-sight distances probed. The GPIPS angular sampling is orders 
of magnitude finer than \planck achieves, so GPIPS could be resolving structures that are
too small to be resolved by \planck. Yet many of the features found here in the GPIPS 
2-D (e.g., Figure~\ref{fig_GPA_LB}) and 1-D (e.g., Figure~\ref{fig_Quad_L}) plots span angular sizes that are much larger than the \planck polarization
resolution. Line-of-sight effects could be causing \planck beams to integrate
dust emission over much
longer distances than are probed using GPIPS, again leading to smoothing 
of GPIPS-resolved changes by the \planck observations and analyses. 
GPIPS findings, derived from polarizations of background stars, will
necessarily be limited in the line-of-sight distances that can be probed, as stellar faintness and 
extinction both contribute to detection limitations \citepalias[][and Section~\ref{gaia_analysis}, above]{Clemens12a}. 
However, the presence of the
Galactic bulge in the GPIPS stellar distributions (Figure~\ref{fig_Quad_L} and 
Figure~\ref{fig_N_LB})
shows that the GPIPS far horizon must extend to at least 5-7~kpc from the sun.

The question of exactly why GPIPS and \planck differ in their inferred GPA distributions remains 
unanswered and may have to wait until robust models of the Galactic magnetic field,
at spatial and angular resolutions finer than currently exist, become available for testing.

\subsection{The Anti-Correlation of WGPA and P50}\label{SxP}

The anti-correlation of WGPA and P50 seen in Figure~\ref{fig_corner}.E has been seen
previously in \planck polarization data \citep[e.g., Figure~23 of][]{Planck_XIX_2015}
and perhaps in SCUBA submm polarization data even earlier \citep{Poidevin10}. 
More recent \planck analysis \citep{Planck_XII_2018} used a larger region of the sky over which to 
find a width-polarization relation they argued was of the form $S \times P = $\,constant, 
where $S$ is a measure of polarization position angle dispersion and $P$ is the fractional polarization measured at 353~GHz. 

In \citet{Planck_XII_2018}, an analytic model was constructed using gaussian fluctuations
containing combinations of uniform and random (turbulent) magnetic fields for multiple zones along
the line of sight to test the observed $S \times P$ relation, and a strong anti-correlation was found between $S$ and $P$.
They argue that this anti-correlation is a direct result of magnetic field topology and
should be a general property of the Galactic magnetic field.

\begin{figure}
\includegraphics[width=7.2in]{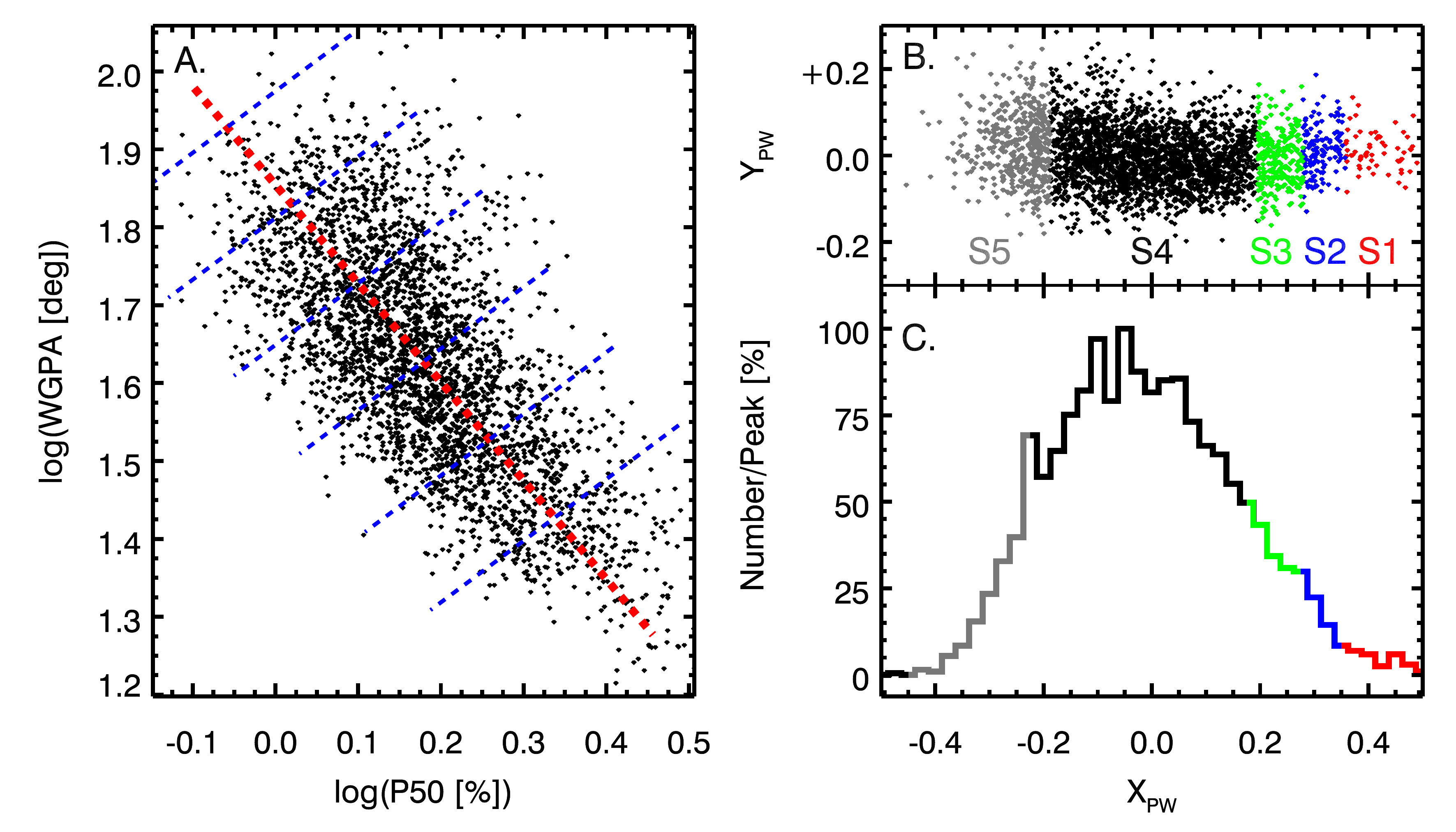}
\caption{The WGPA-P50 anti-correlation and the five selected subsamples of GPIPS FOVs. 
(Left, A-panel): Distribution of GPIPS WGPA and P50 FOV-based data 
values, plotted in base-10 log-log format. Black dots  
correspond to the WGPA and P50 values of the individual GPIPS FOVs. 
The red, dashed line
represents the robust linear regressed line. Dashed blue lines 
oriented perpendicular to the red line indicate notional binning of data into
bins spaced along the red line. 
(Right-upper, B-panel): Same distribution of WGPA and P50 data points as
for the A-panel, after rotating and shifting to make the regressed linear-fit line centered and horizontal (see text).
Data points are colored red,
blue, green, black, and gray to indicate the data subsample to which they 
were assigned - red for the S1 highest-P50/lowest-WGPA FOVs subset through the other colors to gray for S5, representing
the lowest-P50/highest-WGPA FOVs subset.
(Right-lower, C-panel): Histogram of the numbers of GPIPS FOVs falling
into 41 bins along the $X_{PW}$ axis. Histogram colors are the same as
for the data points in the B-panel.
\label{fig_WGPA_P}}
\end{figure}

To examine whether GPIPS was also returning similar results, the plot of WGPA versus
P50, which should be closely equivalent to the $S$ versus $P$ plots in the \planck papers,
was recast in the log-log form used in the \planck papers and is presented as the A-panel of Figure~\ref{fig_WGPA_P}.
There, each black dot is plotted at the log-based location corresponding to the
median UF1 polarization (P50) and the interquartile range (WGPA) for one GPIPS FOV. 
The red
line represents the best fit robust linear regression, resulting in a slope of
$-1.266 \pm 0.014$.
This power law index is similar to the $-1$ value dictated by the $S \times P$ relation.
But, the measured index misses negative unity by almost twenty times its uncertainty. While the GPIPS
data appear to be generally consistent with the \planck findings, there are key
differences. 

\subsubsection{Locating Strong and Weak Magnetic Field Regions}\label{sec_b-zones}

In \citet{Planck_XIX_2015}, the highest PA-dispersed regions were argued to form
boundaries between isolated segments characterized by high-$P$ values and 
low-PA dispersions (their Figure~22.top).
These isolated segments could represent regions of greater magnetic field strength, or at least
greater magnetic field orientation coherence.
The sizes of the isolated \planck segments appear to be a few degrees. Since GPIPS data
exhibit a similar anti-correlation of $S$ and $P$, the question of whether similar isolated segments of
low PA-dispersion are surrounded by high PA-dispersion regions in GPIPS was examined
to try to learn about these potentially magnetic-dominated zones.

The Figure~\ref{fig_WGPA_P} A-panel distribution
was divided into five subsamples of GPIPS FOVs that spanned low- to high-P50 values.
This division was performed in a way that was equivalent to collecting the
A-panel distribution into bins oriented perpendicular to the red line, as suggested
by the dashed blue lines in that panel. 
Such binning respects the $S \times P$ anti-correlation while retaining sensitivity to 
differences in the magnetic field properties of the GPIPS FOVs along the correlation
direction.
The method involved rotating the distribution of 
A-panel points to cause the red line of correlation to become horizontal, resulting in 
the FOV distribution shown in the B-panel of Figure~\ref{fig_WGPA_P}. Through this 
rotation, the new axes ($X_{PW}$ and
$Y_{PW}$) became linear combinations of P50 and WGPA and thereby carried mixed 
units. In the B-panel, the
GPIPS fields are mostly distributed horizontally without much coherent curvature of 
$Y_{PW}$ on $X_{PW}$. As a result, the GPIPS FOVs became primarily distributed 
along the new $x$-axis, which maximally separated FOVs with high-P/low-WGPA from FOVs with
the opposite behavior.
The Figure~\ref{fig_WGPA_P} C-panel shows a normalized histogram, after marginalizing
over $Y_{PW}$, with the
rotated axis $X_{PW}$ as the new indicator of likely magnetic field strength or coherence.

Five subsamples of the rotated GPIPS FOV values were selected. These probed the magnetic nature of the high-P50/low-WGPA FOVs using three of the subsamples, the low-P50/high-WGPA FOVs with one subsample, and collected the remaining (moderate-P50/moderate-WGPA) FOVs into one reference subsample. The most highly-polarized, and least PA-dispersed, subsample of GPIPS 
FOVs were chosen to be the 64 FOVs ($\sim$2\% of the total in the full distribution) with the most positive $X_{PW}$ values in 
Figure~\ref{fig_WGPA_P}.C. This subsample was designated S1 and those FOVs are colored red in the B- and C-panels. 
The 129 next-most highly-polarized GPIPS FOVs ($\sim$4\%) comprised the S2 subsample and are colored blue. The
258 GPIPS FOVs ($\sim$8\%) beyond S2 became the S3 subsample and are colored green. At the other end of the histogram, a subsample of 452 GPIPS FOVs ($\sim$14\% - equal in size to the the total of the S1, S2, and S3 subsamples) containing the least polarized
and most PA dispersed FOVs was designated S5 and its FOVs are indicated in gray. The remaining 2,334 GPIPS FOVs ($\sim$72\%) in
the middle of the distribution were designated S4 and are colored black to identify the non-extreme, more average
behaviors of their FOVs.

\paragraph{Multi-Dimensional Analysis of Coherent High-P/Low-WGPA Structures}\label{hi_P_zones}

\begin{figure}
\includegraphics[width=7.2in, trim= 0in 1in 0in 0in]{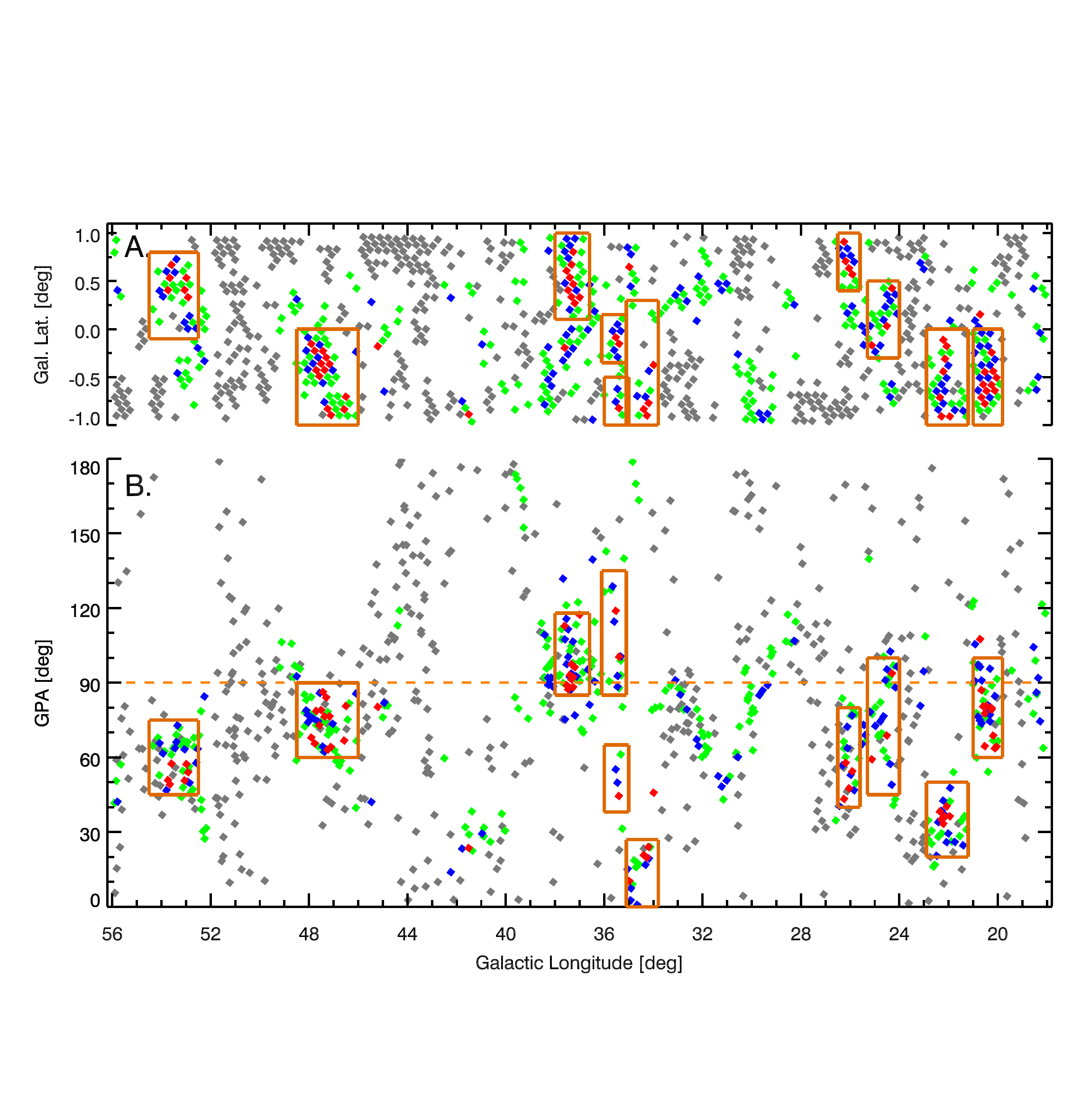}
\caption{Multi-dimensional distributions of the (strong field) S1, S2, S3 and (weak field) S5 
subsamples of GPIPS FOVs. 
(Top, A-panel): Galactic longitude and latitude locations of the subsample 
FOVs. The S1, S2, and S3 (red, blue, and green) FOVs appear to group into distinct structures
while the S5 (gray) FOVs mostly form structures at the latitude limits of GPIPS. 
(Middle, B-panel): Median GPA versus Galactic longitude for the subsample FOVs.
The spatial coherence seen for the S1, S2, and S3 FOVs in the top panel is 
also present as coherent GPA values for each distinct structure in the lower panel. This is not the case
for the S5 (gray) FOVs, which show much less GPA clustering. Light brown rectangles 
indicate the extents of the ten groups described in the text and listed in Table~\ref{tab_struct}.
\label{fig_HiP_LB}}
\end{figure}

Figure~\ref{fig_HiP_LB} offers a two-fold comparison of the relative locations of the S1, S2, and S3 high-P50/low-WGPA selected FOVs (red, green, and blue points, respectively) and the S5 low-P50/high-WGPA selected FOVs (gray points).
The top, A-panel of Figure~\ref{fig_HiP_LB} presents the Galactic longitude and latitude
distribution of those FOVs. 
There appear to be many distinct clumps of multiple GPIPS
FOVs exhibiting high-P50/low-WGPA values. 
The largest of these clumps show central cores comprised of S1 (red) points 
surrounded, or bordered, by S2 (blue) and S3 (green) FOV points. 
{Under the assumption of rough uniformity of gas velocity dispersion and density over FOV sizes,} these could signify zones of high magnetic field strength being embedded in somewhat 
weaker field regions. Alternatively, they could signify zones where the magnetic field is
purely in the plane of the sky, and so maximizes P50 values and minimizes WGPA ones,
while being surrounded by regions of more mixed, less uniform, magnetic field inclination angles.

The S5 (gray) FOV points also show multiple resolved regions,
but these tend to be located mostly at the Galactic latitude limits of the survey region. As these
zones contain somewhat fewer UF1 stars and lower P50 values, the S5 regions could represent
either lower extinction zones, weaker magnetic field zones, or zones with more random magnetic field orientations and/or inclinations.

There does
appear to be a rough anti-correlation of S5 fields with the S1, S2, and S3 fields in 
longitude and latitude space, but concluding that the high-WGPA S5 fields are defining boundaries
around, or between, low-WGPA S1, S2, and S3 fields seems unsupported in these GPIPS data.
Instead, the high-P zones appear to be surrounded by more average (e.g., S4 - not shown in these 
plots for clarity) magnetic field strength (or orientation coherence) regions.

The bottom, Figure~\ref{fig_HiP_LB} B-panel shows GPA values versus Galactic longitude locations of the
FOVs. Many of the same groupings of colors remain coherently located in this plot.
Thus these regions of higher-P50/lower-WGPA appear to be associated with resolved, coherent
magnetic fields, at least locally, that have nearly constant plane of sky orientations (GPAs).

\paragraph{Resolved Strong Field Regions}

In order to obtain
some quantitative evaluation of this clustering behavior, and to locate these potentially strong
magnetic field regions, coherent groupings of points were identified
visually in Figure~\ref{fig_HiP_LB}. Groups needed to exhibit 
agreement in subsample designation (S-number or plotted color), Galactic longitude and latitude, 
and GPA so as to remain identifiable as distinct clumps. Subjectively,
there appear to be about ten such large groups, each of which has multiple high-P50/low-WGPA (strong B-field) S1 (red) FOVs,
many S2 (blue) FOVs, and many S3 (green) FOVs. There are also some lesser groups that
have one or no S1 FOVs. 
The longitude and latitude center locations as well as angular and GPA spans for the ten major groups are summarized in Table~\ref{tab_struct} and are shown as light brown rectangles in Figure~\ref{fig_HiP_LB}.

In Table~\ref{tab_struct}, the first column is a group sequence number, ordered by increasing Galactic
longitude. The second column offers designations, based on the
mean Galactic longitude of the group. The next column 
lists the longitude range spanned by each group, followed by the latitude range 
spanned. The fifth column lists the GPA range spanned. The sixth column lists the
total number of the highest-P S1 (red) and S2 (blue) GPIPS FOVs contained within each group. 
These two extreme FOV type appear to be more clustered in the Figure~\ref{fig_HiP_LB} A- and B-panels than do the S3 (green) FOVs and so were chosen to define the group extents and
contents. 

\begin{deluxetable}{clcccc}
\tabletypesize{\small}
\tablecaption{High-P/Low-WGPA Groups of Potentially Strong Magnetic Fields\label{tab_struct}}
\tablewidth{0pt}
\tablehead{
\colhead{Number} & \colhead{Designation} & \multicolumn{2}{c}{Galactic Ranges of}& \colhead{GPA Range}  & \colhead{N(FOVs)} \\
&&\colhead{Longitude} & \colhead{Latitude} & & \colhead{(S1+S2)}\\
&&\colhead{(\degr)} &\colhead{(\degr)}&\colhead{(\degr)} \\
\colhead{(1)}&\colhead{(2)}&\colhead{(3)}&\colhead{(4)}&\colhead{(5)}&\colhead{(6)}
}
\startdata
1	& L20.5	& [19.8, 21.0] & [$-$1.0, 0.0] & [60, 100] & 22 \\
2	& L22.0	& [21.2, 22.9] & [$-$1.0, 0.0] & [20, 50]  & 16 \\
3	& L24.5	& [24.0, 25.3] & [$-$0.3, $+$0.5] & [45, 100] & 13 \\
4	& L26.0	& [25.6, 26.5] & [$+$0.4, $+$1.0] & [40, 80] & 9 \\
5	& L34.3	& [33.8, 35.1] & [$-$1.0, $+$0.3] & [0, 27] & 7 \\
6	& L35.3A	& [35.0, 36.0] & [$-$1.0, $-$0.5] & [38, 65] & 3 \\
7	& L35.3B	& [35.1, 36.1] & [$-$0.35, $+$0.15] & [85, 135] & 7 \\
8	& L37.5	& [36.6, 38.0] & [$+$0.1, $+$1.0] & [85, 118] & 17 \\
9	& L47.5	& [46.0, 48.5] & [$-$1.0, 0.0] & [60, 90] & 23 \\
10 	& L53.5	& [52.5, 54.5] & [$-$0.1, $+$0.8] & [45, 75] & 15 \\
\enddata
\end{deluxetable}

The total number of S1 and S2 GPIPS FOVs contained within the ten groups is
132. This is about 68\% of the total number of S1+S2 FOVs. 
That is, the highest-P50/lowest-WGPA FOVs appear to be the most clustered class of FOVs.
This would seem to argue that resolved regions with strong magnetic fields are not 
uniformly distributed but instead are localized to perhaps 10-12 major structures.
The average S1+S2 FOV count per group is about 13. For circular
groups, this is equivalent to a diameter projection spanning about 0.4\degr, or
18~pc for an average assumed distance of 2.6~kpc (see Section~\ref{gaia_analysis}). 
While it is possible that these groups represent
just the highest-P50 centers of larger coherent structures, it is difficult to imagine that
the groups found here also happen to be located within structures having sizes of several degrees, 
as the \planck analyses \citep{Planck_XIX_2015} would suggest. 

The remaining 2,334 S4 GPIPS FOVs (72\%), colored black in the right panel of Figure~\ref{fig_WGPA_P},
were also plotted in the same form as for the high and low P50 fields shown in 
Figure~\ref{fig_HiP_LB}. Those S4 fields, exhibiting more typical values of
P50 and WGPA, {\it were} found to be mostly uniformly distributed
in Galactic longitude and latitude, absent the regions already selected by the high and low P50 fields. The
average fields were also distributed somewhat uniformly along the GPA 
plot, though following the trends seen in Figure~\ref{fig_Quad_L}.C, with no groupings as obvious as those listed in Table~\ref{tab_struct}. 

Thus, the S5 (gray; low P50) fields seem to mostly occupy
the survey latitude boundaries, the S1, S2, and S3 high-P50 fields (red, blue, green) are mostly
found in about ten resolved groupings, and the more moderate P50/WGPA FOVs are fairly
uniformly distributed.

The positions and extents of the ten high-P50 groups listed in Table~\ref{tab_struct} were compared to the positions and extents of previously cataloged star formation regions in the GPIPS zone to ascertain whether star formation correlates with high-P50/low-WGPA (strong magnetic field) conditions. The regions tabulated by \citet{Murray2010} that showed strong thermal free-free emission, as detected by WMAP and correlated with GLIMPSE and {\it Midcourse Space Experiment} \msx \citep{Price2001} images, include eleven within the GPIPS zone. Similarly, the catalog of Red \msx Sources \citep[RMS:][]{Urquhart2014} lists fifteen luminous, and presumably massive, young stars in the GPIPS region.
The agreement with the GPIPS group list of Table~\ref{tab_struct} was weak---4 of 11 for the WMAP regions and only 2 of 15 for the RMS objects. Some of these massive star forming regions are also likely at distances too great for GPIPS to have been able to probe. The generally poor matching of high-P/low-WGPA regions with zones of massive star formation may indicate that strong and/or uniform magnetic field conditions are not prevalent in such settings. 

Are these high-P50/low-WGPA groups physical objects with strong magnetic
field strengths, or are they regions where the magnetic field just happens to be 
well-aligned and mostly in the plane of the sky? Establishing magnetic field strengths for the
regions rests on application of the DCF method, and thereby requires high-quality information for
gas velocity dispersions and gas volume densities across the regions, or Zeeman 
Effect high-resolution and high-sensitivity molecular line spectroscopy. These are beyond the
scope of this paper, but should be pursued.

\subsubsection{GPA Distributions Comparison}

\begin{figure}
\includegraphics[height=6.0in, trim= -1in 0in 0in 0in]{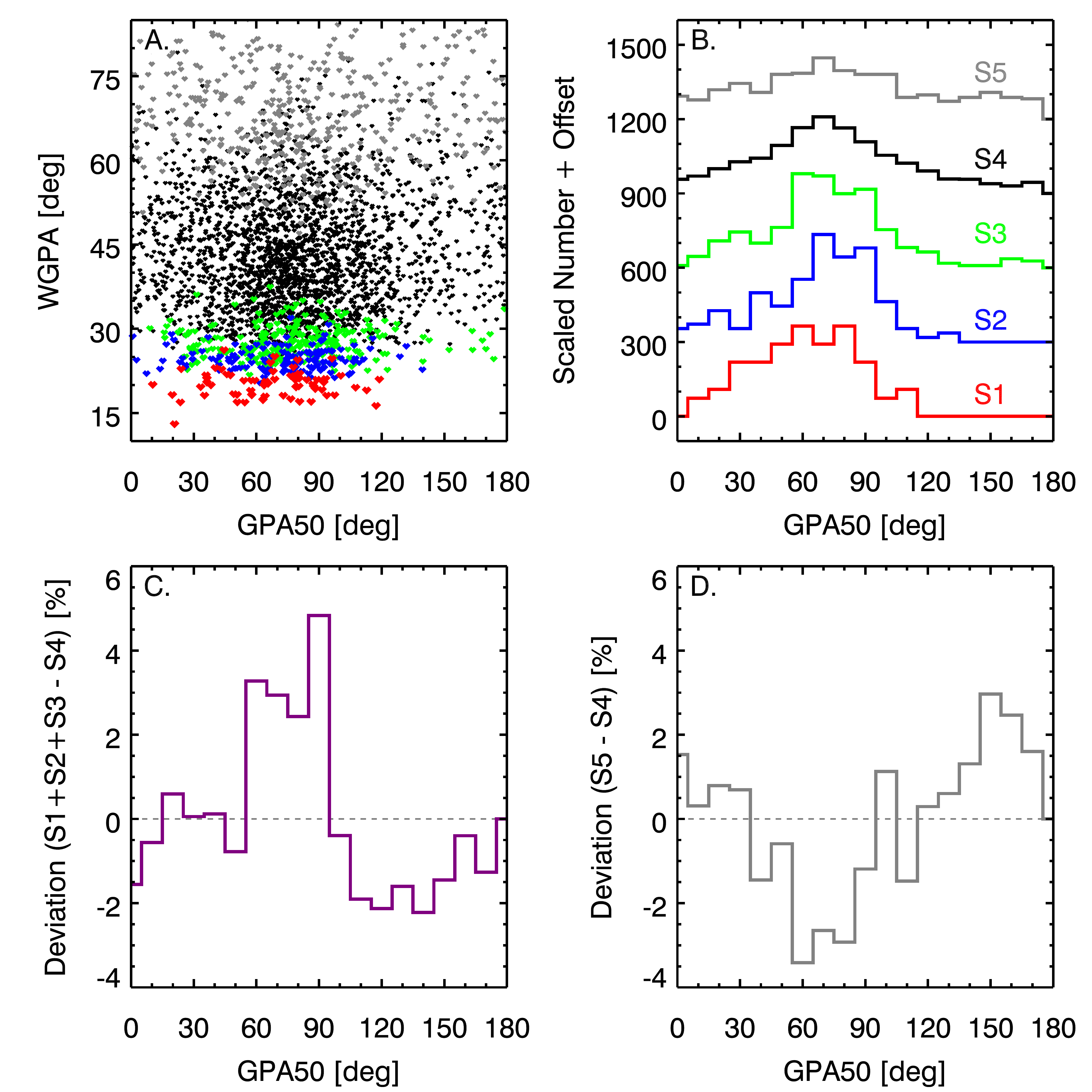}
\caption{Distributions of median GPA values for GPIPS FOVs.
(Upper-Left, A-panel): WGPA versus GPA50 for all GPIPS FOVs, with symbols representing the S1, S2, S3, S4, and S5 subsamples of FOVs 
colored red, blue, green, black, and gray, as in Figure~\ref{fig_WGPA_P}, respectively. 
(Upper-Right, B-panel): GPA50 histograms for each of the subsamples
of GPIPS FOVs. The histograms have been offset successively by 300 and each non-black
histogram was scaled to have the same total FOV counts as in the S4 black histogram, for ease of comparison.
(Lower-Left, C-panel): Difference histogram, in purple, of scaled S1+S2+S3 (red+green+blue) histograms from the B-panel minus the S4 (black) FOV histogram. 
There is an excess of high-P50 FOVs having GPA50 values near 60--90\degr\ and
a deficit of intermediate GPA50 values greater than 90\degr. 
(Lower Right, D-panel):  Difference histogram, in gray, computed from the 
scaled S5 (gray) histogram of the B-panel minus the S4 (black) histogram from that panel.
\label{fig_WGPA_GPA}}
\end{figure}

Do regions of strong (or uniform) magnetic fields, as delineated by the high-P50/low-WGPA FOVs in GPIPS, show mean Galactic polarization position angles that
are distinct from the other, more typical, GPIPS FOVs? To try to answer this question, 
Figure~\ref{fig_WGPA_GPA} was created. Its upper-left, A-panel shows the 
GPIPS FOV values of WGPA versus Galactic polarization position angle GPA50, color-coded into the same S1, S2, and S3 high-P50 (red, blue, green), 
S4 average-P50 (black), and S5 low-P50 (gray) GPIPS FOV classifications, as per the previous discussion.
The overall impression is that the S1, S2, and S3 FOVs occupy lesser values
of WGPA compared to the S4 and S5 points, but this merely reflects the way the FOVs were
chosen for type/color classification. There is a clear asymmetry of FOVs with GPA50, with all subsets of FOVs showing distribution
centers less than GPA50 = 90\degr. The S5 low-P50 (gray) points seem to exhibit a GPA50
spread with the least central concentration---that is, they have the most uniform distribution of GPA50 values.

To examine how the various subsamples of FOVs are 
distributed in GPA50, scaled
histograms were formed and are shown as the upper-right, B-panel in 
Figure~\ref{fig_WGPA_GPA}.
The points in the upper-left, A-panel were binned by GPA50 separately for each of the 
subsamples. The resulting histograms were scaled to the same integrated number of
FOV counts as in the (typical) S4 sample (black histogram; 2,334 FOVs) and successively offset by 300 counts. All five histograms show mostly 
similar behavior: a broad peak near GPA50 = 75\degr, with a slow decay
to lesser GPA50 values and a faster decay to greater GPA50 values. Hence, all GPIPS 
FOVs appear to be drawn from nearly the same parent GPA50 distribution. However, subtle
differences can be seen, especially when comparing the high-P50 S1 (red) and low-P50 S5 (gray)
histograms. The high-P50 S1 histogram appears to have a more concentrated peak region and
no counts beyond GPA50~=~120\degr. The low-P50 S5 histogram has a broader peak region
and a broader distributed region, with significant FOV counts beyond GPA50~=~120\degr.

These small differences were enhanced by forming differences of the high-P50 and low-P50
histograms against the average-P50 histogram.
The S1 (red), S2 (blue), and S3 (green) histograms were added and scaled to the total FOV counts
in the S4 (black) histogram,
and these two resulting histograms were 
differenced to create the purple histogram in the lower-left, 
C-panel of 
Figure~\ref{fig_WGPA_GPA}. Similarly, the S5 low-P50 histogram (gray) was scaled to the S4 (black)
histogram and differenced, to create the gray histogram in the lower-right, D-panel.
In the C-panel, the high-P/low-WGPA FOVs show
an excess of GPIPS FOVs exhibiting near-disk-parallel, 60-90\degr\ GPA50 values, 
relative to the average GPIPS FOVs, and a deficit of FOVs
with GPA50 values in the 110 to 170\degr\ range. In the D-panel, the low-P50/high-WGPA, gray difference
histogram shows features with almost opposite behavior, showing a broad deficit 
for GPA50 values in the range 30-90\degr\ and a slight excess of FOVs, relative to the average, for GPA50 values in the 120-175\degr\ range. 

The C-panel (purple) and D-panel (gray) difference histograms give the impression of being
anti-correlated, but upon plotting the C-panel y-axis values against the D-panel y-axis values and fitting, the linear correlation coefficient was found to be small ($-0.67$ for 19 points) and not significant.
Given the peaked shape of the S4 (black) average FOV sample in the B-panel, the feature seen in the
D-panel (gray) difference histogram likely only indicates that the upper-right, 
B-panel S5 (gray) histogram is 
wider and flatter than the S4 (black) histogram. Thus, in differencing, that which survives to become the S5-S4 (gray) histogram in the D-panel is essentially just 
the negative of the S4 (black) histogram in the B-panel.

In summary, the strong (or uniform) magnetic field candidate zones, traced by the 
high-P50/low-WGPA GPIPS FOVs, appear to have a slight preference for
exhibiting GPA50 values consistent with magnetic fields that are disk-parallel,
while the low-P50/high-WGPA GPIPS FOVs appear to have mostly random magnetic field
orientations. These properties are somewhat in agreement with the findings reported for
\planck polarization data in \citet{Planck_XIX_2015}, though no strong evidence is 
found here for the high-WGPA (weak field) FOVs to be located surrounding the low-WGPA (strong field) FOVs, and the
size scales of the high-P50 groups are much smaller than for the \planck-identified regions.

\subsection{Remaining Questions}

The GPIPS data analyses presented above have revealed new aspects of
the nature of the magnetic field in the cold, dusty, star-forming molecular component
of the interstellar medium in the disk of the first quadrant of the inner Galaxy. In doing
so, questions have arisen about the locations and natures of the structures newly found.

The orientations of NIR polarizations across the GPIPS zone are mostly parallel
to the Galactic midplane, but not perfectly so. An overall offset of about 13-15\degr\ is present.
Resolved regions show orientations that differ by up to 90\degr\ from the
Galactic disk orientation. Is the overall offset merely due to contamination by a few regions showing
large GPA offsets? What are the natures of those GPA offset regions? How do they relate
to star-formation zones, supernovae, and/or large-scale outflows off the disk midplane? 
How would counterparts in external galaxies appear?

How well do magnetic fields follow spiral arms? If material (stars, gas, and dust) flows into and through the arms, how does that affect the spiral magnetic field? Do cool gas and dust clouds acquire magnetic fields in spiral arms or do they retain magnetic fields established at formation? 

Magnetic fields in the Milky Way and external galaxies may be
apportioned into large-scale (uniform), medium-scale (anisotropic random), and small-scale
(random, tangled, or turbulent) aspects \citep[see reviews by][]{Beck15, Haverkorn15}. What can the 
wealth of cold-ISM-probing GPIPS data reveal about the natures of the magnetic field 
on these different size scales in comparison to the corresponding hot-ISM values? 
What are the mean sizes of magnetically-coherent zones or cells \citep[e.g.,][]{Jones92}? 
What is the GPIPS-measured ratio of uniform-to-random magnetic field strength and how does it
change with location?

\clearpage
\section{Summary}\label{summary}

The Galactic Plane Infrared Polarization Survey (GPIPS) was conducted to help reveal the detailed
plane-of-sky orientations of the magnetic field in the cool, dusty, star-forming ISM of the disk of the
Milky Way in the first Galactic quadrant. GPIPS used near-infrared (NIR) observations in the
$H$-band (1.6~\um) to obtain background starlight polarimetry across 76~deg$^2$ of the Galactic midplane. It achieved almost 30~arcsec mean angular sampling for linear polarization measurements. 

GPIPS Data Release 4 (DR4) provides $H$-band in-band stellar magnitudes, linear Stokes parameters,
and derived polarization properties as well as 2MASS, GLIMPSE, \WISE, and \Gaia DR2 cross-references and key data values. 
GPIPS DR4 contains almost 10~million stellar polarization measurements,
of which more than 1~million are of high enough quality to permit individual use.
The remaining almost 9~million will be most useful when considered together in Bayesian or
other analysis methods in conjunction with archival photometric, parallax, and
spectroscopic data.

Here, the high-quality GPIPS data for each of the 3,237 observed $10 \times 10$~arcmin fields-of-view (FOVs) were converted to distribution functions for four key properties: the number of 
stars per FOV; the median polarization percentage; the median polarization position angle; and the interquartile range of the position angle distribution. 
Those FOV-based properties were themselves converted to distribution functions, 
one-dimensional averages (in each of Galactic longitude and latitude), and two-dimensional
distributions of these properties. The median number of high-quality GPIPS stars
per FOV is 252, or about 60 per \planck polarization resolution element. Median NIR polarization percentages are around 1.5\%\ for the $H$-band. 

The median polarization position angle, in Galactic coordinates (GPA), is about 76.7\degr, some
13.3\degr\ offset from that expected for a magnetic field oriented parallel to the Galactic disk
midplane. While none of the four key characterizing properties showed strong variation with Galactic
latitude, longitude variations of GPA were highly significant. Many zones
show GPA deviations of 30\degr\ to 90\degr\ from being disk-parallel.

Because NIR polarization arises due to alignment of dust grains within dusty molecular clouds, which also 
introduce significant extinction, there are distance limits imposed on the detection of GPIPS stars, which favor
sensing the magnetic field in the portion of the first quadrant closer than about 6-7~kpc. This region
includes portions of the Sagittarius and Scutum spiral arms and perhaps the near-side end of the
central Galactic Bar.
This represents an advantage for 
GPIPS in being able to accurately reveal the direction of the magnetic field over limited sightline distances, in comparison to \planck polarimetry, 
which carries contributions spanning vastly greater distances and thereby misses  
small-scale and localized changes in magnetic field properties.

An anti-correlation of the median polarization percentages with the widths of the position 
angle distributions for the FOV-based analysis was revealed and found to be generally similar to that seen in 
\planck analyses \citep[``$S \times p$":][]{Planck_XIX_2015, Planck_XII_2018}. 
There are important differences, however. The angular sizes of the largest ten groups showing
high median polarization percentages (and presumably stronger magnetic field strengths) in GPIPS are nearly an order of magnitude smaller than the
zones seen in \planck. Also, the suggestion that the high polarization zones are surrounded
by low polarization shells or membranes \citep{Planck_XIX_2015} is not supported by 
analysis of the GPIPS data.

\acknowledgments

In addition to the authors, GPIPS observations were conducted 
by S.~Hoq, J.~Shiode, J.~Moreau, M.~Bartlett, C.~Trombley, and M.~Hart.
S.~Hoq and T.~Pillai provided valued manuscript edits and suggested improvements.
This research has made use of the VizieR catalogue access tool, CDS,
 Strasbourg, France (DOI : 10.26093/cds/vizier), originally described
in \citet{Ochsenbein00}.
GPIPS DR4 includes data from the European Space Agency (ESA) mission
{\it Gaia} (\url{https://www.cosmos.esa.int/gaia}), processed by the {\it Gaia}
Data Processing and Analysis Consortium (DPAC,
\url{https://www.cosmos.esa.int/web/gaia/dpac/consortium}). Funding for the DPAC
has been provided by national institutions, in particular the institutions
participating in the {\it Gaia} Multilateral Agreement.
GPIPS DR4 also includes data products from the Two Micron All Sky Survey, 
which is a joint project of the University of Massachusetts and the Infrared 
Processing and Analysis Center/California Institute of Technology, funded by 
NASA and NSF. This publication makes use of data products from the {\it Wide-Field
Infrared Survey Explorer}, which is a joint project of the University of California, Los Angeles,
and the Jet Propulsion Laboratory (JPL)/California Institute of Technology (CalTech), funded
by NASA. This work is based in part on observations made with the {\it Spitzer Space Telescope}, which was operated by JPL/CalTech under a contract with NASA.
This research was conducted 
using the Mimir instrument, jointly developed at Boston University and Lowell 
Observatory and supported by NASA, NSF, and the W.M. Keck Foundation.
Massachusetts NASA Space Grant number 165283 from MIT to Boston University
partially supported C.~Cerny during 2018 summer and is gratefully acknowledged.
The GPIPS effort has been made possible by grants 
AST 06-07500, 09-07790, 14-12269, and 18-14531 from NSF/MPS to Boston University and by grants of 
significant observing time (BU key project status) from the Boston University 
-- Lowell Observatory partnership.

\facility{Perkins}

\clearpage

\appendix

\restartappendixnumbering

\section{Data Quality Evaluations}\label{quality_checks}

Checks and evaluations of data obtained for GPIPS were performed as the data were collected, reduced, and analyzed to develop provisional data products. Observations judged to have not met the quality standards described below were rejected and the FOV(s) were re-observed as needed to obtain the high quality data required for inclusion in the final GPIPS DR4 data products.

Table~\ref{tab_sum} presents a shortened, representative portion of the Field Properties Summary Table, which is contained in the full DR4 file set. The FOV rows selected for presentation in Table~\ref{tab_sum}  highlight FOVs that passed all of the quality controls and a couple of FOVs that did not fully meet one quality test. In the Table, the first column lists the sequential GPIPS FOV number, 
from 0001 to 3237. The second and third columns list the Galactic coordinates of the true centers of the observed FOVs. Note that all image orientations were equatorial. The fourth column lists the angular offset, in arcsec, of the true FOV center direction from the nominal GPIPS grid center direction. Adjacent GPIPS grid centers were designed to provide FOV overlap by up to one arcmin, so acceptable offset values were generally less than about 45~arcsec. The largest center position offset that was judged acceptable was 55~arcsec. The fifth column lists the observing night designations, in  YYYYMMDD form of the UT date. The sixth column identifies the first image number of the observation, and may be matched to the hand-written observing logs (available from the DR4 website) to obtain sky and instrument context regarding specific observations. The seventh column lists the FWHM of the deep photometric image PSF, in arcsec, corrected for the pixel sampling contribution. 

Column eight lists the Secular Amplitude for each observation. This represents the slow variation in stellar brightness levels (above the background sky) for the typically $\sim$300 bright stars per FOV
matched between the first image and each of the remaining 95---118 images for an observation (e.g., Figure 10 of \citetalias{Clemens12a}). Secular Amplitude values exceeding 3\% were judged to indicate contamination by clouds or other problems and those FOVs were re-observed. The ninth column lists the amplitude of the Sky Noise, computed as the root-mean-square (RMS) of the image-to-image star-matched brightness changes, after removal of the secular variation and removal of the modulation by the average polarization for the FOV. This characterization is also described in detail in Appendix B of \citetalias{Clemens12a}. Sky Noise levels above 1.2\% were judged excessive and those FOVs were re-observed. Of the 3,237 FOVs, only one FOV exceeded this criterion and was retained (its Sky Noise value was 1.3\%). The number of images rejected from each observation set for poor PSF shapes is listed in column ten. At least 89 acceptable images per FOV were required for an observation to be included, and the number of accepted images is listed in column eleven. This criterion ensured that multiple images would sample all 16 HWP orientation angles.

Column twelve lists a Pattern Score for each FOV. These were formed by combining seven different quantitative evaluations of the polarization percentage and orientation patterns for each FOV. They were used to assess whether the resulting polarization patterns resembled parallel orientation directions or orientation directions that
aligned tangentially (i.e., swirls), if the polarization percentage increased with offset from the image center, and other negative characteristics. These evaluations uncovered several observing or data reduction failures, including cases of frozen HWP rotation, bad flat-field images, and other issues. A custom software tool, coded in IDL, was created to view the polarization pattern for each FOV observing set and to apply the seven evaluations. These component scores were combined to be the Pattern Score, scaled to range from $-$9 to $+9$. 

Strongly negative Pattern Scores correlated with non-physical, swirl-like orientation patterns that also had polarization percentages rising with radial offset from image centers, neither of which should occur.  Strongly positive Pattern Scores showed
highly-parallel polarization orientations and an absence of data problems. Score values less than
zero indicated possible problems, and those FOVs were re-observed. Some FOVs exhibited complex polarization orientation patterns resulting in mid-range, but mostly positive, Pattern Scores. Re-observation of many of these FOVs recovered the same complex polarization patterns, reinforcing their veracity. First re-observations of negative-scoring FOVs generally did not yield polarization patterns consistent with the initial observations, so these FOVs were re-observed until the patterns became stable in subsequent observations, which resulted in their Pattern Scores becoming positive. 

\movetabledown=25mm
\begin{rotatetable}
\begin{deluxetable}{ccccccccccccc}
\tabletypesize{\footnotesize}
\tablecaption{GPIPS Field Properties Summary Table\label{tab_sum}}
\tablewidth{0in}
\tablehead{
\colhead{GPIPS} &\colhead{Galactic} &\colhead{Galactic} & \colhead{Center} & 
\colhead{UT} & \colhead{First} & \colhead{PSF} & \colhead{Secular} & \colhead{Sky} & 
\colhead{Rejected} & \colhead{Accepted}  & \colhead{Pattern} & 
\colhead{QC} \\
\colhead{FOV}&\colhead{Long.} & \colhead{Lat.} &\colhead{Offset} & 
\colhead{Night} & \colhead{Image} & \colhead{FWHM} & \colhead{Ampl.} &\colhead{Noise} &
\colhead{Image} & \colhead{Image} &  \colhead{Score} &
\colhead{Flag} \\
\colhead{ No.} &\colhead{(\degr)} & \colhead{(\degr)} & \colhead{(arcsec)} &  \colhead{(Y4M2D2)} &
\colhead{No.} & \colhead{(arcsec)} &
\colhead{(\%)} & \colhead{(\%)} &\colhead{Count} & \colhead{Count}  \\
\colhead{(1)}&\colhead{(2)}&\colhead{(3)}&\colhead{(4)}&\colhead{(5)}&\colhead{(6)}
&\colhead{(7)}&\colhead{(8)}&\colhead{(9)}&\colhead{(10)}&\colhead{(11)}&\colhead{(12)}
&\colhead{(13)}
}
\startdata
%
0001&18.082& $-$0.635&5.5&20070919&316&1.196&1.0&0.61&6&96&6&0\\
0002&18.151& $-$0.767&6.1&20080608&2232&1.265&0.0&0.34&2&99&5&0\\
...\\
0460&23.005&0.949&32.7&20120614&2892&1.853&0.0&0.55&0&96&4&0\\
0461&23.078&0.826&2.4&20190627&1329&1.299&0.7& 0.35&0&96&0&2\\
0462&23.143&0.690&15.1&20110609&2757&1.786&0.0&0.40&0&96&7&0\\
...\\
3128&54.851&$-$0.381&10.8&20100605&1290&1.609&0.5&0.39&0&102&4&0\\
3129&54.924&$-$0.514&6.9&20130626&1105&1.485&0.0& 0.45&1&95&1&1\\
3130&54.999&$-$0.646&1.6&20170527&1745&1.658&0.8&0.53&0&96&9&0\\
\enddata
\tablecomments{This is a shortened and selective portion of the full table that is available on 
the GPIPS DR4 website (http://sites.bu.edu/gpips)} 
\end{deluxetable}
\end{rotatetable}

The final column in the Table contains a Quality Control (QC) flag to indicate the overall quality of the final data for each observed FOV. In this column, a zero entry means the observation met all of the criteria for PSF quality, Secular Amplitude, Sky Noise, good image count, and polarization Pattern Score. Values greater than zero represent weighted degrees of departure from the goals. Of the 3,237 FOVs comprising GPIPS, 3,145 (97.2\%) met all of the quality goals, and so have zero value entries in the QC flag column. Three FOVs had QC flag values of two (field numbers 461 [listed in Table~\ref{tab_sum}], 487, and 2115) due to their Pattern Scores being zero. Eighty-two FOVs had QC values of unity, including FOV number 3129, which is also listed in 
Table~\ref{tab_sum}. All of these are due to their Pattern Scores being unity, that is, weakly positive but not as strongly positive (exhibiting highly parallel polarization orientations) as for other fields. Six of the remaining seven FOVs with unity QC values showed moderately higher Secular Amplitudes and center position offsets.

\section{Matching \gaia DR2, 2MASS, GLIMPSE, and \wise Stars to GPIPS Stars}\label{gaia_match}

\setcounter{table}{1}

Stellar data for archival \Gaia DR2, 2MASS, GLIMPSE, and \wise catalog entries were fetched using the VizieR catalog tool \citep{Ochsenbein00} for all point sources contained within the boundaries of each GPIPS FOV, using searches centered at the actual observed coordinates (Table~\ref{tab_sum}). Since GPIPS observations were obtained using sky dithering,
with later sky registration and coadding of the images, the resulting deep photometric and HWP images do not have sharply defined uniform exposure edges nor uniform sizes. To set the search area for each FOV, the PHOTCAT stars with the least and greatest R.A. and decl. values in the FOV were used to delineate the search box extent. To this, an additional 1.5~arcsec wide outer zone was added to extend the field sizes to provide some margin for individual star matching, as described below. The resulting search field sizes averaged about $9.9 \times 9.9$~arcmin, somewhat smaller than the Mimir instrument FOV. The effective field size for the archival catalog searches is about 1\% larger than the effective DR4 FOV size (see Section~\ref{sec_unique}), which should bias the matching statistics only slightly.

Archival stars were matched to GPIPS stars in each FOV for each of the four catalogs using cone searches with maximum relative projected radial offsets of 1.25~arcsec between GPIPS stars and archival catalog stars. This value was selected based on using matches of test FOVs employing a set of trial cone angle values.  Star matching was performed starting from the brightest to the faintest GPIPS stars
in each FOV, and similarly brightest to faintest archival catalog stars, to minimize false matches that might have resulted from the large numbers of faint stars. A few GPIPS stars were found to have potential matches to multiple archival catalog
stars within the radial offset limit. For these cases, color values were computed for each star (e.g., \Gaia g-band minus Mimir $H$-band) and were combined with the radial offset values using gaussian probability functions to select the single most likely stellar match. Stars, once matched, were removed from the remaining GPIPS and archival catalog potential matching pools.

\subsection{Match Rates}

Table~\ref{tab_matches} provides summaries of the numbers and rates of stellar matches of archival catalogs to GPIPS. Additionally, for the POLCAT stars, the matching numbers and fractions for each of the UF1 (and its UF0 subset, listed in italics), UF2, and UF3 star subsets are indicated.
Using Table~\ref{tab_matches}, reverse match rates, for example the fraction of \gaia stars that have GPIPS POLCAT UF1 matches, may be computed from the match numbers divided by the total entries for each of the four archival catalogs across the GPIPS FOVs. These totals are listed in the final row of Table~\ref{tab_matches}.

The greatest match rates of GPIPS to archival catalogs occurs for GLIMPSE. The lowest GLIMPSE match rate is 70.6\%\ for PHOTCATs and the highest is 99.4\%\ for the UF0 subset of UF1 GPIPS stars. For every UF stellar category, the rate of matching to GLIMPSE stars is higher than it is to any of the other three archival catalogs. This is a distinct improvement compared to the GPIPS DR1 matching approach and offers the opportunity to use $H$-band from GPIPS and $M$-band equivalent (4.5~\microns) from GLIMPSE to quantify line-of-sight extinction using the (H-M) colors with the Rayleigh-Jeans Color Excess method \citep[RJCE;][]{Majewski11}.

In Table~\ref{tab_matches}, the \gaia DR2 columns report the matching stellar numbers and rates for GPIPS stars for the case of no application of the parallax evaluation (the ``All" column) and for matching to stars with the potentially useful parallax criteria (the ``Parallax" column). 

\begin{deluxetable}{lccccc}
\tablecaption{Match Properties for GPIPS stars found in Archival Catalogs\label{tab_matches}}
\tablehead{
\colhead{GPIPS}&\multicolumn{2}{c}{{\hspace{0.65cm}\gaia DR2}\hspace{1.7cm}} & \colhead{\hspace{0.5cm}2MASS}\hspace{0.5cm} & \colhead{\hspace{0.5cm}GLIMPSE}\hspace{0.5cm} & \colhead{\hspace{0.5cm}\wise}\hspace{0.5cm} \\
\colhead{Data Set}& \colhead{All} & \colhead{Parallax} \\
\colhead{(1)}&\colhead{(2)}&\colhead{(3)}&\colhead{(4)}&\colhead{(5)}&\colhead{(6)}
}
\startdata
PHOTCAT & 7,732,890 & 5,077,793 & 8,350,602 & 10,955,641 &1,702,876\\
		&49.8\% & 32.7\% & 53.8\% & 70.6\%& 11.0\% \\ [12pt]
POLCAT  & 5,635,761 & 3,858,064 & 7,219,367 & 8,361,968 &1,657,927 \\
		&58.1\% & 39.7\% & 74.4\% & 86.1\% & 17.1\% \\ [6pt]
\ \ \ \ \ \ UF1	& 857,821&641,151&985,415& 1,011,415 & 721,267 \\
			&84.0\% & 62.8\% & 96.5\% & 99.0\% & 70.6\% \\ [6pt]
\ \ \  \ \ \ \  {\it UF0} & {\it 181,096}& {\it 128,983}& {\it 234,215} & {\it 237,449} & {\it 208,154} \\
				&{\it 75.8\%} & {\it 54.0\%} & {\it 98.0\%} & {\it 99.4\%} & {\it 87.1\%} \\ [6pt]
\ \ \ \ \ UF2	& 1,915,221&1,282,912&2,504,421 & 2,691,718 & 674,206 \\
			&66.6\% & 44.6\% & 87.1\% & 93.6\% & 23.4\% \\ [6pt]
\ \ \ \ \ UF3	& 2,862,719&1,934,001 &3,729,531 & 4,658,835 &262,454 \\
			&49.3\% & 33.3\% & 64.2\% &80.2\% & 4.5\% \\[12pt]
Archival Totals&13,879,724&7,750,932&9,198,385&18,331,898&2,109,578\\
\enddata
\end{deluxetable}

Overall, GPIPS POLCAT match rates with the archival catalogs are somewhat greater than PHOTCAT ones, as the latter contains more faint
stars than the former. The overall GPIPS POLCAT match rates are about 58\% for \gaia (40\% for \gaia stars with parallaxes), 74\% for 2MASS, 86\% for GLIMPSE, and 17\% for \WISE. 
\WISE is generally more sensitive than the {\it Spitzer} Infrared Array Camera (IRAC) \citep{Fazio04}, used to conduct the GLIMPSE observations, for directions away from the Galactic plane. But, \WISE\,  has coarser angular resolution than IRAC, resulting in a brighter confusion limit in the Galactic plane and fewer sources per GPIPS FOV than for GLIMPSE in the same wavebands.

In Table~\ref{tab_matches}, there is a trend of increasing GPIPS match rates as UF number decreases, corresponding to increasing apparent brightness. The one interesting exception is the POLCAT UF0 subset matching to \gaia, which shows somewhat lower rates than the larger UF1 sample which contains UF0. This could be caused by the greater $P^\prime$ values in UF0 arising because of their greater dust column densities and thereby greater $A_V$ values. This in turn would affect \gaia matches through the optical $g$-band brightness limit, as these shorter wavelengths would suffer the greatest extinctions. 

The archival catalog matching information was integrated into GPIPS in two ways, one related to the FOV-based data products and one related to the unique star data file.

\subsection{FOV-Based Archival Catalog Stellar Match Data Files}\label{match_files}

For the FOV-based products, the 2MASS and GLIMPSE matching stellar information was directly integrated into the PHOTCAT and POLCAT files for each of the 3,237 GPIPS FOVs. These 2MASS and GLIMPSE values supersede the values listed in the GPIPS DR1 through DR3 listings, as follows. Prior to DR4, only the highest-quality 2MASS data were included in GPIPS files by requiring an ``A" photometry quality rating for each of the $J$-, $H$-, and $K$-band 2MASS magnitudes \citep{Skrutskie06}. For DR4, this criterion was {removed, allowing all 2MASS matches to be included.} 
The fraction of GPIPS POLCAT stars with 2MASS entries thereby increased from about 64\% for DR1 to about 74\% for DR4.
Similarly, prior to DR4, GLIMPSE data were included in GPIPS files only through being matched via their 2MASS designations. For DR4, GLIMPSE point source positions were directly matched to GPIPS stars and were included in GPIPS files. This approach greatly increased the fraction of GPIPS entries with GLIMPSE matches, especially those for which 2MASS matches were lacking due to faintness. 

The \gaia DR2 stellar data were further examined to determine the quality of the parallax information provided in DR2. Stars with possibly useful parallax values for GPIPS applications by potential users were judged to be those for which the parallax SNR ($\pi / \sigma_\pi$) was greater than 0.5 and the parallax value was greater than $-2$~mas. 
As noted earlier in the text, these liberal limits will mainly remove non-detection upper limits and spurious values while avoiding potential population biases \citep{Luri18,BailerJones18}. These do not include the additional criteria on $g$-band mag, $H$-band mag, or $(H - K)$ color uncertainties that were applied to create Figures~\ref{fig_G_para}, \ref{fig_G2_hk_mh}, and \ref{fig_G3_hk_para} and the associated summaries 
presented in Table~\ref{tab_summary}.

For each of  \gaia DR2, 2MASS, GLIMPSE, and \wise, the FOV-based approach generated one data file per FOV for each of these four archival catalogs. In each file, every archival star contained within the FOV search area for that catalog is listed, along with PHOTCAT and POLCAT star identifiers for each matching GPIPS star. This approach provides bidirectional matching information and enables characterizing match success rates. Note that these match rates are somewhat optimistic, in the sense that not every match of an archival star to a GPIPS star will necessarily provide uniformly high quality stellar information. Users are strongly encouraged to evaluate the uncertainties listed for each reported value for matches between GPIPS and the archival entries.

Each \gaia, 2MASS, GLIMPSE, and \wise match file has metadata for the GPIPS FOV number, search center sky direction in R.A. and decl., and field search size in the R.A. and decl. directions. Each match file has one row for each archival star found in the search field. Each lists an R.A.-ordered star number, followed by entries listing the matching POLCAT and/or PHOTCAT star, referenced to the identifying entries in the GPIPS files, or a $-99$ value to indicate no match to a GPIPS star was found. The R.A. and decl. listed in the archival catalog follow next. The 2MASS, GLIMPSE, and \wise match file rows next list the three- or four-band magnitudes and uncertainties ($J$, $H$, and $K$ for 2MASS; 3.6, 4.5, 5.8, and 8.0~\microns\ for GLIMPSE; W1[3.6], W2[4.5], W3[12], and W4[22~\microns] for \wise), followed by the star designation unique to that catalog.

\Gaia DR2 match files contain similar metadata, star numbers, GPIPS star numbers, and R.A. and decl. values as reported in \gaia DR2. Next, each row lists the \Gaia $g$-band magnitude, $g$-band mag. uncertainty, parallax, parallax uncertainty, proper motions and their uncertainties in the R.A. and decl. directions, and the \gaia star designation label.

\subsection{Unique Star Data File Contents and Archival Match Information}\label{unique_data}

As an alternative to the FOV-based data files, the unique star data 
file \citep{Clemens20} collects all GPIPS photometry and polarimetry and all matching stellar data appropriate to each GPIPS star, with data fields for each star
as listed in the example entry shown as Table~\ref{tab_fields}. The unique star file contains entries for 
13,861,329 GPIPS stars found in the 3,237 PHOTCAT files.

\clearpage
\startlongtable
\begin{deluxetable}{lccl}
\tablecaption{Data Fields for Each Star in the Unique Star File\label{tab_fields}}
\tablehead{\colhead{Field Name}&\colhead{Type}&\colhead{Example Values}&\colhead{Description}
}
\startdata
NUM&L\tablenotemark{a}&605776&RA-ordered star serial number\\
DESIG\_GPIPS&S&\multicolumn{2}{l}{\ \ \ \ `GPIPS\_J182558.20-124843.5'}\\
RA\_DEG&D&276.49249&J2000 RA, in degrees\\
DEC\_DEG&D&$-$12.81209&J2000 decl., in degrees\\
GAL\_L&D&18.65135&Galactic longitude, in degrees\\
GAL\_B&D&$-$0.27934&Galactic latitude, in degrees\\
GPSTAR&L&Array[4]:&FOV$\times 10^5 + $ stellar ID No.\\
&L&3405524&Observation 1: FOV 34, star number 5524\\
&L&3500495&Observation 2: FOV 35, star number 495\\
&L&4806180&Observation 3: FOV 48, star number 180\\ 
&L&4900307&Observation 4: FOV 49, star number 307\\
CAT\_FLAG&I&Array[5]:&Array of archive catalog match codes\\
&I&1&a POLCAT star matched, if =1\\
&I&31&a \Gaia DR2 star matched, see Table~\ref{match_quality} for code meaning\\
&I&7&a 2MASS star matched, see Table~\ref{match_quality} for code meaning\\
&I&15&a GLIMPSE star matched, see Table~\ref{match_quality} for code meaning\\
&I&31&a WISE star matched, see Table~\ref{match_quality} for code meaning\\[6 pt]
H\_MAG&D&10.31231&Mimir $H$-band photometric magnitude\\
E\_HMAG&D&0.00039&Internal uncertainty in $H$-band mag.\\
S\_PHOT&D&0.02254&External $H$-band mag. uncert. \citep[see][]{Clemens12c}\\
H\_VAR&I&0&Variability flag for overlapping FOVs (see text)\\
P&D&1.47&Debiased polarization percentage ($P^\prime$)\\
E\_P&D&0.33&Uncertainty in polarization percentage\\
PA\_DEG&D&13.7&Polarization position angle (EPA) E of N, in degrees\\
GPA\_DEG&D&75.9&Galactic position angle (GPA), in degrees\\
E\_PA&D&6.3&Uncertainty in position angles, in degrees\\
Q&D&1.34&Stokes Q (\%), normalized by Stokes I\\
E\_Q&D&0.33&Uncertainty in Stokes Q (\%)\\
U&D&0.69&Stokes U (\%), normalized by Stokes I\\
E\_U&D&0.32&Uncertainty in Stokes U (\%)\\
NHWP&I&64&Number of HWP images in which the star appears (see text)\\
UF&I&1&Usage Flag (1, 2, 3, or $-$99) \citep{Clemens12c}\\[6pt]
\multicolumn{3}{c}{Designations in Archival Catalogs}\\
DESIG\_GAIA&S&\multicolumn{2}{l}{\ \ \ \ `15979648'}\\
DESIG\_2MASS&S&\multicolumn{2}{l}{\ \ \ \ `J18255820-1248437'}\\
DESIG\_GLIMPSE&S&\multicolumn{2}{l}{\ \ \ \ `G018.6514-00.2794'}\\
DESIG\_WISE&S&\multicolumn{2}{l}{\ \ \ \ `J182558.18-124842.9'}\\[6pt]
\multicolumn{3}{c}{\gaia DR2 data values:}\\
PAR&D&0.0056&Parallax, in milli-arcsec\\
E\_PAR&D&0.1948&Uncertainty in parallax, in milli-arcsec\\
GMAG&D&17.1452&$g$-band photometric magnitude\\
E\_GMAG&D&0.0016&Uncertainty in $g$-band mag.\\
PMRA&D&0.128&Proper motion along RA, in mas yr$^{-1}$\\
E\_PMRA&D&0.329&Uncertainty in RA proper motion, in mas yr$^{-1}$\\
PMDEC&D&$-$2.088&Proper motion along Dec, in mas yr$^{-1}$\\
E\_PMDEC&D&0.372&Uncertainty in Dec proper motion, in mas yr$^{-1}$\\[6pt]
\multicolumn{3}{c}{2MASS data values:}\\
J\_2MASS&D&11.975&$J$-band photometric magnitude\\
E\_J2MASS&D&0.033&Uncertainty in $J$-band mag.\\
H\_2MASS&D&10.408&$H$-band photometric magnitude\\
E\_H2MASS&D&0.058&Uncertainty in $H$-band mag.\\
K\_2MASS&D&9.784&$K$-band photometric magnitude\\
E\_K2MASS&D&99.99&Uncertainty in K-band mag. {\it (note upper limit)}\\[6pt]
\multicolumn{3}{c}{GLIMPSE data values:}\\
B36MAG&D &9.137&IRAC 3.6~$\mu$m (``$\sim$L") band photometric magnitude\\
E\_B36MAG&D&0.104&Uncertainty in IRAC 3.6~$\mu$m band mag.\\
B45MAG&D&9.011&IRAC 4.5~$\mu$m (``$\sim$M")  band photometric magnitude\\
E\_B45MAG&D&0.078&Uncertainty in 4.5~$\mu$m band mag.\\
B58MAG&D&8.795&IRAC 5.8~$\mu$m band photometric magnitude\\
E\_B58MAG&D&0.053&Uncertainty in IRAC 5.8~$\mu$m band mag.\\
B80MAG&D&8.767&IRAC 8.0~$\mu$m band photometric magnitude\\
E\_B80MAG&D&99.99&Uncertainty in IRAC 8.0~$\mu$m band mag. {\it (note upper limit)}\\[6pt]
\multicolumn{3}{c}{\WISE data values:}\\
W1MAG&D&8.975&\WISE Band 1 (3.4~$\mu$m; ``$\sim$L") photometric magnitude\\
E\_W1MAG&D&0.025&Uncertainty in \WISE Band 1 mag.\\
W2MAG&D&8.877&\WISE Band 2 (4.6~$\mu$m; ``$\sim$M") photometric magnitude\\
E\_W2MAG&D&0.022&Uncertainty in \WISE Band 2 mag.\\
W3MAG&D&7.733&\WISE Band 3 (12~$\mu$m) photometric magnitude\\
E\_W3MAG&D&0.195&Uncertainty in \WISE Band 3 mag.\\
W4MAG&D&3.969&\WISE Band 4 (22~$\mu$m) photometric magnitude\\
E\_W4MAG&D&0.069&Uncertainty in \WISE Band 4 mag.\\
\enddata
\tablenotetext{a}{Data type: I - short integer; L - long integer; D - double precision real;
S - string. Arrays are indicated by variable type and number of entries.}
\end{deluxetable}
\clearpage

Most of the data fields in the unique star file are conventional and are based on existing GPIPS \citepalias{Clemens12c}, 2MASS, GLIMPSE, \gaia DR2, and \WISE catalogs and can be interpreted using Table~\ref{tab_fields} and Table~\ref{match_quality} as guides. Where no data exist or there were no matching archival catalog data, values and/or uncertainties of 99.99 were entered. Where the uncertainty in polarization percentage $\sigma_P$ was greater than the raw polarization percentage $P_{RAW}$, the tabulated polarization percentage (``P" as representing debiased $P^\prime$ values) was set to zero and the
corresponding position angle uncertainty SPA was set to 180 deg. For \gaia matches, upper limits on
parallax and proper motions are noted with values of 30.0 and uncertainties of 90.0. 

A few of the fields listed in the unique star file are newly introduced here, and warrant discussion.
The GPSTAR array carries the combined FOV and star number in that FOV for the
up to four observations of any one unique star. For the example star shown in
Table~\ref{tab_fields}, there were four observations of this star, as indicated in the 
GPSTAR entries. For stars with only a single observation, the first entry of GPSTAR contains the FOV and star number and the remaining GPSTAR entries are set to $-$1. 

The CAT\_FLAG array provides a quick encoding of information regarding the natures of the
other catalog properties for stars matched to the unique GPIPS star. The first CAT\_FLAG
entry is either unity, if there is matching GPIPS polarization information from one
or more POLCATs, or zero if there is no POLCAT information for this unique PHOTCAT star. 
The second through fifth entries encode quality information related to each archival catalog
match star, as listed in Table~\ref{match_quality}. The score codes in the second
column of Table~\ref{match_quality} were summed for each catalog to create the 
reported value. That is, a
CAT\_FLAG score of 31 in the second array entry identifies a \gaia DR2 matched 
star with reported values of parallax, $g$-band magnitude, as well as RA and decl. 
values of proper motion. Note that these scores do not indicate confidence levels:
instead, they only reports the presence or absence of key quantities. Score maxima
are unity for GPIPS POLCAT, 15 for 2MASS, and 31 for the remaining catalogs.
The CAT\_FLAG encoding is not perfect, however. For the example UF1 star
values presented in Table~\ref{tab_fields}, the 2MASS $K$-band and GLIMPSE 8.0~$\mu$m band uncertainties are
reported as upper limits (with values of 99.99 in their uncertainty fields) while their 
photometric values are still listed in those archival catalogs as well as here.
Users are warned that the quick summaries provided in the CAT\_FLAG array may
miss such details.

\begin{deluxetable}{ccl}
\tablecaption{CAT\_FLAG Entry Encoding\label{match_quality}}
\tablehead{\colhead{Entry}&\colhead{Additive}&\\
\colhead{Number}&\colhead{Score}&\colhead{Description}\\
\colhead{(1)}&\colhead{(2)}&\colhead{(3)}
}
\startdata
1&1&GPIPS POLCAT star was matched\\[6pt]
2&1&\gaia DR2 star was matched\\
&2&E\_PAR not upper limit (90.0)\\
&4&E\_GMAG not upper limit (99.99)\\
&8&E\_PMRA not upper limit (90.0)\\
&16&E\_PMDEC not upper limit (90.0)\\[6pt]
3&1&2MASS PSC star was matched\\
&2&E\_J2MASS not upper limit (99.99)\\
&4&E\_H2MASS not upper limit (99.99)\\
&8&E\_K2MASS not upper limit (99.99)\\[6pt]
4&1&GLIMPSE star was matched\\
&2&E\_B36MAG not upper limit (99.99)\\
&4&E\_B45MAG not upper limit (99.99)\\
&8&E\_B58MAG not upper limit (99.99)\\
&16&E\_B80MAG not upper limit (99.99)\\[6pt]
5&1&\WISE star was matched\\
&2&E\_W1MAG not upper limit (99.99)\\
&4&E\_W2MAG not upper limit (99.99)\\
&8&E\_W3MAG not upper limit (99.99)\\
&16&E\_W4MAG not upper limit (99.99)\\
\enddata
\end{deluxetable}

The H\_VAR field identifies stars with multiple GPIPS observations that exhibit $H$-band
photometric differences in excess of ten times their propagated external photometric difference
uncertainties. For the up to four observations possible for stars in overlapping FOVs,
if any of the (up to) six posible pair-wise photometric differences exceed this $10 \sigma$ 
threshold then the H\_VAR field was set to unity, otherwise it was set to zero.
A total of 18,499 stars have H\_VAR~=~1 values.

In Table~\ref{tab_fields}, the NHWP field most often lists the number of distinct HWP images (up to 16) 
in which a non-overlapping star was detected. Faint stars might fail to be detected in all 16 HWP images for a FOV, so
this field can be useful for culling faint polarization candidates. For overlapping FOVs, NHWP
carries the {\it sum} of all constituent HWP detection counts, resulting in a maximum value for
this field of 64, as is the case for the example star properties listed in Table~\ref{tab_fields}.
For stars with PHOTCAT magnitudes but no PHOTCAT matches, the NHWP fields were set to 99. 
Combining the GPSTAR and NHWP field information will provide adequate insight into whether
a multiply observed star was detected in sufficient HWP images to be useful.

Finally, the UF field carries the 1, 2, and 3 values associated with the UF1, UF2, and UF3
classifications described in the text. Note that for multiply observed stars, this classification
was reassessed after the stellar polarization data were merged from the multiple observation
values. This could result in a lower UF value than appear in the FOV-based single observations
of the star. Also, for stars in the PHOTCAT-based unique star file that do not have GPIPS 
polarization data (e.g., no POLCAT star matches), the UF field was set to $-99$. 

Two potentially useful classes of data are absent in this combined data file. The first is the 
detailed photometric and polarimetric information obtained for the duplicate stars that appear in more than
one GPIPS FOV because of the overlapping nature of the survey. The polarization information
for these stars is merged, as described below, in the unique star file, though FOV-identifying
information for each merged star is in the GPSTAR array. The second class of missing
data are the FOV-based files of archival data, especially of the archival stars in those files that
are {\it not} matched to GPIPS stars. These unmatched stars are fully absent in the unique
star file and are only listed in the FOV-based files. A less vital third class of missing information
involves some of the ancillary polarization information, such as where each star appears on
the Mimir detector FOV, which is used for instrumental polarization correction. Any later 
recalibration of the GPIPS polarization, resulting from updated instrumental polarization
corrections, will need to utilize the FOV-based POLCATs and/or to rebuild the unique star file.
The fourth, minor, missing element is the metadata for each GPIPS FOV, normally
included in the PHOTCATs and POLCATs and supported by the electronic copies of the
hand-written observing logs. The GPSTAR array entries provide 
information about the original FOVs and star numbers in the constituent observations
so that these data may be referenced in the FOV-based data products if needed.

Despite these absences, the DR4 unique star data product is potentially the most useful for accessing GPIPS, as it contains virtually all relevant photometric, polarimetric, and parallax
data for conducting a wide variety of studies with the least data accessing software overhead.

\section{Polarization Signal to Noise Choices and Effects}\label{Appendix_PSNR}

In Figures~\ref{fig_mosaic} through \ref{fig_GPA_vs_P_Hmag} in the text, the Usage Flag (UF) designations introduced in \citetalias{Clemens12c} are seen to reveal important aspects about the nature of these designated stars and the ISM they probe. One aspect of a uniform survey, such
as GPIPS, being applied to a wide range of stellar brightnesses is that there is a nearly continuous
and wide range of resulting linear polarization signal-to-noise ratios present in the sample of stars. Which ratio values to 
include and which to reject for a particular scientific study depends on the goals and
questions posed. For the purposes of this paper, and its primarily FOV-based median polarization
(and magnetic field) characterizations, what is the best balance between the number of
stars selected for analysis and the signal-to-noise criterion applied to select them that will
also accurately reveal the properties of the magnetic field being surveyed?

The UF2 and UF3 classified stars were previously \citepalias{Clemens12c} shown
to be generally unsuitable as individual probes of plane-of-sky magnetic field properties,
though they could serve such purposes with suitable sample- or area-averaging. Addressed here is the
question of whether the UF1 stars are suitable or whether additional, more stringent, 
signal-to-noise criteria must be applied in order to extract meaningful magnetic field properties
from the GPIPS data.

In the following subsections, the UF1 and UF0 polarization properties for the GPIPS-1619 field are compared, the polarization signal-to-noise
distributions for all GPIPS stars are considered, and differences between UF1 and UF0 
WGPA (see Section~\ref{sec_uf1}) properties are examined for all GPIPS FOVs as functions of the numbers of stars in each
FOV as well as the P50 and WGPA values in each field. 
The conclusion from this examination is that
the UF1 stars, those with $m_H < 12.5$~mag and $\sigma_P < 2$\%, are shown to reveal magnetic field properties in the GPIPS-1619 field nearly as well as their UF0 subset. Also, consideration of the
full set of GPIPS FOVs does not change this conclusion.

\subsection{GPIPS-1619 Comparisons}

\begin{figure}
\includegraphics[width=7.25in]{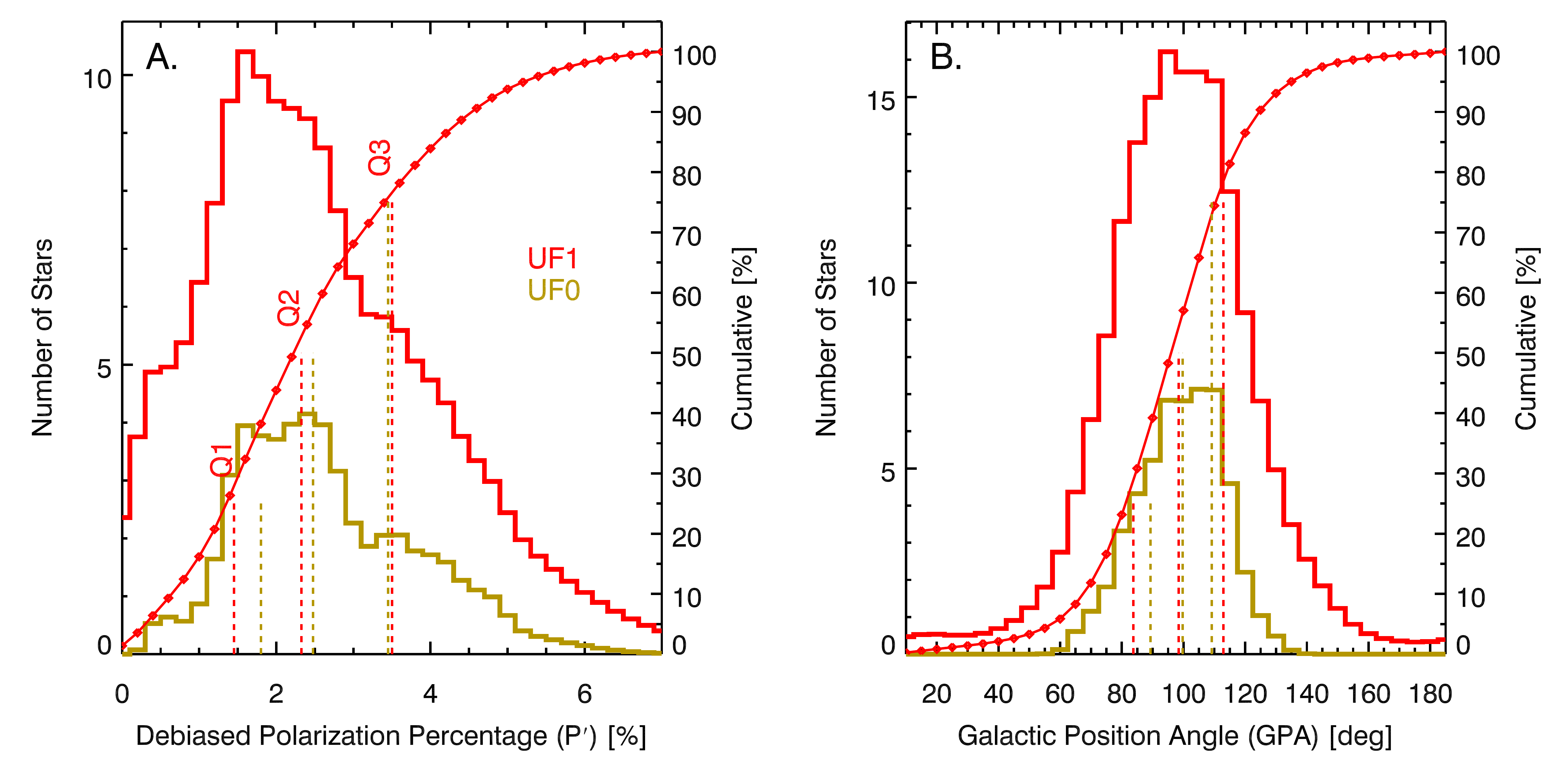}
\caption{Histograms of debiased polarization percentage $P^\prime$ (Left, A-panel) and Galactic position angle GPA
(Right, B-panel) for UF1 stars (red lines) and UF0 stars (those with P$^\prime$SNR~$\ge 3$; yellow lines) in GPIPS-1619 (previously shown as Figure~\ref{fig_UF1_histos_GPA_and_P} for the UF1 stars only). These histograms were
constructed using accumulated gaussian distribution representations for the properties 
of each star. Cumulative probability distributions for the UF1 stars are shown as the smooth red curves that connect points
centered in each bin and are referenced to the right axis labels. Vertical red dashed lines and labels identify
the locations of the three quartile boundaries of the cumulative distributions for the UF1 stars. The similar quartile boundaries for the UF0 stars are shown as the vertical dashed yellow lines.
\label{fig_histos_GPA_and_P}}
\end{figure}

Figure~\ref{fig_histos_GPA_and_P} presents the same histograms of $P^\prime$ and GPA
for the UF1 stars found in the GPIPS-1619 FOV as was previously displayed as 
Figure~\ref{fig_UF1_histos_GPA_and_P}, but here adds histograms in yellow
for the UF0 stars that additionally meet a P$^\prime$SNR~$\ge 3$ criterion. Analysis of the UF0 histograms yields $P^\prime$ quartile boundaries of 1.76, 2.57, and 3.73\% and GPA quartile boundaries of 86.5, 99.2, and 111.6\degr,
which are indicated as vertical, dashed, yellow lines at the quartile locations. These UF0 star polarization properties do not differ greatly from those derived from the larger UF1 set, though the UF0 $P^\prime$ values are biased to greater values through rejection of lesser polarization percentages. 

\subsection{Polarization Signal-to-Noise Distribution Properties}

\vskip 6pt
\begin{figure}
\includegraphics[width=6in, trim= -2in 0in 0in 0in]{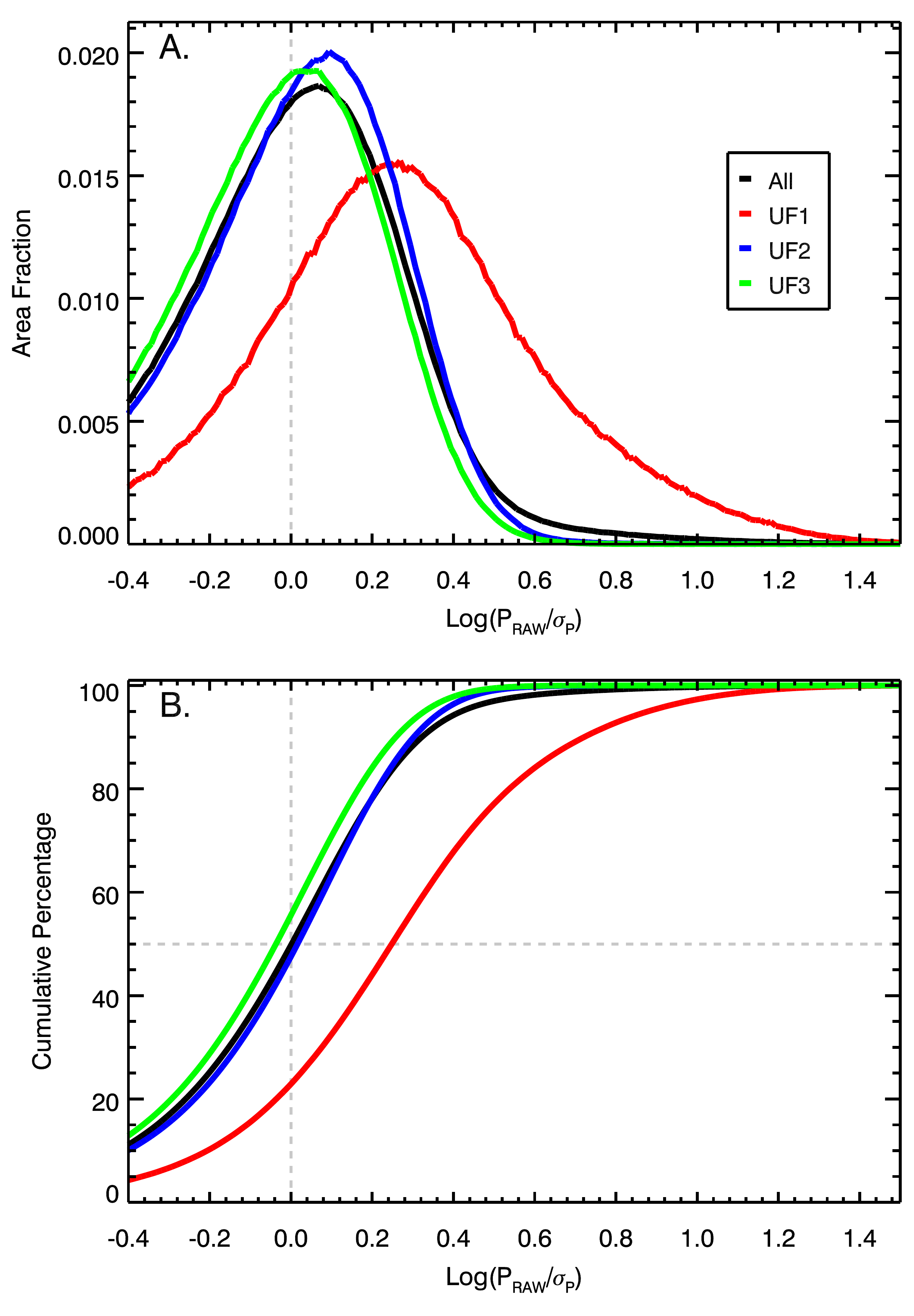}
\caption{Raw polarization percentage signal-to-noise ratio (PSNR $\equiv P_{RAW} / \sigma_P$) distributions of GPIPS POLCAT stars. Horizontal axes show base-10 logarithm of PSNR.
(Top, A-panel) Distributions of PSNR by UF classification subset. Vertical axis is the fractional probability for each GPIPS UF data subset.
The distribution of PSNR for the combination of all GPIPS POLCAT stars is plotted in black. The UF1 distribution is shown in red. The UF2 and UF3 distributions are shown in blue and green, respectively. Vertical dashed line represents unity
PSNR. 
(Bottom, B-panel) Cumulative PSNR distributions, with the same mapping of UF subset type to color. While the UF2 and UF3 stars show median PSNR values near unity, and thus are expected to 
exhibit strong noise bias, the UF1 stars are dominated by significant PSNR ratios.}
\label{fig_PSNR}
\end{figure}

Figure~\ref{fig_PSNR}.A (top panel), shows histograms of all POLCAT stars as a function of raw (not debiased) polarization signal-to-noise (PSNR $\equiv P_{RAW} / \sigma_P$) for the UF1, UF2, UF3 subsets and for all GPIPS stars, as red, blue, green, and black curves, respectively. Figure~\ref{fig_PSNR}.B (lower panel) shows their cumulative distributions. Vertical dashed gray lines indicate where $P_{RAW} / \sigma_P$ is unity. When their raw polarization values are debiased to become $P^\prime$, all stars to the left of the unity lines would become polarization upper limits. Hence, the $P_{RAW} / \sigma_P$ distributions capture better the full range of GPIPS data quality than would the debiased (P$^\prime$SNR) distributions.

In Figure~\ref{fig_PSNR}.A, the UF2, UF3, and combined (``All") GPIPS distributions exhibit peaks that are located somewhat to the right of the PSNR unity line. Their cumulative distributions (B-panel) show that their median values are also close to unity $P_{RAW} / \sigma_P$. That the combined GPIPS distribution follows closely the UF2 and UF3 distributions is because those subsamples dominate in numbers of stars over the UF1 sample (90\% vs 10\%).
The UF1 histogram, however, is shifted to higher PSNR values. Its cumulative distribution (B-panel) shows that 20\% of UF1 stars have $P_{RAW} / \sigma_P$ of unity or less and 80\% of the UF1 stars show $P_{RAW} / \sigma_P$ in excess of unity.

\subsection{Effects of Different {\rm P$^\prime$SNR} Selections}

Next, a
sweep of P$^\prime$SNR criterion choices was performed for the GPIPS-1619 FOV stellar data, followed by
a limited comparison of results for two P$^\prime$SNR criterion choices for the full set of GPIPS FOVs.
The evaluating quantity was the interquartile range WGPA, selected as likely the most 
sensitive to changes in the nature of the polarization properties of the different stellar
subsamples.

\begin{figure}
\includegraphics[width=6.5in]{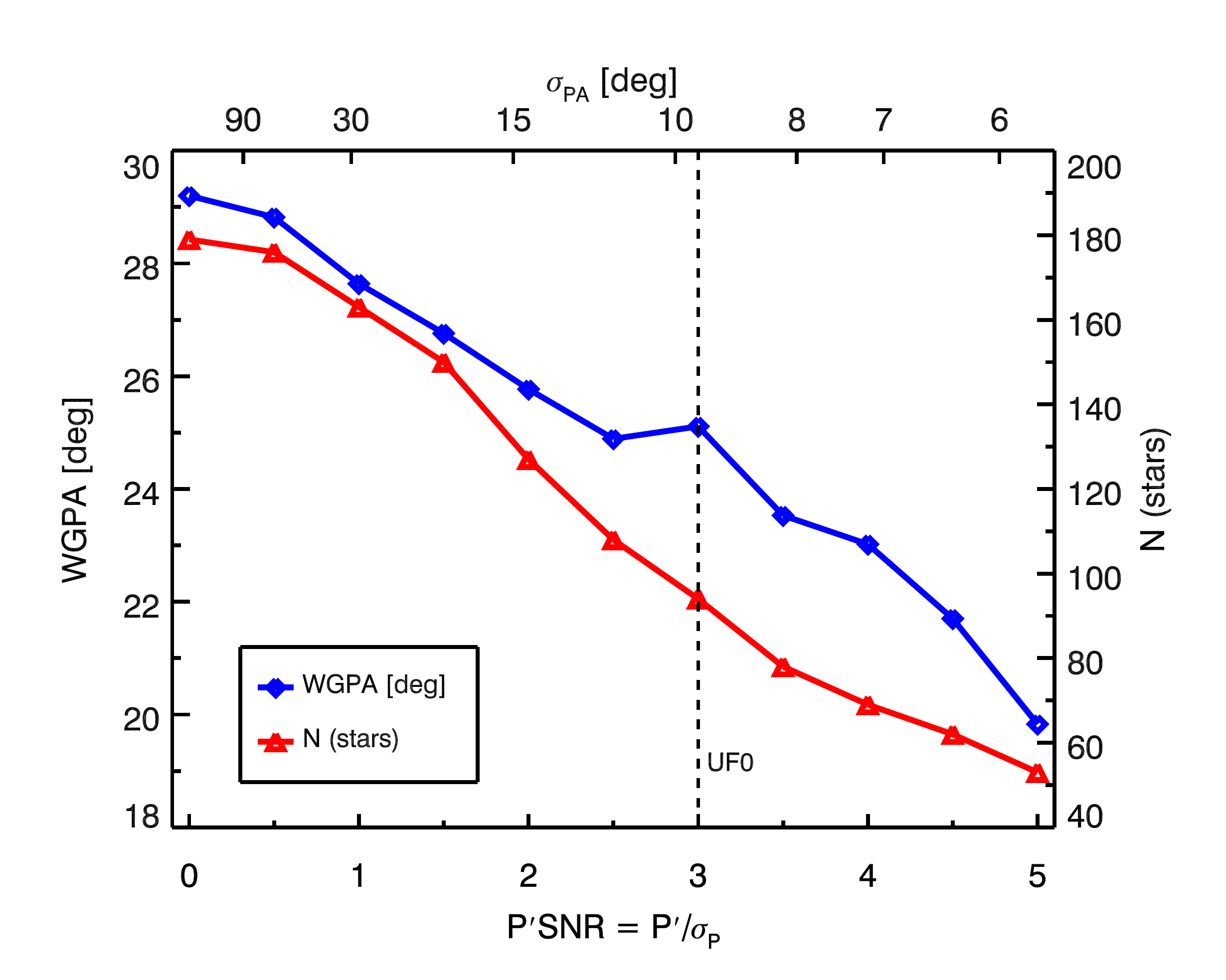}
\caption{Interquartile ranges WGPA, and the numbers of selected stars, versus debiased polarization percentage signal-to-noise P$^\prime$SNR for stars in the GPIPS-1619 FOV. 
WGPA values, computed for P$^\prime$SNR lower limits stepped by 0.5 from zero, are shown as blue diamonds connected by blue line segments and are referenced to the left vertical axis. The numbers of stars in the GPIPS-1619 FOV meeting each P$^\prime$SNR cutoff are shown as the red triangles connected by red line segments and are referenced to the right vertical axis.
The top horizontal axis indicate polarization position angle uncertainties computed from the P$^\prime$SNR limits shown in the bottom axis.}
\label{fig_WGPA_vs_PSNR}
\end{figure}

For the GPIPS-1619 examination, subsets were drawn from the UF1 stars, applying additional P$^\prime$SNR criteria that stepped from zero through the UF0 value of three and on to five. The resulting run of WGPA versus P$^\prime$SNR for the GP1619 FOV is presented in Figure~\ref{fig_WGPA_vs_PSNR}. In the Figure, the blue diamonds and blue connecting lines show the decrease in WGPA values as P$^\prime$SNR is increased. The red triangles and connecting lines show the numbers of stars meeting the P$^\prime$SNR criteria. Both curves fall with increasing P$^\prime$SNR, as fewer stars in the GPIPS-1619 FOV meet the selection criterion. The overall decrease in WGPA, with P$^\prime$SNR ranging from zero to five, is about 32\%\ for this FOV while the number of stars decreases by 70\%. The lower WGPA values might signal greater coherence of the magnetic field in the plane of the sky, but the associated lower star numbers could also be suggesting the introduction of a bias due to, for example, the selected stars only sampling higher extinction lines of sight. Such lines of sight could be limited to only a small, dusty region within the FOV and not representative of the magnetic field properties across the full FOV.

\begin{figure}
\includegraphics[width=7.25in]{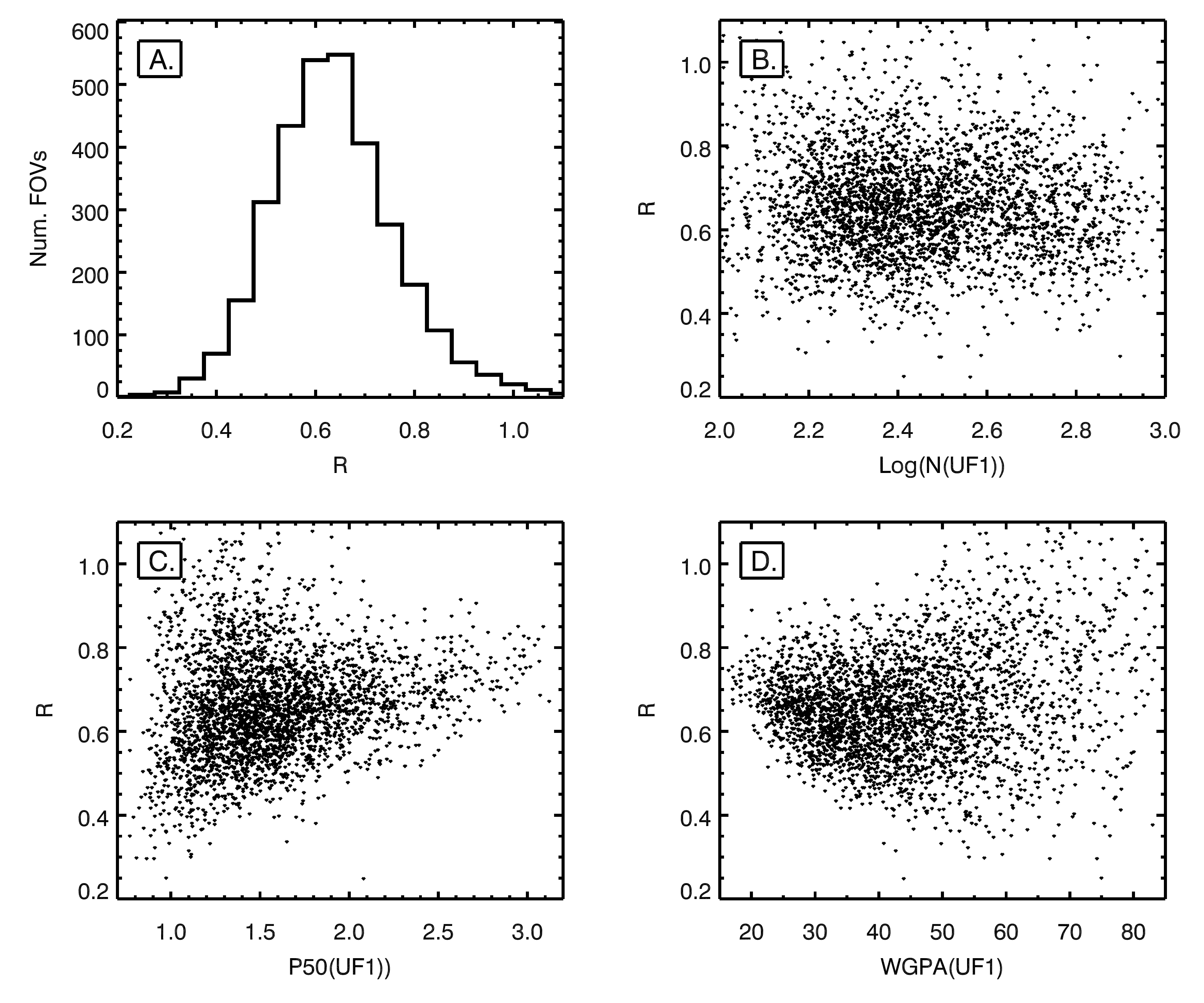}
\caption{Behaviors of the ratio $R$ (WGPA for UF0 stars divided by WGPA for UF1 stars) for each FOV. 
(Top-left, A-panel) Histogram of $R$ values. The mean is 0.66 and the standard deviation is 0.12.
(Top-right, B-panel) $R$ as a function of log(N) of the number of UF1 stars in each FOV.
(Bottom-left, C-panel) $R$ as a function of the median debiased polarization percentage P50 for UF1 stars in each FOV.
(Bottom-right, D-panel) $R$ as a function of WGPA for the UF1 stars in each FOV.
There are very few FOVs with low-$R$ and high-P50 or with low-$R$ and low-WGPA.}
\label{fig_R}
\end{figure}

To uncover the nature of possible biases introduced by the addition of a P$^\prime$SNR criterion to the UF1 selection criteria, a representative ratio $R$ was developed. It was computed from the ratio of the WGPA value found when selecting a stellar subsample of UF1 stars using the UF0 criterion (P$^\prime$SNR~$\ge$~3) to the WGPA value for the UF1 stars alone (e.g., $R \equiv \rm{WGPA}_{UF0} / \rm{WGPA}_{UF1}$). Figure~\ref{fig_R}.A (upper left) shows the histogram of $R$ values obtained for the set of all 3,237 GPIPS FOVs. There is a well-defined peak, near the median, with a mean of 0.66 and standard deviation of 0.12. Figure~\ref{fig_R}.B (upper right) shows the run of $R$ values with the log of the number of UF1 stars in each GPIPS FOV. No trend is apparent. 

The similar plots of $R$ versus P50 (Figure~\ref{fig_R}.C) and WGPA (Figure~\ref{fig_R}.D) show a couple of possible trends. 
In the $R$ versus P50 Figure~\ref{fig_R} C-panel, there is an absence of low $R$ values where P50 values are high, which suggests that UF1 and UF0 stars in such fields are equally sampling the magnetic field orientations there. In the $R$ versus WGPA D-panel, there is also an absence of low $R$ values where WGPA values are small. The same explanation likely applies, that in regions where the magnetic field is more coherent, and WGPA values are therefore small, UF1 values are not greatly different from UF0 ones.

Where $R$ has the greatest spread include low-P50 and high-WGPA FOVs. In those FOVs,
$R$ values can be as small as 0.2, likely because the UF0 GPA histogram has collapsed due
to having too few constituent stars per FOV. Some $R$ values for low-P50 and high-WGPA FOVs exceed unity,
again likely signifying too few UF0 stars are present with which to form meaningful GPA 
histograms. While UF0 stars, and UF1 stars, are both fine for sampling FOVs where P50 is
high and WGPA is low, UF1 stars appear to be better for sampling the low-P50 and/or
high-WGPA FOVs.

\vskip 6pt

The greatest confidence polarization properties will be obtained by selecting the greatest P$^\prime$SNR values, such as those of the P$^\prime$SNR~$\ge~3$ UF0 subset. However, even across the 76~deg$^2$ of the inner Galactic plane surveyed by GPIPS, such stars are relatively rare.
The UF0 stars account for only 2.8\% of the stars measured for polarization and sample the survey
area at 70~arcsec mean separation. Relaxing the P$^\prime$SNR criterion admits some low confidence polarization values, but does not appear to introduce strong log(N), P50, GPA, or WGPA biases relative to the UF0 subset and may avoid possible problems associated with the UF0
stars in the low-P50 FOVs. The UF1 WGPA values are greater than the UF0 ones, but by a mostly
predictable ratio. Also, electing to characterize the Galactic disk magnetic field properties via the UF1 subsample increases the stellar probe numbers by more than a factor of four, leading to halving of
the mean angular sampling separation. For the purposes of the FOV-based median comparisons of polarization (magnetic field orientation) 
properties offered in the text, the UF1 choice seems appropriate. User-based studies must 
carefully assess their own selection criteria against their specific science goals.

The overall conclusion is that the UF1 selection criteria result in FOV-based samples of stars with measured polarization properties that are able to probe and characterize magnetic field orientations adequately and without the possible biases (especially against low-P50/high-WGPA FOVs)  introduced when adding additional P$^\prime$SNR cutoffs.

\section{Alternate Displays for Figures 12 - 15}\label{Appendix_CBLUT}

This section offers representations of Figures~\ref{fig_N_LB} through \ref{fig_WGPA_LB} using color lookup tables better suited to individuals with color vision deficiency (CVD) and to enable others to distinguish finer details than presented in those Figures in the text.

\begin{figure}
\includegraphics[width=7.25in]{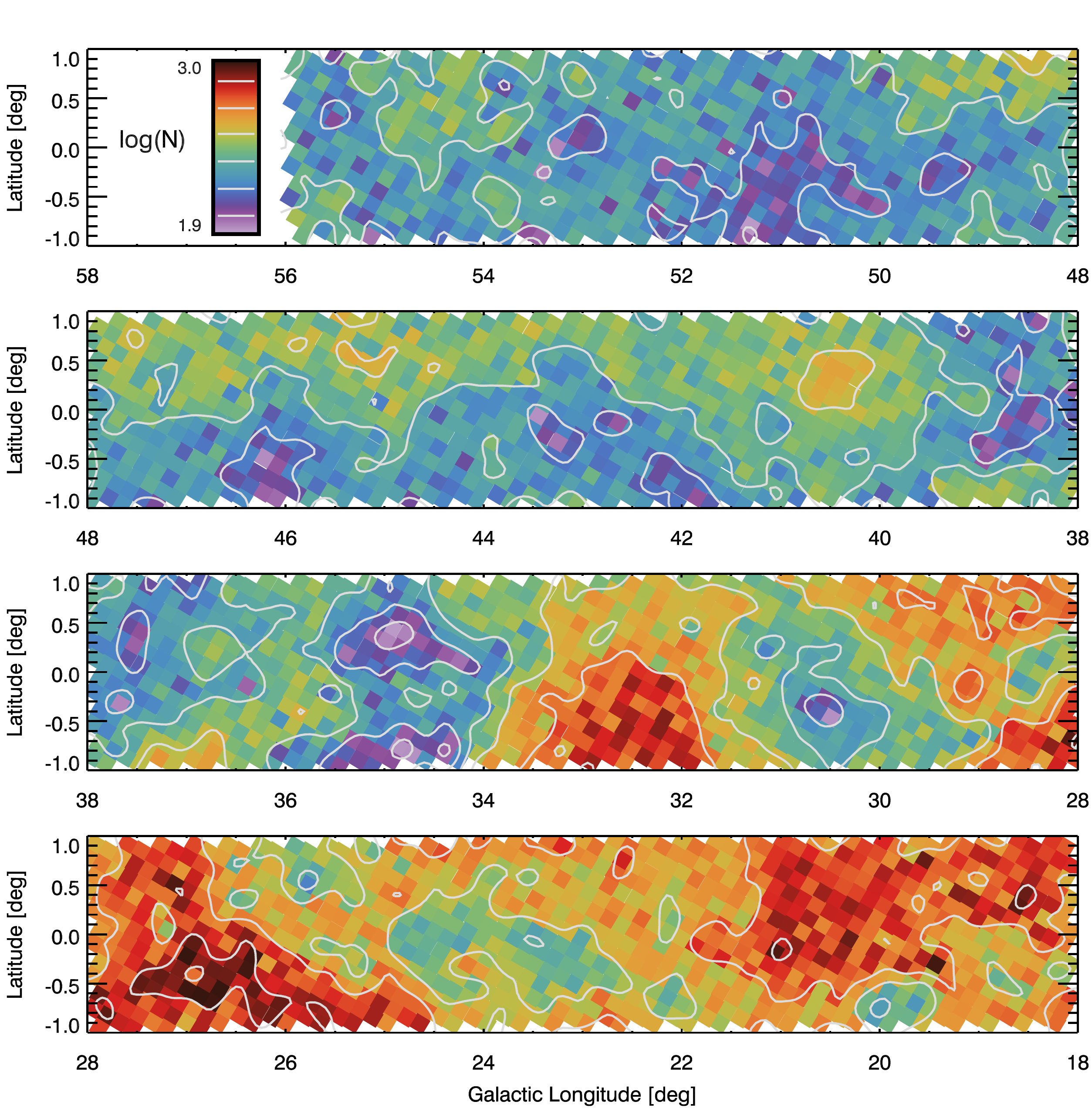}
\caption{Representation of Figure~\ref{fig_N_LB}: False-color representation of the 
base-10 log of the number of UF1 stars in each
of the GPIPS FOVs, versus Galactic longitude and latitude, as four longitude slices. 
Color look-up rectangle, with six nearly equally spaced gray contours, corresponding to 100, 150, 225, 340, 500, and 750 UF1 stars
per FOV, is shown in the top slice. 
The smallest number of UF1 stars in any FOV is 64, the largest is 1,112. 
Note that extinction and the Galactic bulge both likely play roles in the variations in the numbers of 
detected stars with direction.
\label{fig_D5}}
\end{figure}

\begin{figure}
\includegraphics[width=7.25in]{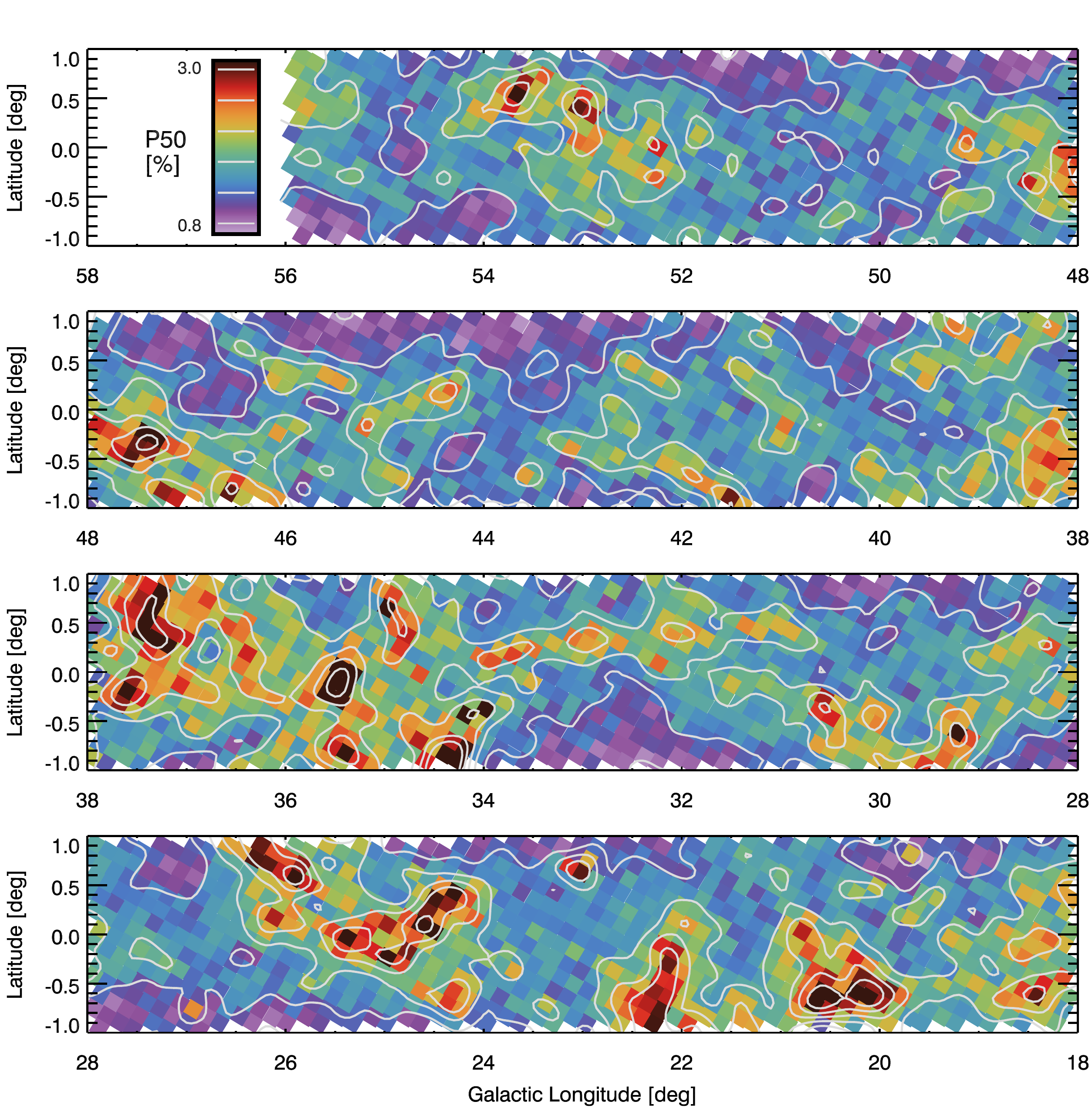}
\caption{Representation of Figure~\ref{fig_P_LB}: Galactic longitude and latitude distribution of the median percentage polarization P50 in each GPIPS FOV. 
Polarization medians range from around 0.8\% to over 3\% in the $H$-band,
with a weak tendency to be higher along the disk midplane. Six gray contour 
levels corresponding to P50 values of 0.9, 1.3, 1.7, 2.1, 2.5, 
and 2.9\% are drawn over the false color panels and inside the color look-up rectangle in the top panel.
\label{fig_D6}}
\end{figure}

\begin{figure}
\includegraphics[width=7.25in]{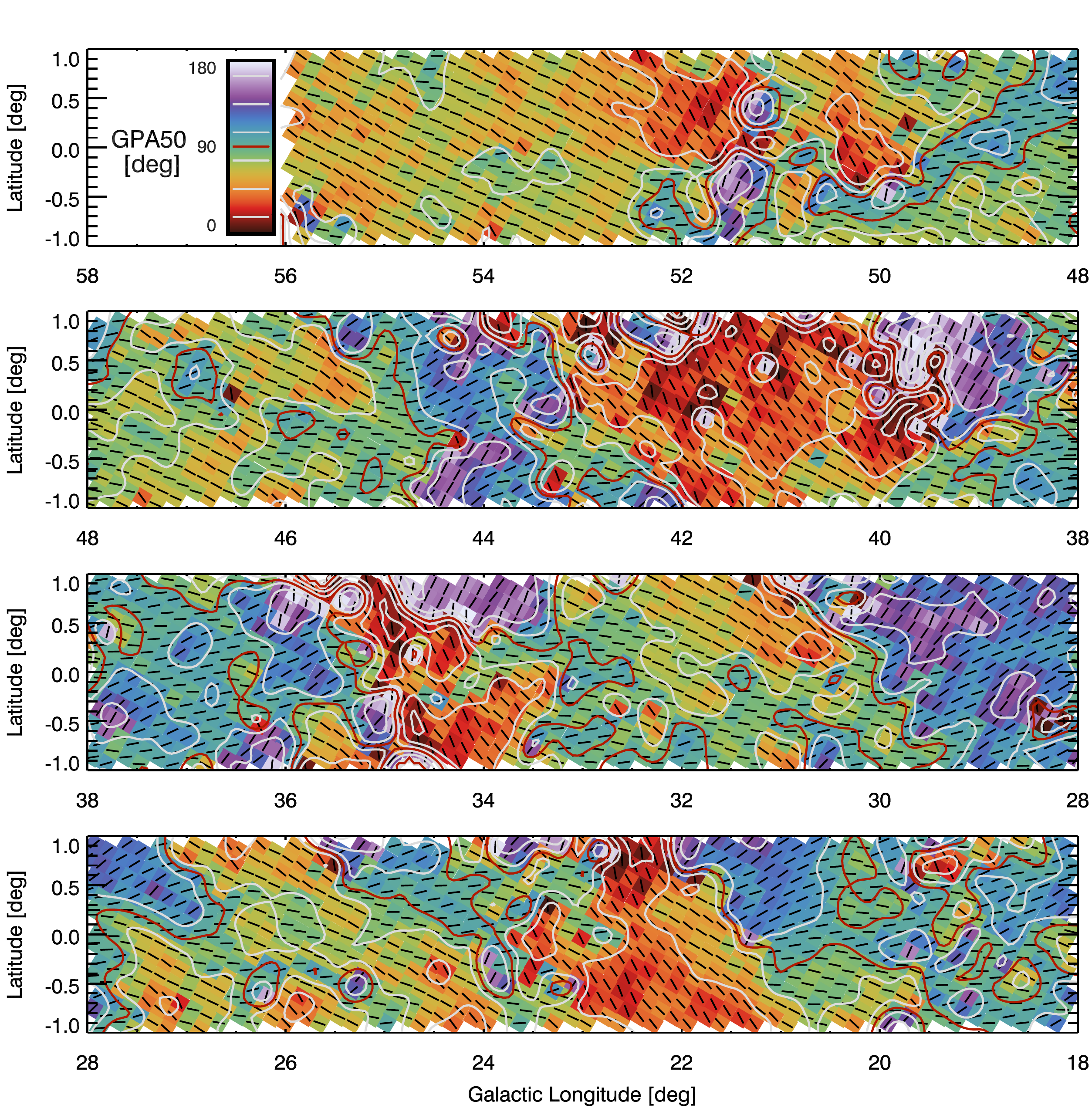}
\caption{Representation of Figure~\ref{fig_GPA_LB}: Galactic longitude and latitude distribution of GPA50 in each GPIPS FOV
in the same four-slice presentation as in the previous Figures. 
Unlike in Figure~\ref{fig_GPA_LB},
look-up table colors here do not wrap around 0\degr\ and 180\degr, so aliasing is more prominent seen. 
Black lines within each FOV box indicate the GPA orientation. Six gray contour 
levels, corresponding to GPA50 values of 15, 45, 75, 105, 135, 
and 165\degr, and one red contour, corresponding to the disk-parallel GPA50 value of 90\degr,
are drawn over the false color panels and inside the color look-up rectangle in the top panel.
\label{fig_D7}}
\end{figure}

\begin{figure}
\includegraphics[width=7.25in]{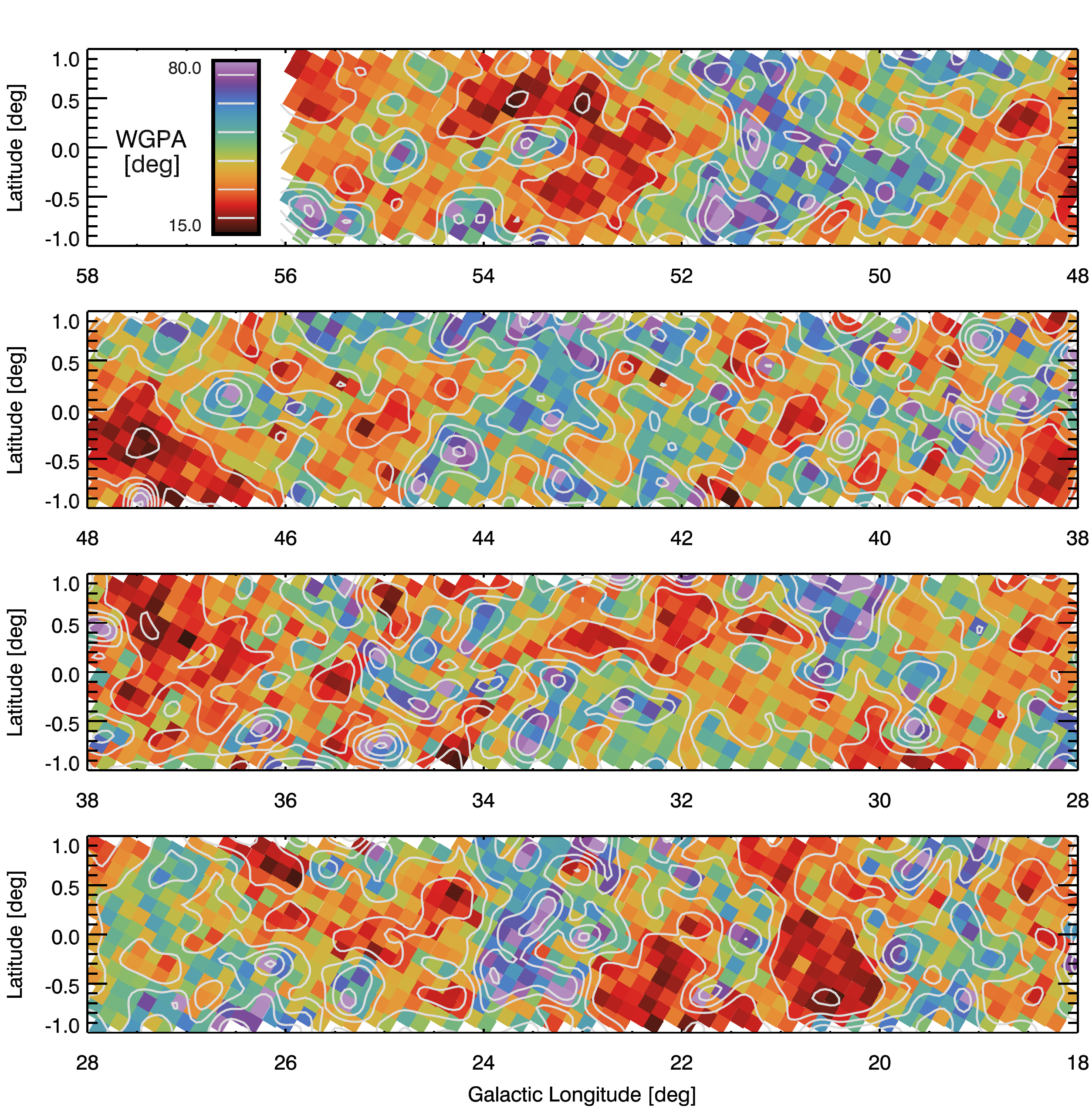}
\caption{Representation of Figure~\ref{fig_WGPA_LB}: Galactic longitude and latitude distribution of WGPA, the interquartile ranges of the
GPA distributions in each of the GPIPS FOVs. The color look-up rectangle in
the uppermost panel indicates that WGPA values range from a low of about
15\degr\ (red colors) to a high of about 80\degr\ (dark blue colors). Six gray contour 
levels corresponding to WGPA values of 20, 31, 42, 54, 64, 
and 75\degr\ are drawn over the false color panels and inside the color look-up rectangle in the top panel. Note that the value to color mapping in this Figure is inverted from previous Figures to highlight 
the low-WGPA regions in darker red colors.
\label{fig_D8}}
\end{figure}

Production of Figures~\ref{fig_D5} through \ref{fig_D8} was based on concepts described on website of Paul~Tol (https://personal.sron.nl/$\sim$pault/\#fig:scheme\_rainbow\_discrete). In particular, the 34 discretes RGB triplet values shown in Figure~19 on that website were copied and interpolated to 245 levels to create a rainbow-like color look up table (CLUT).  New versions of Figures~\ref{fig_N_LB} through \ref{fig_WGPA_LB} were constructed using this CLUT (with stretches and inversions as needed to agree best with the original Figure versions). To evaluate the effectiveness of the new CLUT, side-by-side comparisons of the new figures with the old figures were performed using the Color Oracle (https://colororacle.org/index.html) tool set to Deuteranopia, which modifies displayed colors to simulate that most common CVD. The new CLUT was judged to do better at retaining discrimination of displayed value to color than was done by the original CLUT. The reproductions of the Figure~\ref{fig_N_LB} through
\ref{fig_WGPA_LB} using the new, CVD-friendly CLUT are shown in the following.

\vskip 6pt

\vskip 6pt
\clearpage

\vskip 6pt

\vskip 6pt

\end{document}